\documentclass[aps,prd,groupedaddress,showpacs]{revtex4}
\usepackage{epsf,epsfig,graphics,graphicx,amssymb,amsthm}

\begin{document}

\title{New results on vacuum fluctuations: Accelerated detector versus inertial detector in a quantum field}

\author{I-Chin Wang}

\affiliation{
Institute of Physics, Academia Sinica, Taipei 11529, Taiwan\\
}

\begin{abstract}

    We investigate the interaction between a moving detector and a quantum field, especially about how the
    trajectory of the detector would affect the vacuum fluctuations when the detector is moves in a quantum field (the Unruh effect).
    We focus on two moving detectors system for future application in quantum teleportation.
    We find that the trajectory of a uniformly accelerated detector in Rindler space cannot be extended to
    a trajectory in which a detector moves at constant velocity. Based on our previous work, we redo the calculations
    and find that a term is missing from the past calculations, and we also find that there are some restrictions on the
    values for the parameters in the solutions. In addition, without inclusion of the missing term, the variance from the quantum field for
    the inertial detector will be zero and is unlikely in such a system. When all these points are combined, there is a difference in the two-point
    correlation function between the inertial detector and the accelerated detector in the early-time region.
    The influence of proper acceleration can be seen in the two-point correlation functions.
    This might play a role in the quantum teleportation process and be worth studying thoroughly. \\

\end{abstract}

 \maketitle

\date{\today}
%{\today}

%%%%%%%%%%%%%%%%%%%%%%%%%%%%%%%%%%%%%%%%%%%%%%%%%%%%%%%%%%%%%%%%%%%%%%%%%%%%%%%%
\section{Introduction}
%%%%%%%%%%%%%%%%%%%%%%%%%%%%%%%%%%%%%%%%%%%%%%%%%%%%%%%%%%%%%%%%%%%%%%%%%%%%%%%%

The Unruh effect was originally proposed for the study of Hawking
radiation near a black hole~\cite{Un76}, and it was found that an
uniformly accelerated detector would experience a thermal bath at
temperature $T_{U}=\frac{\hbar a}{2\pi c k_{B}}$, where $a$ is the
proper acceleration. It involves the interaction between the
background quantum field and a moving detector which has constant
acceleration. It is also known that the accelerating charges emit
radiation. In the
literature~\cite{Grove86,RSG91,Un92,MPB93,T86GF88,H93,AM94,MP96},
many physicists were interested in learning whether there were any
differences between the Unruh effect and the radiation from
accelerated charges in the quantum field; for example, was the
emitted radiation part of the energy flux in the Unruh effect?
Later, this question was extended to the atom system and to whether
an accelerated atom emits radiated energy? What was the connection
to the Unruh effect? Hu and his collaborators worked on the minimal
coupling model~\cite{HR00,R96,RHA96,RHK97} and considered what roles
the equilibrium condition and nonequilibrium condition play in the
accelerated detector~\cite{JH2002,HR00,R96,RHK97}. In recent years,
the kinds of influences between the moving detector and a background
quantum filed have been applied to certain quantum teleportation
processes. However, the difference between an inertial trajectory
and the uniformly accelerated trajectory for a moving detector is
not obvious enough. If we could clearly see the effect about proper
acceleration, it would be helpful to the understanding of some
important systems, for example, the atomic optical and particle
quantum filed systems.

Based on the work in Refs.~\cite{HR00,R96,RHA96,RHK97,JH2002,Lin03},
 we follow here the work of Lin and co-workers~\cite{LH2005,LCH2008,LCH2015}, and we recheck the computation of
a uniformly accelerated detector(UAD)~\cite{LH2005}. It was
originally thought that the solution of a UAD~\cite{LH2005} could be
applied to the inertial detector case directly by taking the limit
such that the proper acceleration $a\rightarrow 0$ (i.e. an inertial
detector moves at a constant velocity and therefore has zero proper
acceleration). However, when we check past results for two-point
correlation functions for a UAD, we find that the previous solution
cannot be applied to the inertial detector case by taking the limit
$a\rightarrow 0$. As an inertial detector moves at constant
velocity, we need to apply a real inertial trajectory. In fact, the
original goal for our work was to apply the previous
results~\cite{LH2005,LCH2008} in  certain quantum teleportation
processes, but some errors occurred when we did it, so we have to
recheck our setup and calculations. Therefore, we start from the
beginning, where we apply a real inertial trajectory and another
uniformly accelerating trajectory for the moving detector, solving
the solutions for these two trajectories and then comparing the
difference on the two-point correlation functions between the
inertial and
uniformly accelerated detectors.\\

We find that a term is missing from previous two-point correlation
functions and that it is about the vacuum fluctuations of the moving
detector. Without this term, we would have a zero variance from the
quantum field for the inertial detector . This was not noticed
before. When we include this term, the strange zero variance issue
disappears and the values of the two-point correlation functions
also change. This change makes the difference between the inertial
and accelerated detector more clear than in the previous results.
Since we apply perturbations to solve the equations, assumptions
about the perturbation method also set an allowed region for the
values of the parameters in the solutions. This restriction also
affects the solutions for the two-point correlation functions. When
these considerations are included, the difference between the
inertial
detector and accelerated detector is more obvious than ever.\\

To apply our present result to future work about Quantum
teleportation, we consider two moving Unruh-DeWitt detectors in our
present model~\cite{LCH2008}: Alice and Bob. We assume that Alice is
static in space and Bob is moving in space. Bob's worldline has two
different choices: one is the trajectory for a uniformly
acceleration, while the other is for a constant velocity motion. For
simplicity, we focus here on the moving detector Bob and study the
interaction between the quantum field and the internal degrees of
freedom $Q$ for detector Bob. We compute the two-point functions
$\langle QQ \rangle_{v}$ and $\langle \dot{Q}\dot{Q} \rangle_{v}$
for the two different trajectories(i.e.,s$\langle QQ \rangle_{v}$ is
the two-point function
 vacuum fluctuation of the internal degrees of freedom for Bob),
and we then compare the plots of $\langle QQ \rangle_{v}$ and
$\langle \dot{Q}\dot{Q} \rangle_{v}$ for the inertial and uniformly
accelerating detectors. We find that these two different types of
detectors have different effects on the curves for the two-point
functions $\langle QQ \rangle_{v}$ and $\langle \dot{Q}\dot{Q}
\rangle_{v}$ in the early-time region. Meanwhile, we also write
detailed calculations and point out some key points in the
calculations about obtaining $\langle
QQ \rangle_{v}$ and $\langle \dot{Q}\dot{Q} \rangle_{v}$.\\

To demonstrate the importance of the allowed region on the values of
the parameters in this model, we choose some improper values for the
parameters in the model and calculate the two-point correlation
functions. Improper values would lead to different trends for the
two-point functions $\langle QQ \rangle_{v}$ and make the Unruh
effect unclear.\\

This paper is organized as follows. In Sec. II we set up and
introduce the model and the method, and some detailed derivations
are placed in the appendixes. In Sec. III we focus on one detector
and investigate the moving detector Bob in a quantum field and
whether it moves at a constant acceleration or a constant velocity.
We solve the solutions for two different trajectories for the
detector Bob and compute the two-point correlation functions of
internal degree of freedom $Q$ of Bob. We then discuss the allowed
values for the parameters in the solutions and do the numerical
plots for the two-point functions of $Q$. Later, we compare the
plots and determine the difference between the inertial detector and
accelerated detector. Sec. IV is the summary.

%%%%%%%%%%%%%%%%%%%%%%%%%%%%%%%%%%%%%%%%%%%%%%%%%%%%%%%%%%%%%%%%%%%%%%%%%%%%%%%
\section{Model}\label{Model}
%%%%%%%%%%%%%%%%%%%%%%%%%%%%%%%%%%%%%%%%%%%%%%%%%%%%%%%%%%%%%%%%%%%%%%%%%%%%%%%

We consider two Unruh-DeWitt detectors Alice and Bob that are at
different spatial points and in different states of motion. Each
detector has an internal degree of freedom $Q$ that interacts with a
common scalar field $\Phi$. Assuming that Alice is static and Bob is
moving (Bob could be uniformly accelerated or could move at a
constant velocity; we will calculate the solutions for these two
cases later). The trajectories for Alice and Bob are
$z^{\mu}_{A}(t)$ and $ z^{\mu}_{B}(\tau)$, respectively. The action
for this setup is as follows:
\begin{eqnarray}
S&=&-\int d^{4}x\sqrt{-g}\frac{1}{2}\partial_\mu
\Phi(x)\partial^{\mu} \Phi(x)+\int d\tau_{A}\,\, {m_{0}\over
2}\left[\left(\partial_{A}Q_{A}\right)^2
    -\Omega_{0}^2 Q_{A}^2\right]+\int d\tau_{B}\,\, {m_{0}\over
2}\left[\left(\partial_{B}Q_{B}\right)^2
    -\Omega_{0}^2 Q_{B}^2\right] \nonumber\\ & &
    + \lambda_0 \int d^4 x \int dt \,\,
    Q_{A}(t)\Phi (x)\delta^4\left(x^{\mu}-z_{A}^{\mu}(t)\right)+ \lambda_0 \int d^4 x \int d\tau \,\,
    Q_{B}(\tau)\Phi
    (x)\delta^4\left(x^{\mu}-z_{B}^{\mu}(\tau)\right),\nonumber\\
  \label{Stot1}
\end{eqnarray}
where $Q_{A}$ and $Q_{B}$ are the internal degrees of freedom for
the detectors Alice and Bob. They are assumed to be two identical
harmonic oscillators with mass $m_{0}=1$ with the bare natural frequency $\Omega_{0}$\cite{LCH2008}. \\
If we assume that the coupling between the detectors and the field
is turned on at the moment when $t=\tau=0 $ ($t$ is the proper time
for Alice and $\tau$ is the proper time for Bob), the state of this
combined system is a direct product of a quantum state
$|q_{A},q_{B}\rangle$ for Alice's and Bob's detectors $Q_{A}$ and
$Q_{B}$ and Minkowski vacuum $|0_{M}\rangle$ for the field $\Phi$,
\begin{equation}
  \left|\right. \psi(0)\left.\right> =
  \left|\right. q^{}_A, q^{}_B\left.\right> \otimes
  \left|\right. 0_M \left.\right>. \label{initstat}
\end{equation}

Here $\left|\right. q^{}_A, q^{}_B \left.\right>$ is taken to be a
squeezed Gaussian state with minimal uncertainty, represented in the
Wigner function as
\begin{eqnarray}
  \rho(Q_A,P_A,Q_B,P_B) & = & \exp -{1\over
  8}\left[
  {\beta^2\over\hbar^2}\left( Q_A + Q_B\right)^2 +
  {1\over \alpha^2}\left( Q_A - Q_B\right)^2 \right.\nonumber\\ &  &
  +\left.
  {\alpha^2\over\hbar^2}\left( P_A - P_B\right)^2 +
  {1\over \beta^2}\left( P_A + P_B\right)^2 \right],
\label{initGauss}
\end{eqnarray}
where $Q_A$ and $Q_B$ can be entangled by properly choosing the
parameters $\alpha$ and $\beta$.

After quantizing the field $\Phi$ and the internal degrees of
freedom $Q_{A}$, $Q_{B}$ in the Heisenberg picture(as shown in
Appendix A), the mode functions to the  first order $O(\lambda_{0})$
for $\Phi$, $Q_{A}$ and $Q_{B}$ are as follows:
\begin{eqnarray}
    \left( \partial_{\tau_i}^2 + \Omega_0^2\right)q_i^{(j)}(\tau_i) &=&
      \lambda_0 f^{(j)}(z_i^\mu (\tau_i)), \label{eomqAB1}\\
    \left( \partial_t^2 - \nabla^2 \right)f^{(j)}(x) &=& \lambda_0
      \left[\int_0^{\infty} dt\, q_A^{(j)}\delta^4(x -z^{}_A(t))
     +\int_0^{\infty} d\tau\, q_B^{(j)}\delta^4 (x-z^{}_B(\tau)) \right],
     \label{feAB1}\\
  \left(\partial_{\tau_i}^2 + \Omega_0^2\right)q_i^{(+)}(\tau_i,{\bf k}) &=&
      \lambda_0 f^{(+)}(z_i^\mu(\tau_i), {\bf k}), \label{eomq+1} \\
    \left( \partial_t^2 - \nabla^2 \right)f^{(+)}(x,{\bf k}) &=& \lambda_0
      \left[\int_0^{\infty} dt\, q_A^{(+)}(t,{\bf k})\delta^4(x
      -z^{}_A(t))\right.\nonumber\\
     & & \left.+\int_0^{\infty}d\tau\,q_B^{(+)}(\tau,{\bf k})\delta^4(x-z^{}_B(\tau))
     \right] \label{fe+1} .
\end{eqnarray}
In future work, we would like to study issues pertaining to quantum
teleportation after we obtain the solutions for $\hat{Q_{A}}$ and
$\hat{Q_{B}}$ in this model. In this work, for simplicity and a
clear picture, we will first look solely at the two-point functions
of the internal degrees of freedom $Q$ of the moving detector Bob.
We will consider two different kinds of trajectories for Bob:(i) Bob
is uniformly accelerated and (ii) Bob moves at a constant velocity.
By calculating the solutions and the two-point correlation functions
for $Q$ of the detector Bob under these two types of trajectories,
we will understand the features of acceleration and inertial motion
and be able to apply these results to future applications.

%%%%%%%%%%%%%%%%%%%%%%%%%%%%%%%%%%%%%%%%%%%%%%%%%%%%%%%%%%%%%%%%%%%%%%%%%%%%%%%%%%
\section{TWO-POINT FUNCTIONS OF THE INTERNAL DEGREES OF FREEDOM $Q$ FOR A MOVING DETECTOR}
%%%%%%%%%%%%%%%%%%%%%%%%%%%%%%%%%%%%%%%%%%%%%%%%%%%%%%%%%%%%%%%%%%%%%%%%%%%%%%%%%%

We now focus on the moving detector Bob. For simplicity, we consider
only the moving detector Bob and temporarily ignore the static
detector Alice in the action $S$ in Eq.~(\ref{Stot1}). The only
action that has a $Q_{B}$ part is then the following

\begin{eqnarray}
S&=&-\int d^{4}x\sqrt{-g}\frac{1}{2}\partial_\mu
\Phi(x)\partial^{\mu} \Phi(x)+\int d\tau_{B}\,\, {m_{0}\over
2}\left[\left(\partial_{B}Q_{B}\right)^2
    -\Omega_{0}^2 Q_{B}^2\right] \nonumber\\ & &
    + \lambda_0 \int d^4 x \int d\tau_{B} \,\,
    Q_{B}(\tau_{B})\Phi
    (x)\delta^4\left(x^{\mu}-z_{B}^{\mu}(\tau_{B})\right).
  \label{Stot1B}
\end{eqnarray}

The Heisenberg equations for the operators and the fields are
written in \cite{LH2005}(we take $\hat{Q}_{B}=\hat{Q}$ from now on,
and we also take $m_{0}=1$ in a later
 numerical calculation) are shown in Appendix B(including the solutions of the mode functions and
 the definitions of the states for the quantum field $\Phi$ and the internal degrees of freedom $\hat{Q}$). Later, we start
 with the two-point correlations function $\langle
Q(\tau-\tau_{0})Q(\tau''-\tau''_{0})\rangle_{v}$ for the moving
detector Bob.

%%%%%%%%%%%%%%%%%%%%%%%%%%%%%%%%%%%%%%%%%%%%%%%%%%%%%%%%%%%%%%%%%%%%%%%%%%%%%%%%%%%%%%%%%%%%%%%%%%%%%%
\subsection*{A. Trajectory 1: Two-point function for UAD}\label{traj1}
%%%%%%%%%%%%%%%%%%%%%%%%%%%%%%%%%%%%%%%%%%%%%%%%%%%%%%%%%%%%%%%%%%%%%%%%%%%%%%%%%%%%%%%%%%%%%%%%%%%%%%

The two-point correlation function $\langle
Q(\tau-\tau_{0})Q(\tau''-\tau''_{0})\rangle_{v}$ of the internal
degrees of freedom for the UAD Bob, which is along the trajectory
$z^{\mu}_{B}=(a^{-1}\sinh a\tau, a^{-1}\cosh a\tau, 0, 0)$ with
$a\neq0$, is as follows:

\begin{eqnarray}
& &\langle Q(\tau-\tau_{0})Q(\tau''-\tau''_{0})\rangle_{v}\nonumber\\
&=&\hbar\int^{\infty}_{-\infty}\frac{d^{3}k}{(2\pi)^{3}2\omega}q^{(+)}(\tau;\textbf{k})q^{(-)}(\tau;\textbf{k})\nonumber\\
&=&\frac{\hbar}{2\omega}\int
\frac{d^{3}\vec{k}}{(2\pi)^{3}}\frac{\lambda_{0}}{m_{0}}\sum_{j=+,-}\int_{\tau_{0}}^{\tau}d\tau'c_{j}e^{w_{j}(\tau-\tau')}f_{0}^{(+)}(z(\tau'),\vec{k})\nonumber\\
& &
\cdot\frac{\lambda_{0}}{m_{0}}\sum_{j=+,-}\int_{\tau_{0}}^{\tau''}d\tau'''c^{\ast}_{j'}e^{w^{\ast}_{j'}(\tau''-\tau''')}f^{\ast(+)}_{0}(z(\tau'''),\vec{k})\,,
\end{eqnarray}.

The mode functions $q^{(\pm)}(\tau;{\bf k})$ for $\hat{Q}_{v}(\tau)$
[which are obtained in Eq.~(\ref{Qv}) in Appendix B ]are
\begin{equation}
  q^{(\pm)}(\tau;{\bf k}) = {\lambda_0\over m_0 }\sum_{j=+,-}
  \int^\tau_{\tau_0} d\tau' c_j e^{w_j(\tau-\tau')}
  f^{(\pm)}_0(z(\tau');{\bf k}).\label{q+1}
\end{equation}

When a Fourier Transform of $f^{(+)}_{0}$ is performed,
\begin{equation}
f_{0}^{(+)}(z(\tau'),\vec{k})=\int d\kappa
e^{-i\kappa\tau'}\varphi_{\vec{k}}(\kappa)\,,
\end{equation}

the above two-point function is expressed as
\\

\begin{eqnarray}
& &\langle Q(\tau-\tau_{0})Q(\tau''-\tau''_{0})\rangle_{v}\nonumber\\
&=&\frac{\lambda_{0}^{2}}{m_{0}^{2}}\sum_{j,j'=+,-}\int_{\tau_{0}}^{\tau}d\tau'c_{j}e^{w_{j}(\tau-\tau')}\int
d\kappa
e^{-i\kappa\tau'}\int_{\tau''_{0}}^{\tau''}d\tau'''c^{\ast}_{j'}e^{w^{\ast}_{j'}(\tau''-\tau''')}\int
d\kappa' e^{-i\kappa'\tau'''}\nonumber\\
& &\cdot F\,,\label{2p}
\end{eqnarray}
where $F$ is defined as
\begin{equation}
F\equiv\frac{\hbar}{2\omega}\int_{-\infty}^{\infty}\frac{d^{3}k}{2\pi}\varphi_{\vec{k}}(\kappa)\varphi^{\ast}_{\vec{k}}(\kappa')\,.
\end{equation}

The Fourier factor $\varphi_{\vec{k}}(\kappa)$ is
\begin{equation}
\varphi_{\vec{k}}(\kappa)=\int^{\infty}_{-\infty}\frac{d\tau}{2\pi}e^{-i\omega
z^{0}(\tau)+i\vec{k}\cdot\vec{z}(\tau)}\,,
\end{equation}
and  $F$ is then in the following form:
\begin{eqnarray}
F
&=&\frac{\hbar}{(2\pi)^{3}}\int_{0}^{2\pi}d\phi\int_{-1}^{1}d(\cos\theta)\int_{0}^{\infty}\frac{\omega^{2}d\omega}{2\omega}\int_{-\infty}^{+\infty}\frac{dt}{2\pi}\int_{-\infty}^{+\infty}\frac{dt'}{2\pi}
e^{i\kappa
t-i\kappa't'-i\omega(z^{0}(t)-z^{0}(t')+i\omega\cos\theta|\vec{z}-\vec{z'}|)}\nonumber\\
&=&\frac{\hbar}{2\pi}\int_{0}^{\infty}
d\omega\int_{-\infty}^{+\infty}
\frac{dt}{2\pi}\int_{-\infty}^{+\infty}\frac{dt'}{2\pi}e^{i\kappa
t-i\kappa't'-i\omega\left[z^{0}(t)-z^{0}(t')\right]}\frac{\sin(\omega|\vec{z}(t)-\vec{z}(t')|)}{|\vec{z}(t)-\vec{z}(t')|}
\nonumber\\
&=&\frac{\hbar}{(2\pi)^{4}}\lim_{\epsilon\rightarrow
0}\int_{-\infty}^{+\infty} dt\int_{-\infty}^{+\infty}
dt'\frac{e^{i\kappa(t-\frac{i\epsilon}{2})-i\kappa'(t'+\frac{i\epsilon}{2})}}{|\vec{z}(t-\frac{i\epsilon}{2})-\vec{z}(t'+\frac{i\epsilon}{2})|^{2}-\left[z^{0}(t-\frac{i\epsilon}{2})-z^{0}(t+\frac{i\epsilon}{2})\right]^{2}}\nonumber\\
&=&\frac{\hbar}{(2\pi)^{4}}\lim_{\epsilon\rightarrow
0}\int_{-\infty}^{+\infty} dt\int_{-\infty}^{+\infty}
dt'\frac{a^{2}e^{\frac{\epsilon}{2}(\kappa+\kappa')+i\kappa
t-i\kappa'
t'}}{-4\sinh^{2}(\frac{a}{2}((t-t')-i\epsilon))}\nonumber\\
&=&\frac{\hbar}{(2\pi)^{4}}\lim_{\epsilon\rightarrow
0}\int_{-\infty}^{+\infty} dT\int_{-\infty}^{+\infty}
d\Delta\frac{a^{2}e^{\frac{\epsilon}{2}(\kappa+\kappa')+i(\kappa-\kappa')T+\frac{i\Delta}{2}(\kappa+\kappa')}}{-4\sinh^{2}(\frac{a}{2}(\Delta-i\epsilon))}\nonumber\\
&=&\frac{\hbar
a^{2}}{(2\pi)^{3}}\delta(\kappa-\kappa')\lim_{\epsilon\rightarrow
0}\int_{-\infty}^{+\infty}
d\Delta\frac{e^{\kappa\epsilon}e^{i\kappa\Delta}}{-4\sinh^{2}\frac{a}{2}(\Delta-i\epsilon)}\,\,\,.
\end{eqnarray}

On the third line of the equation, we take $t\rightarrow
t-\frac{i\epsilon}{2}$ to suppress the contribution from
high-frequency modes of the field, and this is equal to setting a
finite time resolution of the system. On the fifth line of the
equation, we take $T\equiv\frac{t+t'}{2}$ and $\Delta\equiv t-t'$,
where the integral is the double complex integral. Note that there
are poles at $\frac{a}{2}(\Delta-i\epsilon)=\pm i\bar{n}$ where
$\bar{n}=0, 1, 2, 3, ... , \infty$. We then plug $F$ back into the
two-point function $\langle
Q(\tau-\tau_{0})Q(\tau''-\tau''_{0})\rangle_{v}$ and
perform the integration of $\tau$. Thus, we have the form\\
\\

\begin{eqnarray}
&&\langle Q(\tau-\tau_{0})Q(\tau''-\tau''_{0})\rangle_{v}\nonumber\\
&=&\frac{\lambda_{0}^{2}}{m_{0}^{2}}\sum_{j,j'=+,-}\int_{\tau_{0}}^{\tau}d\tau'c_{j}e^{w_{j}(\tau-\tau')}\int_{-\infty}^{+\infty}
d\kappa
e^{-i\kappa\tau'}\int_{\tau_{0}'}^{\tau''}d\tau'''c^{\ast}_{j'}e^{w^{\ast}_{j'}(\tau''-\tau''')}\int_{-\infty}^{+\infty}
d\kappa' e^{i\kappa'\tau'''}\nonumber\\
& &\cdot\frac{\hbar
a^{2}}{(2\pi)^{3}}\delta(\kappa-\kappa')\lim_{\epsilon\rightarrow
0}\int_{-\infty}^{+\infty}
d\Delta\frac{e^{\kappa \epsilon}e^{i\kappa\Delta}}{-4\sinh^{2}\frac{a}{2}(\Delta-i\epsilon)}\nonumber\\
&=&\frac{\lambda_{0}^{2}}{m_{0}^{2}}\frac{\hbar}{(2\pi)^{2}}\sum_{j,j'=\pm}(\frac{a^{2}}{2\pi})\lim_{\epsilon\rightarrow 0}\left[\int_{0}^{\infty}d\kappa e^{-i\kappa(\tau_{0}-\tau_{0}'')}\int_{-\infty}^{+\infty} d\Delta\frac{e^{\kappa\epsilon}e^{i\kappa\Delta}}{-4\sinh^{2}\frac{a}{2}(\Delta-i\epsilon)}\right.\nonumber\\
& &\,\,\,\,\,\,\left.+\int_{-\infty}^{0}d\kappa
e^{-i\kappa(\tau_{0}-\tau_{0}'')}\int_{-\infty}^{+\infty}
d\Delta\frac{e^{\kappa\epsilon}e^{i\kappa\Delta}}{-4\sinh^{2}\frac{a}{2}(\Delta+i\epsilon)}\right]\nonumber\\
& &\cdot\frac{c_{j}c^{\ast}_{j'}(e^{w_{j}(\tau-\tau_{0})}-e^{i\kappa(\tau_{0}-\tau)})(e^{w^{\ast}_{j'}(\tau''-\tau''_{0})}-e^{i\kappa(\tau''-\tau''_{0})})}{(w_{j}+i\kappa)(w^{\ast}_{j'}-i\kappa)} \nonumber\\
,\label{delint}
\end{eqnarray}

\begin{figure}[htbp]
\centerline{\psfig{file=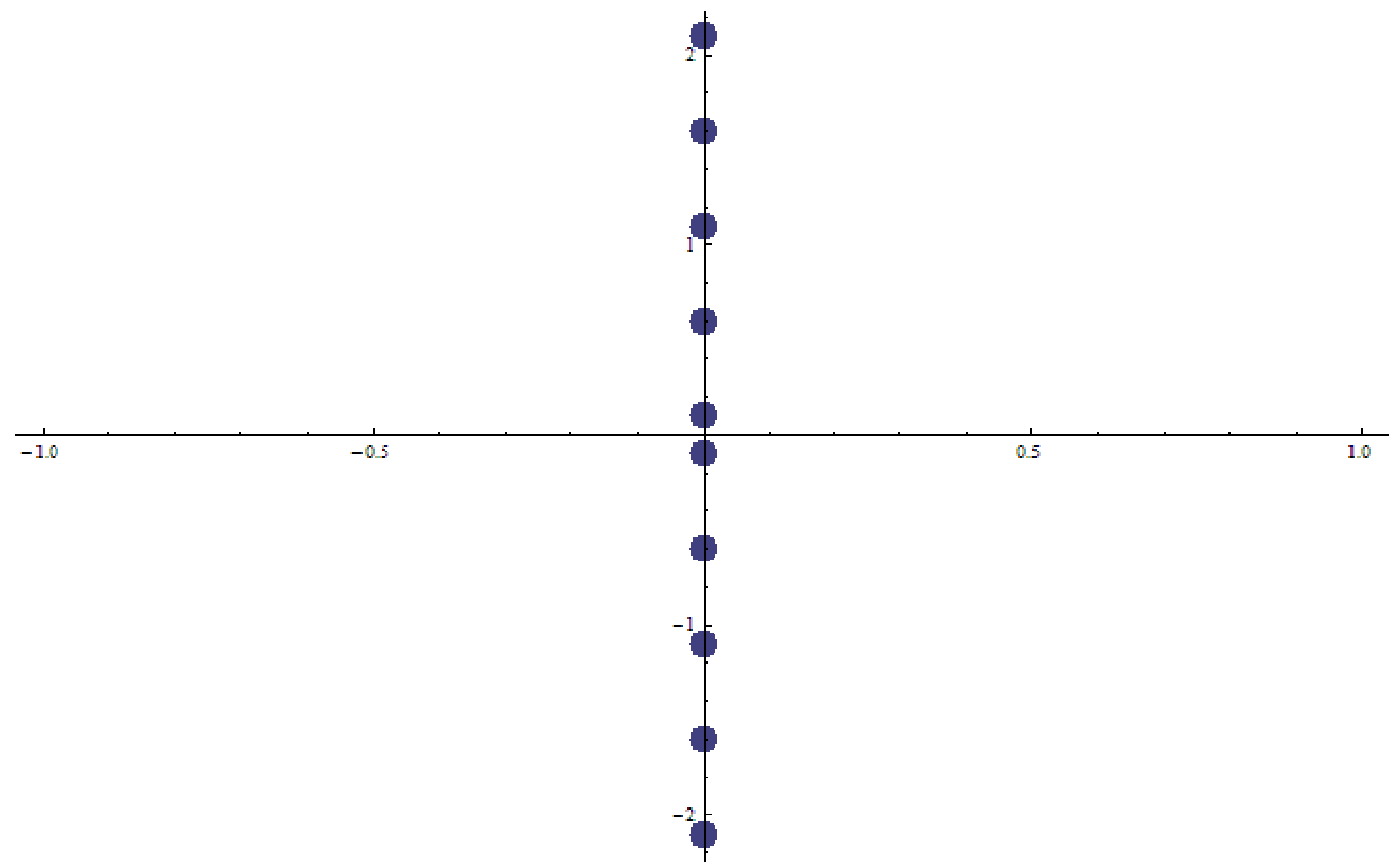, width=14cm}} \caption{Poles
on the complex $\Delta$ plane at $\Delta=i(\epsilon+\frac{2\pi
\bar{n}}{a})$ on the upper complex plane and at
$\Delta=-i(\epsilon+\frac{2\pi \bar{n}}{a})$ on the lower complex
plane. Where $\bar{n}=0,1,2,3,...,\infty$, the poles move to $\pm
i\infty$ as $a\rightarrow0$.} \label{deltapole}
\end{figure}

The integrals on the second equal sign in Eq.$(\ref{delint})$ are
the tricky part (the double complex integrals). Note that inside the
square brackets in Eq.$(\ref{delint})$, the $\kappa$ and $\Delta$
double integrals are split into two parts because the sign of
$\kappa$ determines the contour integration is performed on the
upper half complex plane or on the lower half plane. The first term
in Eq.$(\ref{B2int})$ is the $\Delta$ contour integral circle on the
upper plane when $\kappa>0$, while the second term $\Delta$ is the
contour integral circle on the lower plane when $\kappa<0$. The main
 different from previous result is that they did not
separate the double complex integral and handle the part separately.
If we did not notice this point, we would have just one term and
would ignore the other term, as was done in the past.

For the double integrations, we have the following two parts:

\begin{eqnarray}
\int_{0}^{\infty}d\kappa
e^{-i\kappa(\tau'-\tau''')}\int_{-\infty}^{+\infty}
d\Delta\frac{e^{\kappa\epsilon}e^{i\kappa\Delta}}{-4\sinh^{2}\frac{a}{2}(\Delta-i\epsilon)}+\int_{-\infty}^{0}d\kappa
e^{-i\kappa(\tau'-\tau''')}\int_{-\infty}^{+\infty}
d\Delta\frac{e^{\kappa\epsilon}e^{i\kappa\Delta}}{-4\sinh^{2}\frac{a}{2}(\Delta+i\epsilon)}\label{B2int}\,.
\end{eqnarray}

There are poles in the denominator
$\sinh^{2}\frac{a}{2}(\Delta-i\epsilon)$
 at $\Delta=i(\epsilon+\frac{2\pi \bar{n}}{a})$ and poles in the other denominator
$\sinh^{2}\frac{a}{2}(\Delta+i\epsilon)$
 at $\Delta=-i(\epsilon+\frac{2\pi \bar{n}}{a})$. We may use the identity

\begin{equation}
\csc^{2}\pi
x=\frac{1}{\pi^{2}}\sum_{n=-\infty}^{\infty}\frac{1}{(x-n)^{2}}\,\,,
\end{equation}
 and the relation $\sinh^{2}x=-\sin^{2}(ix)$ to expand those poles.

We expand $\sinh x$ in the first integral such that

\begin{eqnarray}
\frac{-1}{4\sinh^{2}\frac{a}{2}(\Delta-i\epsilon)}=\frac{1}{4\sin^{2}(\frac{ia\Delta}{2}+\epsilon)}
=\frac{-1}{a^{2}}\sum_{n=-\infty}^{\infty}\frac{1}{(\Delta-i\epsilon+i2\pi
n/a)^{2}}\,.
\end{eqnarray}
In the first integral of Eq. $(\ref{B2int})$, only the poles
$n=0,-1,-2,\cdots,-\infty$ are inside the contour. Let $\bar{n}=-n$
in the first integral and rewrite it in the following way:
\begin{eqnarray}
& &\int_{0}^{\infty}d\kappa
e^{-i\kappa(\tau'-\tau''')}\int_{-\infty}^{+\infty}
d\Delta\frac{e^{\kappa\epsilon}e^{i\kappa\Delta}}{-4\sinh^{2}\frac{a}{2}(\Delta-i\epsilon)}\nonumber\\
&=&\int_{0}^{\infty}d\kappa
e^{-i\kappa(\tau'-\tau''')}\int_{-\infty}^{+\infty}
d\Delta\sum_{\bar{n}=0}^{\infty}\frac{e^{\kappa\epsilon}e^{i\kappa\Delta}}{-a^{2}\left[\Delta-i(\epsilon+2\pi
\bar{n}/a)\right]^{2}}\label{sinh1}\,,
\end{eqnarray}
where the given terms $2\pi$ and $n$ are absorbed into $\epsilon$.
Similarly, the second integral in Eq.$(\ref{B2int})$ is expanded in
the same way:
\begin{eqnarray}
\frac{-1}{4\sinh^{2}\frac{a}{2}(\Delta+i\epsilon)}=\frac{1}{4\sin^{2}(\frac{ia\Delta}{2}-\epsilon)}
=\frac{-1}{a^{2}}\sum_{n=-\infty}^{\infty}\frac{1}{(\Delta+i\epsilon+i2\pi
n/a)^{2}}\,.
\end{eqnarray}
Only the poles $n=0,1,2,\cdots , \infty$ are inside the contour of
the second integral. Therefore, the second integral is rewritten as
\begin{eqnarray}
& &\int_{-\infty}^{0}d\kappa
e^{-i\kappa(\tau'-\tau''')}\int_{-\infty}^{+\infty}
d\Delta\frac{e^{\kappa\epsilon}e^{i\kappa\Delta}}{-4\sinh^{2}\frac{a}{2}(\Delta+i\epsilon)}\nonumber\\
&=&\int_{-\infty}^{0}d\kappa
e^{-i\kappa(\tau'-\tau''')}\int_{-\infty}^{+\infty}
d\Delta\sum_{n=0}^{\infty}\frac{e^{\kappa\epsilon}e^{i\kappa\Delta}}{-a^{2}\left[\Delta+i(\epsilon+2\pi
 n/a)\right]^{2}}\label{sinh2}\,,
\end{eqnarray}

%%%%%%%%%%%%%%%%%%%%%%%%%%%%%%%%%%%%%%%%%%%%%%%%%%%%%%%%%%%%%%%%%%%%%%%%%%%%%%%%%%%%%%%%%%%%%%%%%%%%%%
\begin{figure}[htbp]
\centerline{\psfig{file=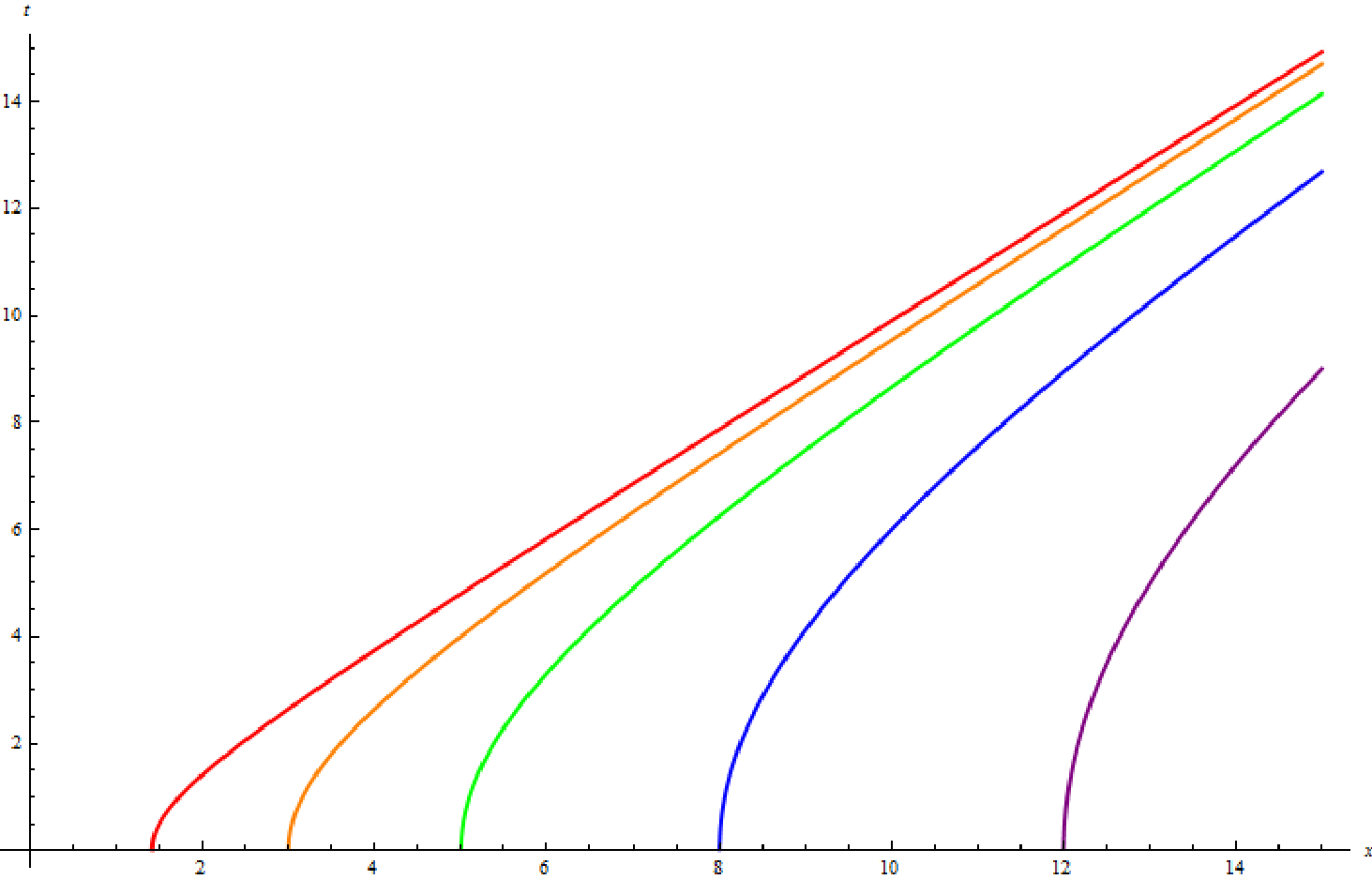, width=14cm}} \caption{The plot
of the worldline $z^{\mu}_{B}=(a^{-1}\sinh a\tau , a^{-1}\cosh a\tau
, 0, 0)$ for differing proper acceleration values $a$. As the proper
acceleration a$\rightarrow 0$, the worldline is shifted far away.
The proper acceleration of the red line is the largest among all
lines, while the proper acceleration of the purple line is the
smallest among all lines.} \label{accLine}
\end{figure}
%%%%%%%%%%%%%%%%%%%%%%%%%%%%%%%%%%%%%%%%%%%%%%%%%%%%%%%%%%%%%%%%%%%%%%%%%%%%%%%%%%%%%%%%%%%%%%%%%%%%%%

In our spacetime diagram $a\leq1$ (we assume the light speed $c=1$,
so the proper acceleration $a\leq1$). Note that when $a\rightarrow
0$ the poles on the complex $\Delta$ plane move to $\infty$ and the
arc integration $\int_{arc}d\Delta f(\Delta)$ in Cauchy's integral
formula is no longer $0$(the arc cannot include the poles when the
poles move to $\infty$); the integral is ill defined. And if we
think carefully, back to Fig.~\ref{accLine} and check the worldline
for the detector Bob, we can see that as $a\rightarrow 0$ , Bob is
very far from origin, and thus Bob cannot exchange the signal with
Alice in a finite time interval. This situation is not the setup
that we want (we need Alice and Bob to be separated by a proper
distance so that they can exchange signals with each other in a
reasonable time interval and we can study the properties in the
quantum teleportation process in such a setup in the future work).
Therefore, the result when $a\rightarrow0$ in Eq. $(\ref{B2int})$ is
not the actual setup that can be extended to the case in which Alice
and Bob have a finite distance between them and exchange signals
when Bob moves at a constant velocity.

Plugging Eq.$(\ref{sinh1})$ and $(\ref{sinh2})$ back into
Eq.$(\ref{delint})$, we have

\begin{equation}
\frac{-1}{a^{2}}\lim_{\epsilon\rightarrow
0}\left(\int_{0}^{\infty}d\kappa\int_{-\infty}^{\infty}d\Delta\sum_{\bar{n}=0}^{\infty}\frac{e^{\kappa\epsilon}e^{i\kappa\Delta}}{[\Delta-i(\epsilon+\frac{2\pi
\bar{n}}{a})]^{2}}+\int_{-\infty}^{0}d\kappa\int_{-\infty}^{\infty}d\Delta\sum_{\bar{n}=0}^{\infty}\frac{e^{\kappa\epsilon}e^{i\kappa\Delta}}{[\Delta+i(\epsilon+\frac{2\pi
\bar{n}}{a})]^{2}}\right)\,,\label{Delta+-}
\end{equation}

Performing the $\Delta$ integration(taking $\epsilon=0$ in the end)
, we then have the following

\begin{eqnarray}
&
&\frac{-1}{a^{2}}\left(\int_{0}^{\infty}d\kappa\sum_{\bar{n}=0}^{\infty}(2\pi
i)i\kappa e^{i\kappa i(+\frac{2\pi
\bar{n}}{a})}+\int_{-\infty}^{0}d\kappa\sum_{\bar{n}=0}^{\infty}(-2\pi
i)i\kappa e^{i\kappa i(-\frac{2\pi \bar{n}}{a})}\right)\nonumber\\
&=&\frac{-1}{a^{2}}\left(\int_{0}^{\infty}d\kappa\sum_{\bar{n}=0}^{\infty}(2\pi
i)i\kappa e^{i\kappa i(\frac{2\pi
\bar{n}}{a})}-\int_{-\infty}^{0}d\kappa\sum_{\bar{n}=0}^{\infty}(2\pi
i)i\kappa e^{-i\kappa i(\frac{2\pi \bar{n}}{a})}\right)\,.
\end{eqnarray}

Plugging the above result back into Eq. $(\ref{delint})$, we obtain
the following terms:

\begin{eqnarray}
&
&\frac{-1}{a^{2}}\left(\int_{0}^{\infty}d\kappa\sum_{\bar{n}=0}^{\infty}(2\pi
i)i\kappa e^{i\kappa i(\frac{2\pi
\bar{n}}{a})}-\int_{-\infty}^{0}d\kappa\sum_{\bar{n}=0}^{\infty}(2\pi
i)i\kappa e^{-i\kappa i(\frac{2\pi
\bar{n}}{a})}\right)e^{-i\kappa(\tau_{0}-\tau''_{0})}\nonumber\\
&
&\cdot\frac{C_{j}C^{\ast}_{j'}(e^{w_{j}(\tau-\tau_{0})}-e^{i\kappa(\tau_{0}-\tau)})
(e^{w^{\ast}_{j'}(\tau''-\tau''_{0})}-e^{i\kappa(\tau''-\tau''_{0})})}{(w_{j}+i\kappa)(w^{\ast}_{j'}-i\kappa)}\nonumber\\
&=&\frac{-1}{a^{2}}\left[\int^{\infty}_{0}
d\kappa\sum^{\infty}_{\bar{n}=0}X^{\bar{n}}\emph{F}(\kappa)-\int^{0}_{-\infty}d\kappa\sum^{\infty}_{\bar{n}=0}X^{-\bar{n}}\emph{F}(\kappa)\right]\,,\label{Delta+-1}
\end{eqnarray}

where $X^{\bar{n}}=e^{-\kappa\frac{2\pi}{a}}$ and
$\emph{F}(\kappa)=(2\pi
i)i\kappa\cdot\frac{C_{j}C^{\ast}_{j'}(e^{w_{j}(\tau-\tau_{0})}-e^{i\kappa(\tau_{0}-\tau)})(e^{w^{\ast}_{j'}(\tau''-\tau''_{0})}-e^{i\kappa(\tau''-\tau''_{0})})e^{-i\kappa(\tau_{0}-\tau''_{0})}}
{(w_{j}+i\kappa)(w^{\ast}_{j'}-i\kappa)}$\,.

Note that since there are many poles on the imaginary axis of the
$\kappa$ complex plane, it is difficult to do a contour integration
in the form of Eq.$(\ref{Delta+-1})$. To avoid the difficulty of
such contour integral, we can reshape Eq. $(\ref{Delta+-1})$ in the
following way to avoid the poles:

\begin{eqnarray}
& &\frac{-1}{a^{2}}\left[\int^{\infty}_{0}
dk\sum^{\infty}_{\bar{n}=0}X^{\bar{n}}-\int^{0}_{-\infty}dk\sum^{\infty}_{\bar{n}=0}X^{-\bar{n}}\right]\emph{F}(\kappa)\nonumber\\
&=&\frac{-1}{a^{2}}\left[\int_{0}^{\infty}d\kappa\frac{1}{1-X}-\int_{-\infty}^{0}d\kappa\frac{1}{1-X^{-1}}\right]\emph{F}(\kappa)\nonumber\\
&=&\frac{-1}{a^{2}}\left[\int_{0}^{\infty}d\kappa\frac{1}{1-X}-\int_{-\infty}^{0}d\kappa\frac{X}{X-1}\right]\emph{F}(\kappa)\nonumber\\
&=&\frac{-1}{a^{2}}\left[\int_{0}^{\infty}d\kappa\frac{1}{1-X}+\int_{-\infty}^{0}d\kappa\frac{X-1+1}{1-X}\right]\emph{F}(\kappa)\nonumber\\
&=&\frac{-1}{a^{2}}\left[\int_{0}^{\infty}d\kappa\frac{1}{1-X}+\int_{-\infty}^{0}d\kappa\frac{1}{1-X}+\int_{-\infty}^{0}d\kappa\frac{X-1}{1-X}\right]\emph{F}(\kappa)\nonumber\\
&=&\frac{-1}{a^{2}}\left[\int_{-\infty}^{\infty}\frac{d\kappa
\emph{F}(\kappa)}{1-X}-\int_{-\infty}^{0}d\kappa
\emph{F}(\kappa)\right]\,.\label{trick1}
\end{eqnarray}

When we use the above trick (we rewrite the integral region from
$-\infty$ to $\infty$ and include the poles inside the contour in
the first integral ) , the difficult contour integral from $0$ to
$\infty$ is prevented.

We then put the above results back into Eq. $(\ref{delint})$, and
then the two-point function in Eq. $(\ref{delint})$ is thus reshaped
as
\begin{eqnarray}
& &\langle Q(\tau-\tau_{0})Q(\tau''-\tau''_{0})\rangle_{v}\nonumber\\
&=&\frac{\lambda_{0}^{2}\hbar}{m_{0}^{2}(2\pi)^{2}}\left[\sum_{j,j'=\pm}C_{j}C^{\ast}_{j'}\int_{-\infty}^{\infty}\frac{\kappa
d\kappa
e^{-i\kappa(\tau_{0}-\tau''_{0})}(e^{w_{j}(\tau-\tau_{0})}-e^{i\kappa(\tau_{0}-\tau)})(e^{w^{\ast}_{j'}(\tau''-\tau''_{0})}-e^{i\kappa(\tau''-\tau''_{0})})}
{(1-e^{-2\pi\kappa/a})(w_{j}+i\kappa)(w^{\ast}_{j'}-i\kappa)}\right.\nonumber\\
&
&\left.-\sum_{j,j'=\pm}C_{j}C^{\ast}_{j'}\int_{-\infty}^{0}\frac{\kappa
d\kappa
e^{-i\kappa(\tau_{0}-\tau''_{0})}(e^{w_{j}(\tau-\tau_{0})}-e^{i\kappa(\tau_{0}-\tau)})(e^{w^{\ast}_{j'}(\tau''-\tau''_{0})}-e^{i\kappa(\tau''-\tau''_{0})})}
{(w_{j}+i\kappa)(w^{\ast}_{j'}-i\kappa)}\right]\,,\label{2p2term}
\end{eqnarray}
and we have two terms in the two-point correlation function. Note
that the first term is the old result in the previous work, while
the second term is the new result that is missing from the previous
work.

The steps above are the key points in the calculations. To avoid the
difficult $\kappa$ integration (many poles on the imaginary axis of
the $\kappa$ plane) we thus reshape Eq. $(\ref{delint})$ into the
form of Eq. $(\ref{2p2term})$. When we compare this new result to
the previous result, we find that the second term in Eq.
$(\ref{2p2term})$ is missing from the previous results. This is a
careless accident that can happen when we deal with the double
complex integral in Eq. $(\ref{delint})$. Also, remember that when
$a=0$ the result of this two-point function is no longer true,
because this result corresponds to the situation in which Bob is
very far away, so Bob and Alice cannot exchange messages within a
reasonable time interval. The mathematical reason for this is shown
in the part of $\Delta$ contour integration, where
 the denominator $\sinh^{2}\frac{a}{2}(\Delta-i\epsilon)$ of the
$\Delta$ integration on the last line of Eq. $(\ref{delint})$ is no
longer a hyperbolic sine function as $a=0$. This corresponds to
Fig.~\ref{accLine}, in which the world line of Bob shifts to very
far away as $a=0$, and this is not the  setup that we want.

If we compare the difference of this new result for the two-point
correlation function more easily to the previous one \cite{LH2005}),
the two-point correlation function is expressed in the following
form:

\begin{eqnarray}
& &\langle Q(\tau-\tau_{0})Q(\tau''-\tau''_{0})\rangle_{v}\nonumber\\
&=&\frac{\lambda_{0}^{2}\hbar}{(2\pi)^{2}m_{0}^{2}}\sum_{j,j'}\int_{0}^{\infty}\frac{\kappa
d\kappa}{1-e^{-2\pi\kappa/a}}\frac{c_{j}c_{j'}^{\ast}e^{-i\kappa(\tau_{0}-\tau''_{0})}}{(w_{j}+i\kappa)(w^{\ast}_{j'}-i\kappa)}
(e^{w_{j}(\tau-\tau_{0})}-e^{-i\kappa(\tau-\tau_{0})})(e^{w^{\ast}_{j'}(\tau''-\tau''_{0})}-e^{i\kappa(\tau''-\tau''_{0})})\nonumber\\
&=&\frac{\lambda_{0}^{2}\hbar}{m_{0}^{2}(2\pi)^{2}}\left[\sum_{j,j'=\pm}C_{j}C^{\ast}_{j'}\int_{-\infty}^{\infty}\frac{\kappa
d\kappa
e^{-i\kappa(\tau_{0}-\tau''_{0})}(e^{w_{j}(\tau-\tau_{0})}-e^{i\kappa(\tau_{0}-\tau)})(e^{w^{\ast}_{j'}(\tau''-\tau''_{0})}-e^{i\kappa(\tau''-\tau''_{0})})}
{(1-e^{-2\pi\kappa/a})(w_{j}+i\kappa)(w^{\ast}_{j'}-i\kappa)}\right.\nonumber\\
&
&\left.-\sum_{j,j'=\pm}C_{j}C^{\ast}_{j'}\int_{-\infty}^{0}\frac{\kappa
d\kappa
e^{-i\kappa(\tau_{0}-\tau''_{0})}(e^{w_{j}(\tau-\tau_{0})}-e^{i\kappa(\tau_{0}-\tau)})(e^{w^{\ast}_{j'}(\tau''-\tau''_{0})}-e^{i\kappa(\tau''-\tau''_{0})})}
{(w_{j}+i\kappa)(w^{\ast}_{j'}-i\kappa)}\right]\nonumber\\
&=&\langle QQ\rangle_{v1}-\langle QQ\rangle_{v2}\,.\label{2p2term1}
\end{eqnarray}

The first term $\langle QQ\rangle_{v1}$ is the old result from the
previous work(\cite{LH2005}), and the second term $\langle
QQ\rangle_{v2}$ is the missing term( a new term). The advantage of
the two-point function being reshaped  in the above manner
 is that it can be compared and computed more easily since the poles on the imaginary axis
are now included inside the contour and the $\kappa$ integrations
can be done. Also, we can easily compare this new result to the old
result(\cite{LH2005}) and see the differences between the new and
old results more clearly.

\begin{figure}[htbp]
\centerline{\psfig{file=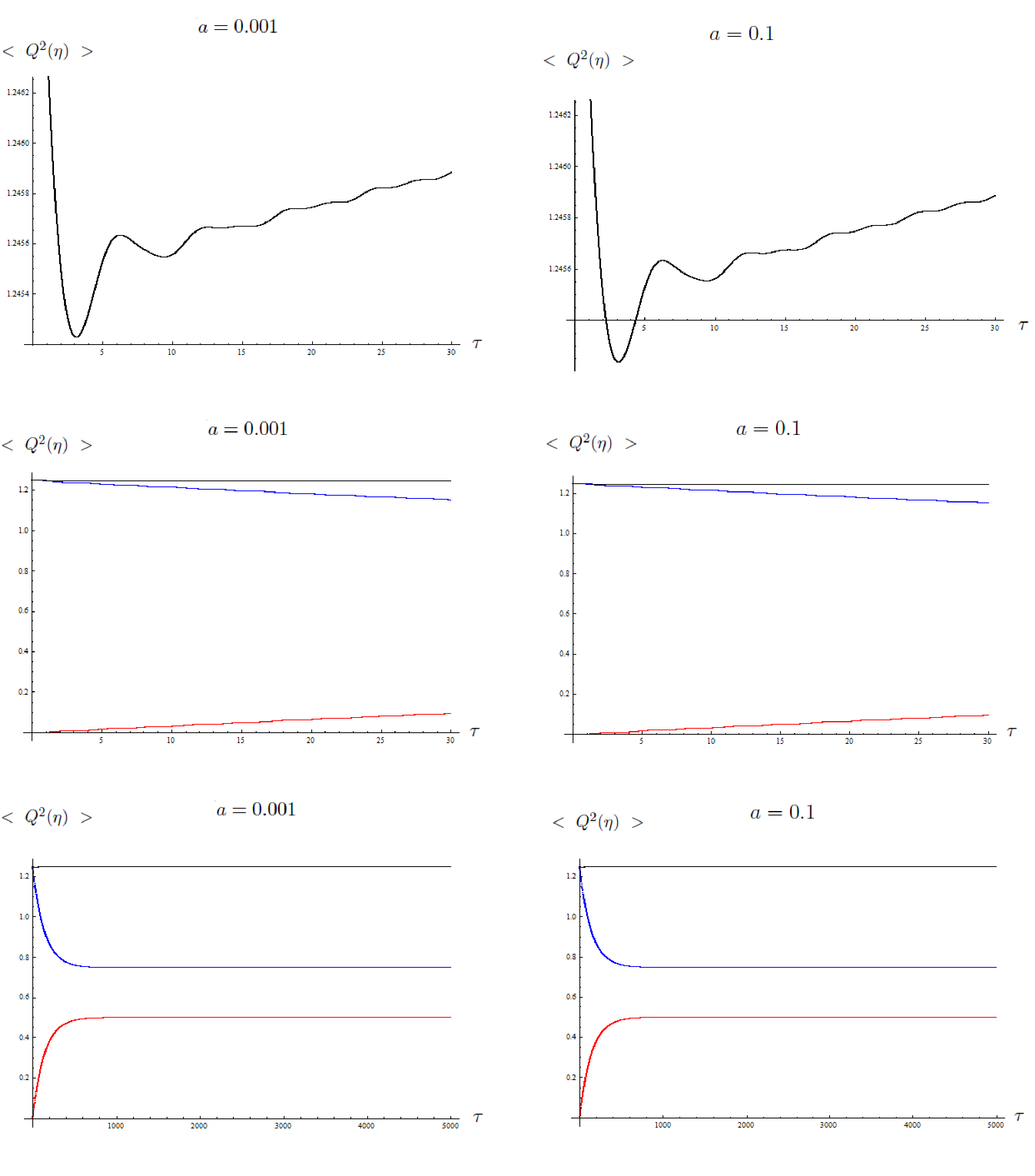, width=18cm}} \caption{The plots for
$\langle Q^{2}(\eta)\rangle_{v1}$[red line; Eq. $(\ref{accv1})$ with
$\Lambda_{0}$ excluded ], $-\langle Q^{2}(\eta)\rangle_{v2}$[blue
line; Eq. $(\ref{accv2})$ with $\Lambda_{0_{v2}}$ excluded ], and
the sum $\langle Q^{2}(\eta)\rangle_{v}$ [ i.e., black line $\langle
Q^{2}(\eta)\rangle_{v1}$ $-\langle Q^{2}(\eta)\rangle_{v2}$]. Here
$\Omega=1.0$, $\lambda_{0}=0.3$(which is $\gamma=0.00358$), and
$m_{0}=\hbar=1$. Note that $-\langle Q^{2}(\eta)\rangle_{v2}$ is
larger than $\langle Q^{2}(\eta)\rangle_{v1}$. The black line[i.e.,
$\langle Q^{2}(\eta)\rangle_{v}$] oscillates at the beginning and
arrives at the saturated value later. When $t=30$, the curve for the
proper acceleration $a=0.1$ arrives at the value $1.24589$, while
the curve for a smaller proper acceleration $a=0.001$ arrives at the
value $1.24588$; the difference is only $0.00001$. When $t=5000$,
curves for both small or large proper acceleration values arrive at
the same final magnitude $1.24772$.} \label{A_m}
\end{figure}

\begin{figure}[htbp]
\centerline{\psfig{file=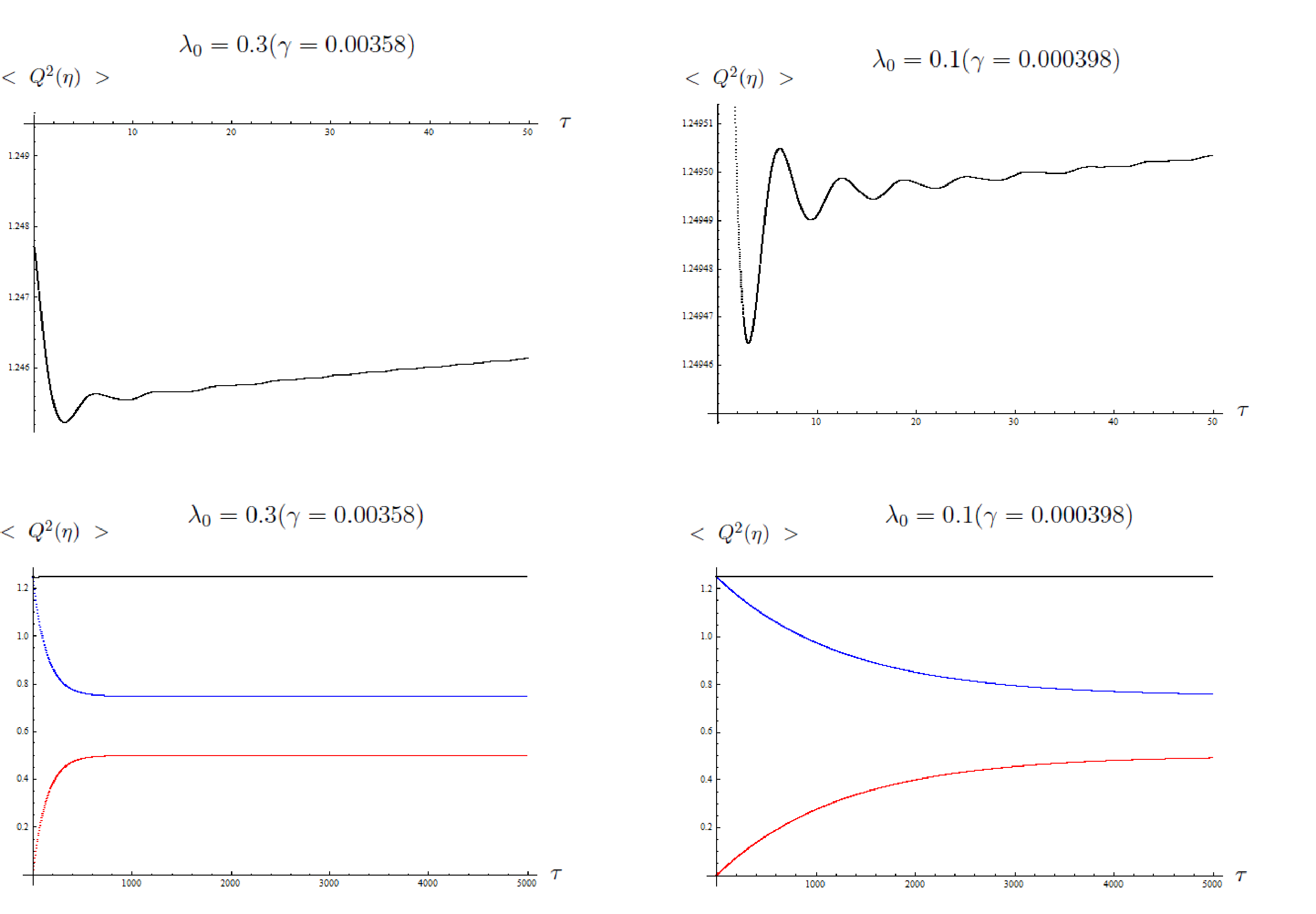, width=19cm}} \caption{The decay
parameter $\gamma$($\gamma=\frac{\lambda_{0}^{2}}{8\pi m_{0}}$). The
plots for $\langle Q^{2}(\eta)\rangle_{v1}$[red line; Eq.
$(\ref{accv1})$ with $\Lambda_{0}$ excluded ], $-\langle
Q^{2}(\eta)\rangle_{v2}$[blue line; Eq. $(\ref{accv2})$ with
$\Lambda_{0_{v2}}$ excluded ], and the sum $\langle
Q^{2}(\eta)\rangle_{v}$[black line, which is $\langle
Q^{2}(\eta)\rangle_{v1}$ $-\langle Q^{2}(\eta)\rangle_{v2}$]. Here
$\Omega=1.0$, $a=0.001$, and $m_{0}=\hbar=1$. The decay parameters
are different in the plots. The black lines oscillate at the
beginning and then arrive at different saturated values later at
proper time $\tau=5000$ for different $\lambda_{0}$. When $\tau=50$,
the final value for $\gamma=0.000398$ is 1.2495, while when
$\tau=50$ the final value for $\gamma=0.00358$ is 1.24613. When
$\tau=5000$, the final value for $\gamma=0.000398$ is 1.24974, while
when $\tau=5000$ the final value for $\gamma=0.00358$ is 1.24772. A
large $\lambda_{0}$($\lambda_{0}=0.3$) has a larger value than a
smaller $\lambda_{0}$($\lambda_{0}=0.1$), while a smaller
$\lambda_{0}$ arrives at the same saturated value later than a
larger $\lambda_{0}$($\lambda_{0}=0.3$). The decay parameter
$\gamma$ affects the saturated time.} \label{B}
\end{figure}

\begin{figure}[htbp]
\centerline{\psfig{file=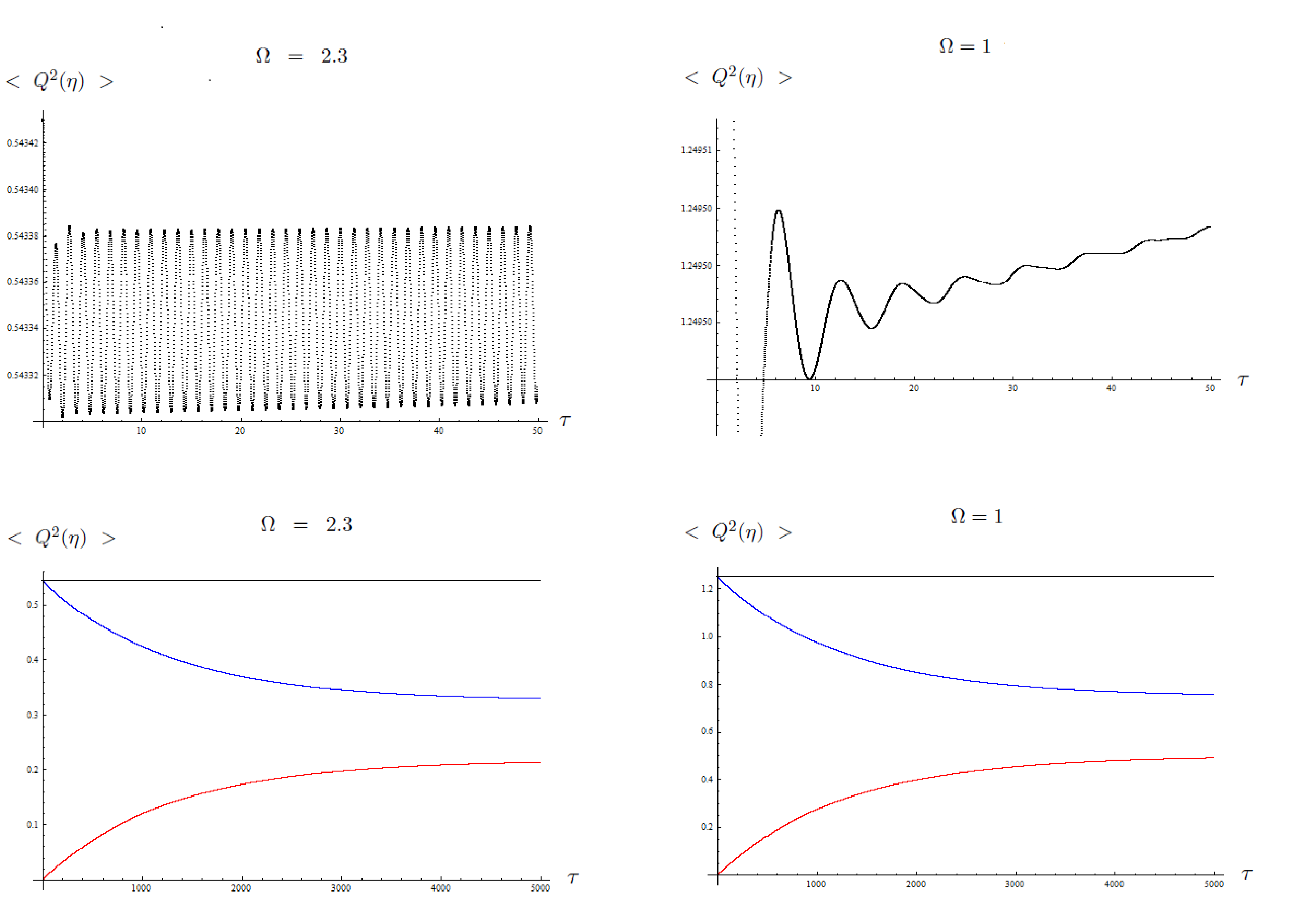, width=19cm}} \caption{The frequency
$\Omega$. The plot for $\langle Q^{2}(\eta)\rangle_{v1}$[red line;
Eq. $(\ref{accv1})$ with $\Lambda_{0}$ excluded ] and $-\langle
Q^{2}(\eta)\rangle_{v2}$[blue line; Eq. $(\ref{accv2})$ with
$\Lambda_{0_{v2}}$ excluded ], and the sum $\langle
Q^{2}(\eta)\rangle_{v}$[black line; which is $\langle
Q^{2}(\eta)\rangle_{v1}$ $-\langle Q^{2}(\eta)\rangle_{v2}$]. Here
$a=0.001$, $\lambda_{0}=0.1$($\gamma=0.000398$), and
$m_{0}=\hbar=1$. Two values of $\Omega$ are chosen($\Omega=2.3$ and
$1.0$). The curves arrive at different saturated values for
different $\Omega$. The smaller $\Omega$($\Omega=1.0$) has a higher
saturated value than the bigger $\Omega$($\Omega=2.3$) at a later
time $\tau=5000$. When $\tau=5000$, the final value for $\Omega=2.3$
is $\langle Q^{2}(\eta)\rangle_{v}=0.543429$, while the final value
for $\Omega=1.0$ is $\langle Q^{2}(\eta)\rangle_{v}=1.24974$. The
frequency parameter $\Omega$ affects the final saturated value.
Also, a bigger $\Omega$ has more vibrations in the same $\tau$
region and is more active.} \label{C}
\end{figure}

Here we simply write the results of $\langle QQ\rangle_{v1}$, and
$-\langle QQ\rangle_{v2}$ in Eq. $(\ref{2p2term1})$ below, the
detailed derivations is written in Appendix D.

\begin{eqnarray}
& &\langle QQ\rangle_{v1}\nonumber\\
&=&\frac{2\hbar\gamma}{\pi
m_{0}\Omega^{2}}\theta(\eta)Re\{(\Lambda_{0}-\ln\frac{a}{\Omega})e^{-2\gamma\eta}\sin^{2}\Omega\eta\nonumber\\
&
&+\frac{a}{2}e^{-(\gamma+a)\eta}\left[\frac{F_{\gamma+i\Omega}(e^{-a\eta})}{\gamma+i\Omega+a}\left(\frac{-i\Omega}{\gamma}\right)e^{-i\Omega\eta}+
\frac{F_{-\gamma-i\Omega}(e^{-a\eta})}{\gamma+i\Omega-a}\left(\left(1+\frac{i\Omega}{\gamma}\right)e^{i\Omega\eta}-e^{-i\Omega\eta}\right)\right]\nonumber\\
&
&-\frac{1}{4}\left[\left(\frac{i\Omega}{\gamma}+e^{-2\gamma\eta}\left(\frac{i\Omega}{\gamma}+1-e^{-2i\Omega\eta}\right)\right)(\psi_{\gamma+i\Omega}
+\psi_{-\gamma-i\Omega})\right.\nonumber\\
&
&\,\,\,\,\,\,\left.-\left(\frac{-i\Omega}{\gamma}+e^{-2\gamma\eta}\left(\frac{i\Omega}{\gamma}+1
-e^{-2i\Omega\eta}\right)\right)i\pi\coth\frac{\pi}{a}(\Omega-i\gamma)\right]\}\,.\label{accv1}
\end{eqnarray}

%%%%%%%%%%%%%%%%%%%%%%%%%%%%%%%%%%%%%%%%%%%%%%%%%%%%%%%%%%%%%%%%%%%%%%%%%%%%%%%%%%%%%%%%%%%%%%%%%%%%%%%%%%%

\begin{eqnarray}
& &-\langle QQ\rangle_{v2}\nonumber\\
&=&\frac{-2\hbar\gamma}{\pi
m_{0}}Re\{\Lambda_{0_{v2}}-\frac{e^{-2\gamma(\tau-\tau_{0})}}{8\Omega^{2}}\left[(1-\frac{i\Omega}{\gamma}-e^{2i\Omega(\tau-\tau_{0})})\cdot(i\pi+2\log(\gamma-i\Omega)+2\Gamma(0,-(\gamma-i\Omega)(\tau-\tau_{0})))\right.\nonumber\\
& &\left.
+(1+\frac{i\Omega}{\gamma}-e^{-2i\Omega(\tau-\tau_{0})})\cdot(-i\pi-2\log(\gamma+i\Omega)+2\Gamma(0,-(\gamma+i\Omega)(\tau-\tau_{0})))\right]
\nonumber\\
&
&-\frac{i}{8\Omega\gamma}\left[-i\pi-2\log(\frac{\gamma+i\Omega}{\gamma-i\Omega})+2\Gamma(0,(\gamma+i\Omega)(\tau-\tau_{0}))-2\Gamma(0,(\gamma-i\Omega)(\tau-\tau_{0}))\right]\}\,,\label{accv2}
\end{eqnarray}

%%%%%%%%%%%%%%%%%%%%%%%%%%%%%%%%%%%%%%%%%%%%%%%%%%%%%%%%%%%%%%%%%%%%%%%%%%%%%%%%%%%%%%%%%%%%%%%%%%%%%%%%%%%

where $\Lambda_{0}$ and $\Lambda_{0_{v2}}$ are the terms containing
the divergent parts [$\Gamma(0,0)$ and $\log(0)$] as
$\tau''\rightarrow\tau$ and $\tau''_{0}\rightarrow\tau_{0}$ , and
are absorbed into the renormalized constant or coefficient in the
experiment. Figure~\ref{A_m}-~\ref{C} are the numerical results for
$\langle QQ\rangle_{v1}$ and $-\langle QQ\rangle_{v2}$. In the
plots, the red line is the term $\langle QQ\rangle_{v1}$(i.e., the
old result) and the blue line is the term $-\langle
QQ\rangle_{v2}$(i.e., the missing term), while the black line is the
sum $\langle QQ\rangle_{v1}-\langle QQ\rangle_{v2}$. The
contributions from the vacuum fluctuations of the two-point function
$\langle QQ\rangle_{v}$(i.e., the black line) for the internal
degrees of freedom $Q$ begin with a relatively high value, then
oscillate and
reach to a saturated value at a later time.\\

In fig.~\ref{A_m}, we change the proper acceleration $a$ and keep
the other parameters the same. We can see that for $a=0.1$ and
$a=0.001$, the black curves have the same shape, but the values are
slightly different. The value of the two-point function $\langle
QQ\rangle_{v}$ for the $a=0.1$ curve is just higher than the
$a=0.001$ curve for only a very small number, $0.00001$. If we think
that a uniformly accelerated detector would experience a different
thermal radiance-a different temperature in the background(the Unruh
effect)-this different background would produce different vacuum
fluctuations for $\langle QQ\rangle_{v}$. Thus, we can see that
although the difference of the effect from proper acceleration from
$a=0.1$ to
$a=0.001$ UAD is small, it is indeed present.\\

In Fig. ~\ref{B}, we change the value of the coupling constant
$\lambda_{0}$, or, say we change the decay parameter $\gamma$. This
is because the definition $\gamma=\frac{\lambda_{0}^{2}}{8\pi
m_{0}}$ and we also use perturbations method in these computations.
Therefore, $\lambda_{0}<1$ is the basic assumption for perturbation
(i.e., $\lambda_{0}$ is the expansion parameter). The allowed region
for $\gamma$ is then $\gamma<0.039$. In the previous
work\cite{LH2005}, we chose $\gamma=0.1$, which is equal to
$\lambda_{0}=1.585$. This value is too big and obviously violates
the basic assumption of the perturbation, making the perturbative
solutions inconsistent with the perturbation method. According to
our experience, a safe choice is to make the expansion parameter
$\lambda_{0}\approx 0.1$. This is why we choose $\lambda_{0}=0.1$
and $0.3$(corresponding to $\gamma=0.000398$ and
$0.00358$) in our numerical plots.\\

In Fig.~\ref{C}, we alter the value of frequency $\Omega$(i.e., the
frequency for the internal degrees of freedom of the detector) and
keep the other parameters the same. We choose $\Omega=2.3$ and
$1.0$, and the magnitude of the two-point function $\langle
QQ\rangle_{v}$ for $\Omega=2.3$ is larger than in the $\Omega=1.0$
case. Also, in the same $\tau$ interval, the curve for the
$\Omega=2.3$ case has more oscillations than the curve for the
$\Omega=1.0$ case does.\\

%%%%%%%%%%%%%%%%%%%%%%%%%%%%%%%%%%%%%%%%%%%%%%%%%%%%%%%%%%%%%%%%%%%%%%%%%%%%%%%%%%%%%%%%%%%%%%%%%%%%%%%%%%%

As in the calculations of the two-point function $\langle
QQ\rangle$, we also compute the two-point function
$\langle\dot{Q}\dot{Q}\rangle$, and the result is listed below. As
shown above, there is an extra term
$-\langle\dot{Q}\dot{Q}\rangle_{v2}$ in our new result which is
missing from the previous result~\cite{LH2005}.

%%%%%%%%%%%%%%%%%%%%%%%%%%%%%%%%%%%%%%%%%%%%%%%%%%%%%%%%%%%%%%%%%%%%%%%%%%%%%%%%%%%%%%%%%%%%%%%%%%%%%%%%%%%%

\begin{eqnarray}
& &\langle\dot{Q}\dot{Q}\rangle_{v1}\nonumber\\
&=&\frac{2\hbar\gamma}{\pi
m_{0}\Omega^{2}}\theta(\eta)Re\{(\Lambda_{1}-\ln\frac{a}{\Omega})\Omega^{2}+(\Lambda_{0}-\ln\frac{a}{\Omega})e^{-2\gamma\eta}
(\Omega\cos\Omega\eta-\gamma\sin\Omega\eta)^{2}\nonumber\\
&
&+\frac{a}{2}(\gamma+i\Omega)^{2}e^{-(\gamma+a)\eta}\left[\frac{F_{\gamma+i\Omega}(e^{-a\eta})}{\gamma+i\Omega+a}\left(\frac{i\Omega}{\gamma}\right)e^{-i\Omega\eta}+
\frac{F_{-\gamma-i\Omega}(e^{-a\eta})}{\gamma+i\Omega-a}\left(\left(1-\frac{i\Omega}{\gamma}\right)e^{i\Omega\eta}-e^{-i\Omega\eta}\right)\right]\nonumber\\
&
&+\frac{1}{4}(\gamma+i\Omega)^{2}\left[\left(\frac{i\Omega}{\gamma}+e^{-2\gamma\eta}\left(\frac{i\Omega}{\gamma}-1+e^{-2i\Omega\eta}\right)\right)(\psi_{\gamma+i\Omega}
+\psi_{-\gamma-i\Omega})\right.\nonumber\\
&
&\,\,\,\,\,\,\left.-\left(\frac{-i\Omega}{\gamma}+e^{-2\gamma\eta}\left(\frac{i\Omega}{\gamma}-1
+e^{-2i\Omega\eta}\right)\right)i\pi\coth\frac{\pi}{a}(\Omega-i\gamma)\right]\}\,.\label{accPB2v1}
\end{eqnarray}

%%%%%%%%%%%%%%%%%%%%%%%%%%%%%%%%%%%%%%%%%%%%%%%%%%%%%%%%%%%%%%%%%%%%%%%%%%%%%%%%%%%%%%%%%%%%%%%%%%%%%%%%%%%%

\begin{eqnarray}
& &-\langle\dot{Q}\dot{Q}\rangle_{v2}\nonumber\\
&=&\frac{-2\hbar\gamma}{\pi
m_{0}}\theta(\eta)Re\{\tilde{\Lambda}_{0_{v2}}+\frac{e^{-2\gamma(\tau-\tau_{0})}}{8\Omega^{2}}
\left[((\gamma^{2}+\Omega^{2})(1-\frac{i\Omega}{\gamma})-(\gamma-i\Omega)^{2}e^{2i\Omega(\tau-\tau_{0})})\right.\nonumber\\
&
&\left.\cdot(-i\pi+2\log(\gamma-i\Omega))+((\gamma^{2}+\Omega^{2})(1+\frac{i\Omega}{\gamma})-(\gamma+i\Omega)^{2}e^{-2i\Omega(\tau-\tau_{0})})
\cdot(i\pi+2\log(\gamma+i\Omega))\right]\nonumber\\
& &+\frac{ie^{-(\gamma+i\Omega)(\tau-\tau_{0})}}{4\Omega\gamma}
\left[(\gamma-i\Omega)(\frac{-1}{\tau-\tau_{0}}+e^{(\gamma-i\Omega)(\tau-\tau_{0})}(\gamma-i\Omega)\Gamma(0,(\gamma-i\Omega)(\tau-\tau_{0})))-
(\gamma+i\Omega)\right.\nonumber\\
&
&\left.\cdot(\frac{-1}{\tau-\tau_{0}}+e^{(\gamma+i\Omega)(\tau-\tau_{0})}(\gamma+i\Omega)\Gamma(0,(\gamma+i\Omega)(\tau-\tau_{0})))\right]\nonumber\\
& &+
\frac{e^{-\gamma(\tau-\tau_{0})}}{4\Omega^{2}}\left[((\gamma+i\Omega)e^{-i\Omega(\tau-\tau_{0})}-\frac{\gamma^{2}+\Omega^{2}}{\gamma}e^{i\Omega(\tau-\tau_{0})})
((\gamma+i\Omega)(i\pi-\Gamma(0,-(\gamma+i\Omega)(\tau-\tau_{0})))\right.\nonumber\\
& & \cdot
e^{-(\gamma+i\Omega)(\tau-\tau_{0})}-\frac{1}{\tau-\tau_{0}})
+((\gamma-i\Omega)e^{i\Omega(\tau-\tau_{0})}-\frac{\gamma^{2}+\Omega^{2}}{\gamma}e^{-i\Omega(\tau-\tau_{0})})
\cdot(e^{(-\gamma+i\Omega)(\tau-\tau_{0})}(-\gamma+i\Omega)\nonumber\\
&
&\left.\cdot(i\pi+\Gamma(0,(-\gamma+i\Omega)(\tau-\tau_{0})))-\frac{1}{\tau-\tau_{0}})\right]
+\frac{i}{8\Omega\gamma}\left[(\gamma-i\Omega)^{2}(2\log(\gamma-i\Omega)+i\pi)\right.\nonumber\\
& &\left.-(\gamma+i\Omega)^{2}(2\log(\gamma-i\Omega)+3i\pi)\right]\}
\,,\label{accPBv2}
\end{eqnarray}

where $\Lambda_{1}$, $\Lambda_{0}$, and $\tilde{\Lambda}_{0_{v2}}$
are the terms  contains those divergent parts $\Gamma(0,0)$ and
$\log(0)$ as $\tau''\rightarrow\tau$ and
$\tau''_{0}\rightarrow\tau_{0}$ , and they are absorbed into the
renormalized constant or coefficient in the experiment.

The above numerical results are plotted in the following
Figs.~\ref{A_PB}-~\ref{C_PB}. The red line is the term
$\langle\dot{Q}\dot{Q}\rangle_{v1}$, while the green line is the
missing term $-\langle\dot{Q}\dot{Q}\rangle_{v2}$ and the
black line is the sum $\langle\dot{Q}\dot{Q}\rangle_{v1}-\langle\dot{Q}\dot{Q}\rangle_{v2}$.\\

In Fig~\ref{A_PB} , we vary the value $a$ and find the magnitude of
$\langle\dot{Q}\dot{Q}\rangle_{v}$ for different $a$($a=0.1$ or
$0.001$) at the same time that $\tau$ is unchanged. The effect of
the proper acceleration $a$ is not obvious. The curve at first is
arising slightly and then decreasing and oscillating. The trend for
$\langle\dot{Q}\dot{Q}\rangle_{v}$ is decreasing and different from
$\langle QQ \rangle_{v}$, which is slightly increasing.\\

In Fig.~\ref{B_PB}, we vary the decay parameter $\gamma$ (which is
equivalent to varing the coupling constant $\lambda_{0}$). The curve
for the two-point function $\langle\dot{Q}\dot{Q}\rangle_{v}$ also
rises slightly in the beginning and then oscillates and decreases to
a saturated value. The difference is that the curve for
$\gamma=0.00358$ arrives at the saturated value earlier than
$\gamma=0.000398$. At the same time $\tau$, the value of the
two-point function $\langle\dot{Q}\dot{Q}\rangle_{v}$ for different
$\gamma$ is also slightly different. The reason for this is that a
higher $\gamma$ value for the two-point function
curve decays to the same value faster than a lower $\gamma$ curve does.\\

In Fig.~\ref{C_PB}, we alter the frequency of the internal degrees
of freedom for the detector. It is obvious that the magnitude of
$\langle\dot{Q}\dot{Q}\rangle_{v}$ changes significantly when the
internal frequency $\Omega$ is altered. The trend of both curves is
the same in that at first it has a small rise and then it decays and
oscillates to a saturated value. However, a large $\Omega$ has more
oscillations in its decay curve. A small $\Omega$ is less active
than a large $\Omega$. And also, a small $\Omega$ curve has a much
lower
saturated value than a large $\Omega$ curve.\\

The red lines in the plots represent the old results for the
two-point functions $\langle\dot{Q}\dot{Q}\rangle_{v}$ which are
displayed as a dotted line in Fig. $2$ of Ref. ~\cite{LH2005}. The
old results do not have the missing term
$-\langle\dot{Q}\dot{Q}\rangle_{v2}$ (the green line). The black
line is the sum
$\langle\dot{Q}\dot{Q}\rangle_{v1}-\langle\dot{Q}\dot{Q}\rangle_{v2}$,
and it gradually drops to a steady value at late time. This differs
from the old result. The
two-point functions $\langle\dot{Q}\dot{Q}\rangle_{v}$ in the old result increase gradually to a steady value.\\

Comparing the above plots for the two-point function $\langle
\dot{Q}\dot{Q}\rangle_{v}$ to the plots for the two-point function
$\langle QQ\rangle_{v}$ , we find that the difference between
$\langle QQ\rangle_{v}$ and $\langle \dot{Q}\dot{Q}\rangle_{v}$ is
that the proper acceleration parameter $a$ affects the trend of the
oscillating curve in the early-time region (whether it is slightly
increasing or decreasing). The coupling constant $\lambda_{0}$
affects how soon the curve of $\langle \dot{Q}\dot{Q}\rangle$
arrives at the saturated value as shown in Fig.~\ref{B_PB}. The
frequency $\Omega$ affects how large the final saturated value for
$\langle \dot{Q}\dot{Q}\rangle$ will be. As shown in Fig.
~\ref{C_PB}, a smaller
$\Omega$ has a smaller saturated value.\\

Next, we will discuss the allowed region for the value of the
coupling constant $\lambda_{0}$, which is about the decay constant $\gamma$. This part was not noticed before. \\

%%%%%%%%%%%%%%%%%%%%%%%%%%%%%%%%%%%%%%%%%%%%%%%%%%%%%%%%%%%%%%%%%%%%%%%%%%%%%%%%%%%%%%%%%%%%%%%%%%%%%%
%%%%%%%%%%%%%%%%%%%%%%%%%%%%%%%%%%%%%%%%%%%%%%%%%%%%%%%%%%%%%%%%%%%%%%%%%%%%%%%%%%%%%%%%%%%%%%%%%%%%%%

%%%%%%%%%%%%%%%%%%%%%%%%%%%%%%%%%%%%%%%%%%%%%%%%%%%%%%%%%%%%%%%%%%%%%%%%%%%%%%%%%%%%%%%%%%%%%%%%%%%%%%
%%%%%%%%%%%%%%%%%%%%%%%%%%%%%%%%%%%%%%%%%%%%%%%%%%%%%%%%%%%%%%%%%%%%%%%%%%%%%%%%%%%%%%%%%%%%%%%%%%%%%%

%%%%%%%%%%%%%%%%%%%%%%%%%%%%%%%%%%%%%%%%%%%%%%%%%%%%%%%%%%%%%%%%%%%%%%%%%%%%%%%%%%%%%%%%%%%%%%%%%%%%%%
%%%%%%%%%%%%%%%%%%%%%%%%%%%%%%%%%%%%%%%%%%%%%%%%%%%%%%%%%%%%%%%%%%%%%%%%%%%%%%%%%%%%%%%%%%%%%%%%%%%%%%

%Fig6
%%%%%%%%%%%%%%%%%%%%%%%%%%%%%%%%%%%%%%%%%%%%%%%%%%%%%%%%%%%%%%%%%%%%%%%%%%%%%%%%%%%%%%%%%%%%%%%%%%%%%%
\begin{figure}[htbp]
\centerline{\psfig{file=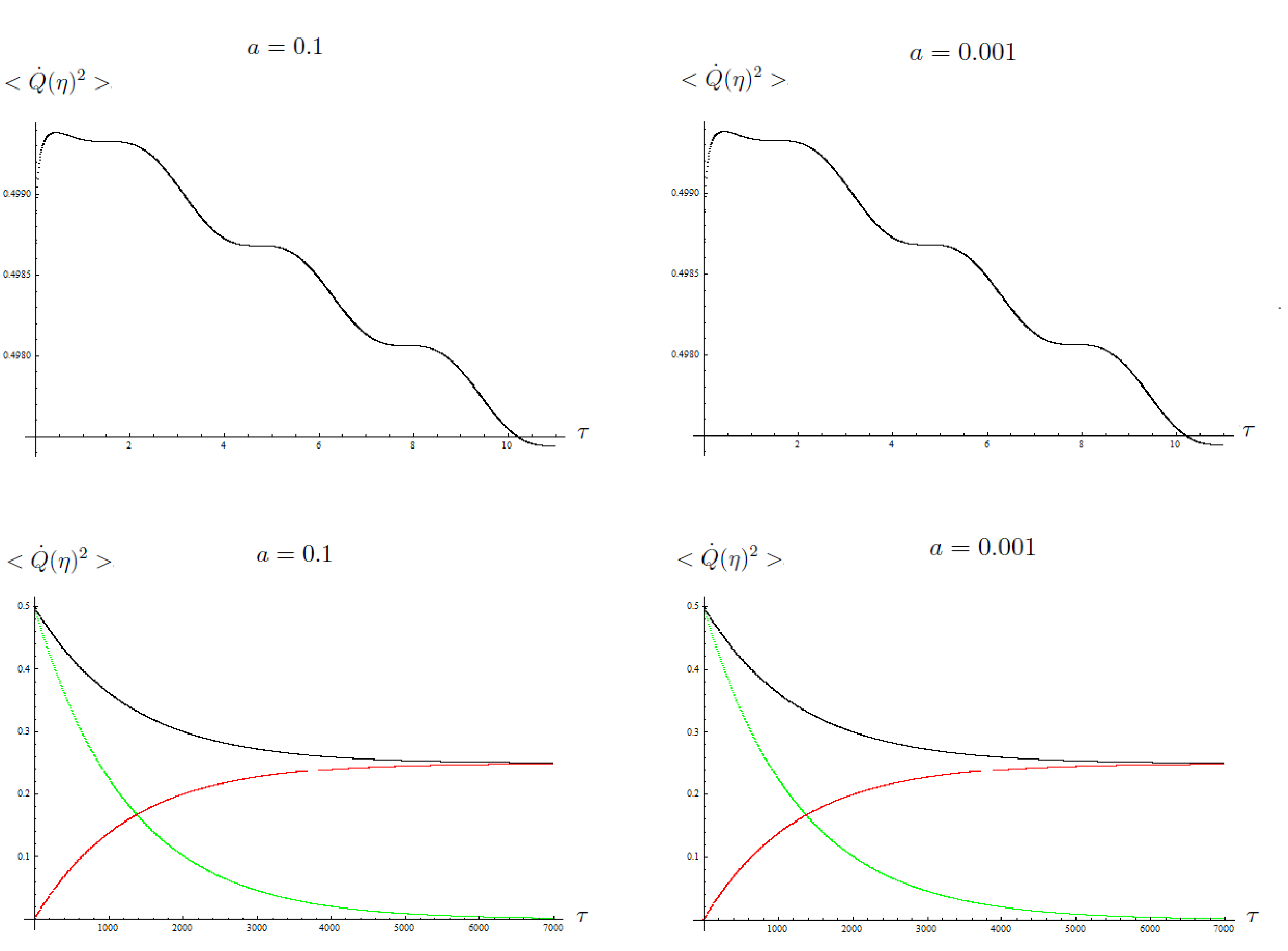, width=18cm}} \caption{The proper
acceleration $a$. The plots for $\langle
\dot{Q}^{2}(\eta)\rangle_{v1}$[red line; Eq. $(\ref{accPB2v1})$ with
$\Lambda_{1}$ excluded ], $-\langle
\dot{Q}^{2}(\eta)\rangle_{v2}$[green line; Eq. $(\ref{accPBv2})$
with $\tilde{\Lambda}_{0_{v2}}$ excluded ], and the sum $\langle
\dot{Q}^{2}(\eta)\rangle_{v}$[black line; which is $\langle
\dot{Q}^{2}(\eta)\rangle_{v1}$ $-\langle
\dot{Q}^{2}(\eta)\rangle_{v2}$]. Here $\Omega=1.0$,
$\lambda_{0}=0.1$(which is $\gamma=0.000398$), and $m_{0}=\hbar=1$.
The green line $-\langle \dot{Q}^{2}(\eta)\rangle_{v2}$ is larger
than $\langle \dot{Q}^{2}(\eta)\rangle_{v1}$ in the early-time
region. The black line oscillates at the beginning and later arrives
at the saturated value at around $\tau\approx6500$. For $\tau=11$
both the $a=0.1$ and $a=0.001$ curves arrive at the same value
$\langle \dot{Q}^{2}(\eta)\rangle_{v}=0.497442$. When $\tau=7000$,
the $a=0.1$ and $a=0.001$ curves both arrive at the same value
$\langle \dot{Q}^{2}(\eta)\rangle_{v}=0.250765$. For $\langle
\dot{Q}^{2}(\eta)\rangle_{v}$, the difference between the proper
accelerations $a=0.1$ and $a=0.001$ is not obvious, as shown in the
plots. However, the trend of the curve for the two-point function
$\langle \dot{Q}^{2}(\eta)\rangle_{v}$ is decreasing and differs
from the $\langle Q^{2}(\eta)\rangle_{v}$ plots.} \label{A_PB}
\end{figure}
%%%%%%%%%%%%%%%%%%%%%%%%%%%%%%%%%%%%%%%%%%%%%%%%%%%%%%%%%%%%%%%%%%%%%%%%%%%%%%%%%%%%%%%%%%%%%%%%%%%%%%

%Fig7
%%%%%%%%%%%%%%%%%%%%%%%%%%%%%%%%%%%%%%%%%%%%%%%%%%%%%%%%%%%%%%%%%%%%%%%%%%%%%%%%%%%%%%%%%%%%%%%%%%%%%%
\begin{figure}[htbp]
\centerline{\psfig{file=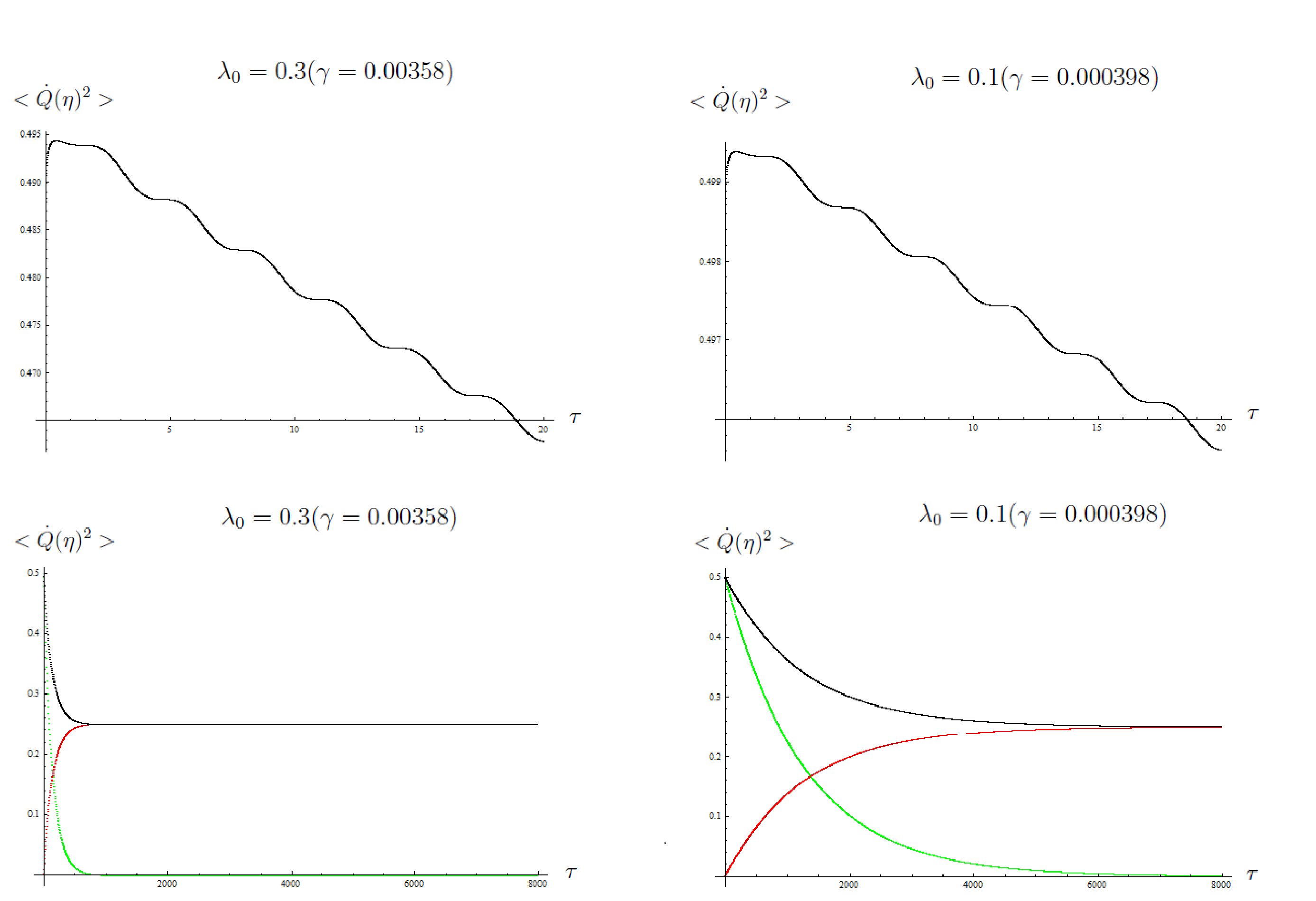, width=19cm}} \caption{The decay
parameter $\gamma$(i.e., the coupling constant $\lambda_{0}$). The
plots for $\langle \dot{Q}^{2}(\eta)\rangle_{v1}$[red line; Eq.
$(\ref{accPB2v1})$ with $\Lambda_{1}$ excluded ], $-\langle
\dot{Q}^{2}(\eta)\rangle_{v2}$[green line; Eq. $(\ref{accPBv2})$
with $\tilde{\Lambda}_{0_{v2}}$ excluded ], and the sum $\langle
\dot{Q}^{2}(\eta)\rangle_{v}$[black line, which is $\langle
\dot{Q}^{2}(\eta)\rangle_{v1}$ $-\langle
\dot{Q}^{2}(\eta)\rangle_{v2}$]. Here $\Omega=1.0$, $a=0.1$, and
$m_{0}=\hbar=1$. The coupling constants $\lambda_{0}$ differ in
these plots. When $\tau=20$, $\langle
\dot{Q}^{2}(\eta)\rangle_{v}=0.4956$ for $\gamma=0.000398$ and
$\langle \dot{Q}^{2}(\eta)\rangle_{v}=0.462788$ for
$\gamma=0.00358$.  At a late time when $\tau=8000$, $\langle
\dot{Q}^{2}(\eta)\rangle_{v}=0.250238$ for $\gamma=0.000398$ and
$\langle \dot{Q}^{2}(\eta)\rangle_{v}=0.248185$ for
$\gamma=0.00358$. A larger $\gamma$ has a higher $\langle
\dot{Q}^{2}(\eta)\rangle_{v}$ value than a smaller decay parameter
$\gamma$. The lines oscillate in early-time region and then arrive
at a saturated value at a late time. A larger $\gamma$ decays faster
than a smaller $\gamma$, and the trend of the $\langle
\dot{Q}^{2}(\eta)\rangle_{v}$ curve is decreasing except a very
short, small rise at the beginning.} \label{B_PB}
\end{figure}
%%%%%%%%%%%%%%%%%%%%%%%%%%%%%%%%%%%%%%%%%%%%%%%%%%%%%%%%%%%%%%%%%%%%%%%%%%%%%%%%%%%%%%%%%%%%%%%%%%%%%%

%Fig8
%%%%%%%%%%%%%%%%%%%%%%%%%%%%%%%%%%%%%%%%%%%%%%%%%%%%%%%%%%%%%%%%%%%%%%%%%%%%%%%%%%%%%%%%%%%%%%%%%%%%%%
\begin{figure}[htbp]
\centerline{\psfig{file=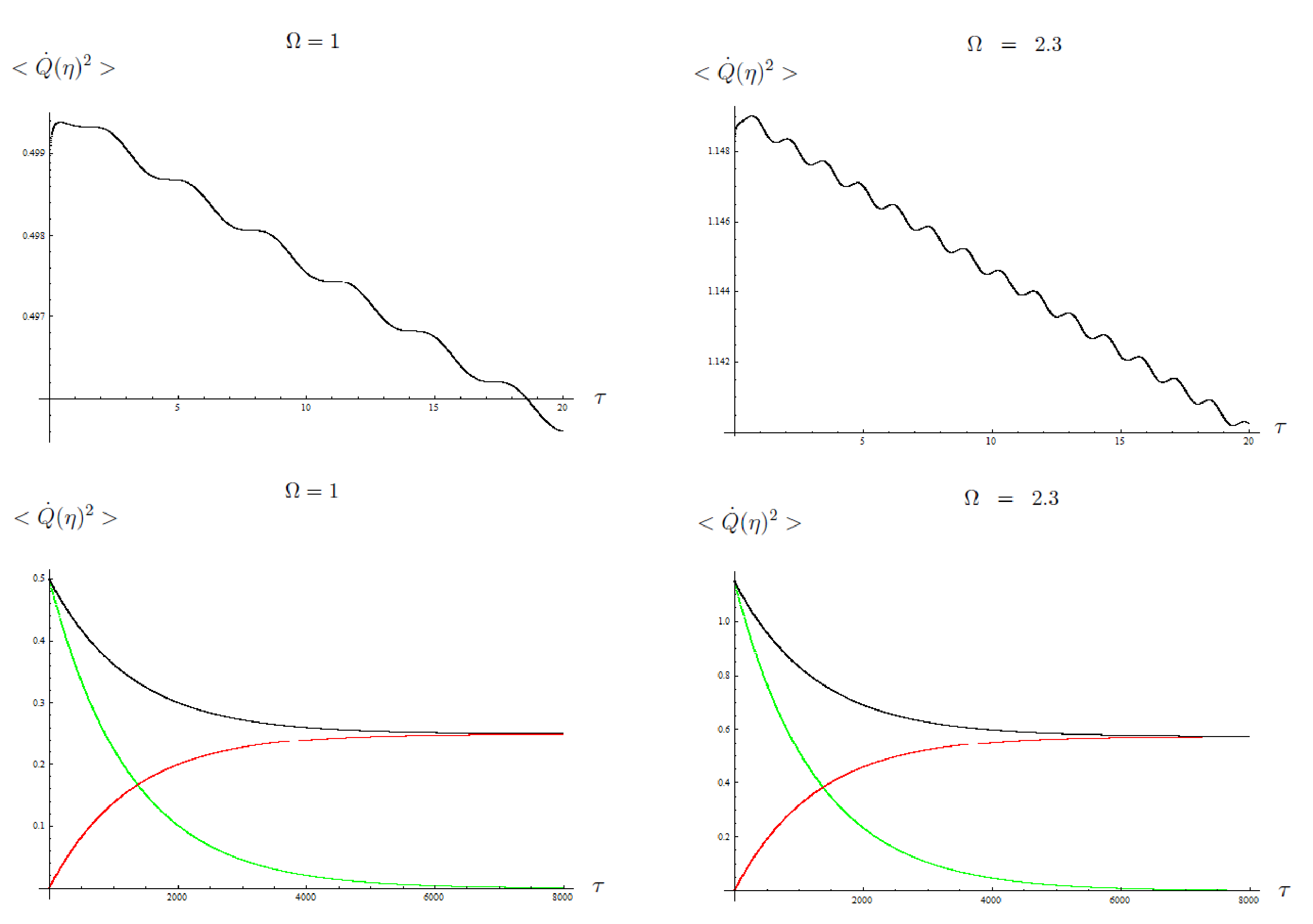, width=19cm}} \caption{The
frequency parameter $\Omega$. The plots for $\langle
Q^{2}(\eta)\rangle_{v1}$[red line; Eq. $(\ref{accPB2v1})$ with
$\Lambda_{1}$ excluded ], and $-\langle
Q^{2}(\eta)\rangle_{v2}$[green line, Eq.$(\ref{accPBv2})$ with
$\tilde{\Lambda}_{0_{v2}}$ excluded ], and the sum $\langle
Q^{2}(\eta)\rangle_{v}$[black line, which is $\langle
Q^{2}(\eta)\rangle_{v1}$ $-\langle Q^{2}(\eta)\rangle_{v2}$]. Here
$a=0.1$, $\gamma=0.000398 (\lambda_{0}=0.1)$ and $m_{0}=\hbar=1$. A
larger $\Omega$ has a higher value for the two-point function
$\langle Q^{2}(\eta)\rangle_{v}$. The black line has a small rise in
the beginning and then decreases to a saturated value. A larger
$\Omega$ has more oscillations in the same time interval as a
smaller $\Omega$. The magnitude of $\Omega$ alters the intensity of
the two-point function $\langle Q^{2}(\eta)\rangle_{v}$.}
\label{C_PB}
\end{figure}
%%%%%%%%%%%%%%%%%%%%%%%%%%%%%%%%%%%%%%%%%%%%%%%%%%%%%%%%%%%%%%%%%%%%%%%%%%%%%%%%%%%%%%%%%%%%%%%%%%%%%%
%Fig9
\begin{figure}[htbp]
\centerline{\psfig{file=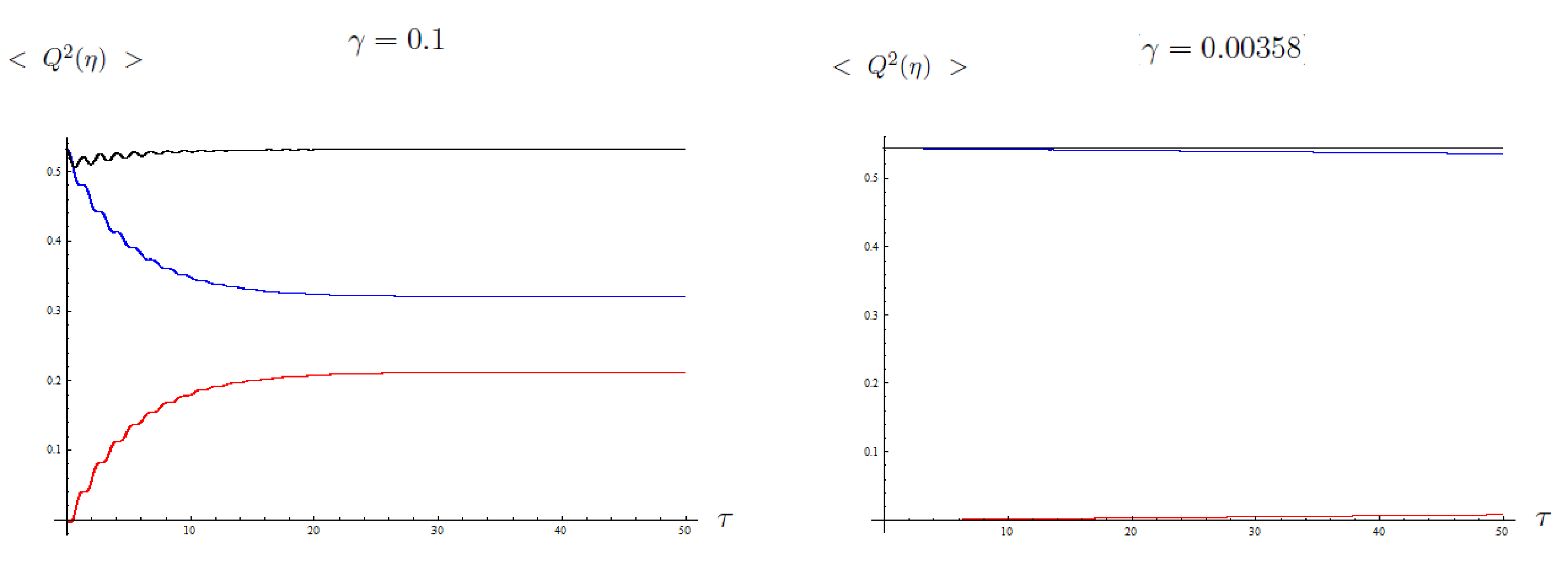, width=19cm}} \caption{The
plots for the proper and improper $\gamma$ values. The plots for
$\langle Q^{2}(\eta)\rangle_{v1}$[red line, Eq. $(\ref{accPB2v1})$
with $\Lambda_{1}$ excluded ], $-\langle
Q^{2}(\eta)\rangle_{v2}$[green line, Eq. $(\ref{accPBv2})$ with
$\tilde{\Lambda}_{0_{v2}}$ excluded], and the sum $\langle
Q^{2}(\eta)\rangle_{v}$[black line, which is $\langle
Q^{2}(\eta)\rangle_{v1}$ $-\langle Q^{2}(\eta)\rangle_{v2}$]. Here
$a=0.001$, $\Omega=2.3$ and $m_{0}=\hbar=1$. $\gamma=0.1$ is the
improper $\gamma$ value, while $\gamma=0.000398$ is the proper
value.} \label{impro}
\end{figure}

%%%%%%%%%%%%%%%%%%%%%%%%%%%%%%%%%%%%%%%%%%%%%%%%%%%%%%%%%%%%%%%%%%%%%%%%%%%%%%%%%%%%%%%%%%%%%%%%%%%%%%
\subsection*{B. Allowed region for $\gamma$ : Proper and improper value for $\gamma$ concerning the contribution of the missing term $-\langle
QQ\rangle_{v2}$}\label{numeTest}
%%%%%%%%%%%%%%%%%%%%%%%%%%%%%%%%%%%%%%%%%%%%%%%%%%%%%%%%%%%%%%%%%%%%%%%%%%%%%%%%%%%%%%%%%%%%%%%%%%%%%%

The $\gamma$ value has some restrictions. In Sec.II Appendix B, we
expand the mode function $f^{(\pm)}$ and $q^{(\pm)}$ by the order of
$\lambda_{0}$ , then use the perturbative method to obtain the
leading order solutions for $q^{(\pm)}$. Later, we use the leading
order solution to compute $\langle Q^{2}(\eta)\rangle_{v}$ to the
first order $O(\lambda_{0})$. Thus, $\lambda_{0}$ is the expansion
parameter and it is supposed to be smaller than $1$. While the decay
parameter $\gamma$ is defined as $\gamma=\frac{\lambda_{0}^{2}}{8\pi
m_{0}}$. Therefore, the value of $\gamma$ has
an allowed region which corresponds $\lambda_{0}<1$.\\

In our previous work~\cite{LH2005}, we took $\gamma=0.1$. This value
corresponds to $\lambda_{0}\approx 1.58$ and is apparently larger
than $1$. In such a case, the perturbative solution is no longer
consistent with our assumption(i.e., $\lambda_{0}$ is smaller than
$1$) if we take $\gamma=0.1$. Let us see what will happen if we take
$\gamma=0.1$ in
the numerical calculations.\\

In Fig. ~\ref{impro}, we do the numerical calculations for two cases
in which $\gamma=0.1$(i.e., equal to $\lambda_{0}\approx1.58$) and
$\gamma=0.000398$ (i.e., equal to $\lambda_{0}=0.1$), with the other
parameters  $\Omega=2.3$ and $a=0.001$ being the same. In the
$\gamma=0.1$ plot, the missing term $\langle
Q^{2}(\eta)\rangle_{v2}$ (the blue line) becomes unimportant very
soon. It drops quickly and then the total effect (the black line) is
dominated by the red line $\langle Q^{2}(\eta)\rangle_{v1}$. The
trend of the black line is similar to the red line: both lines are
increasing. If we lose the second term $\langle
Q^{2}(\eta)\rangle_{v2}$ in our analytic calculations and then would
like to do the numerical integrations at the very beginning as a
double check to get a consistent result(meaning the black line and
red lines are similar and follow the same trend since the numerical
calculation in the beginning does  not neglect the missing term and
 thus will give us the black line),
 one must set the parameter  value at $\gamma=0.1$(i.e., $\lambda_{0}\approx 1.58$). For $\gamma=0.1$,
 the numerical result (black line) will show the same trend
  as the red line[because the second term $-\langle Q^{2}(\eta)\rangle_{v2}$ is unimportant for such a $\gamma$ value].
  In short, we must pick a value for $\gamma$ such that it would make the blue line value small and unimportant.
  The value $\gamma=0.1$ fits this goal. And, we would have thought that our analytic results were correct
  because both analytic and numerical calculations gave us similar curves for the two-point function $\langle Q^{2}(\eta)\rangle_{v}$.
  However, this is just an improper $\gamma$ value giving us a misleading result.
  Thus, we must be careful about the allowed and not allowed regions for the
  parameters in the solutions when we do the numerical calculations as a double check.   \\

Therefore, if one drops the missing term and chooses an allowed
$\gamma$ value, as is shown in Fig. ~\ref{A_m} and ~\ref{B} (i.e.,
$\gamma=0.000398$) , the trends of the black and the red lines look
different-the red line is increasing (with some ripples on it), but
the black line looks quite flat. Without noticing the missing term,
one may give up this value for $\gamma$ and go back to the not
allowed value of $\gamma=0.1$ and think that the numerical test is
consistent with one's analytic result (i.e., dropping the missing
term). This case may give one's a misleading double check that ones
analytic and numerical computations are consistent in terms of the
$\gamma$ value($\gamma=0.1$), so one's analytic results are correct.
But in fact the
contribution of the missing term is suppressed in a not allowed $\gamma$ region. \\

In addition, it is important to be careful when choosing a value of
$a$. If we set the light speed $c=1$, then the proper acceleration
$a$ is smaller than $1$ (i.e., $a<1$). In our present numerical
calculations, we choose the proper acceleration to be $a=0.1$ or
$a=0.001$, both of which are smaller than $1$. In our previous work,
we chose $a=1$, which is not a good choice if we assume that the
light speed $c=1$.
The value $a=1$ is too large. \\

Next, we will continue to consider the inertial detector case.

%%%%%%%%%%%%%%%%%%%%%%%%%%%%%%%%%%%%%%%%%%%%%%%%%%%%%%%%%%%%%%%%%%%%%%%%%%%%%%%%%%%%%%%%%%%%%%%%%%%%%%
\subsection*{C. Trajectory 2 : Inertial detector }\label{traj2}
%%%%%%%%%%%%%%%%%%%%%%%%%%%%%%%%%%%%%%%%%%%%%%%%%%%%%%%%%%%%%%%%%%%%%%%%%%%%%%%%%%%%%%%%%%%%%%%%%%%%%%

We will now calculate the two-point correlation functions $\langle
Q^{2}(\eta)\rangle_{v}$ and $\langle \dot{Q}^{2}(\eta)\rangle_{v}$
for the inertial detector along the trajectory
$\tilde{z}^{\mu}_{B}=(\gamma\tau, \gamma v\tau+x_{a}+d, 0, 0)$. This
trajectory is for an observer moving at constant velocity and has a
finite distance $d$ away from the other static detector. This part
differs from that in the previous work that one applies the UAD
result and takes the limit $a\rightarrow0$ to be the result for an
inertial detector. In this paper, we have already found that in such
a limit $a\rightarrow0$ the
UAD Bob is shifted to very far away and cannot exchange the signal with the inertial detector Alice in a reasonable time interval.\\

To get the two-point correlation functions for the inertial detector
Bob, we simply need to plug the inertial trajectory
$\tilde{z}^{\mu}_{B}=(\gamma\tau, \gamma v\tau+x_{a}+d, 0, 0)$ into
Eq. $(\ref{2p})$. Then the difference is that $z^{0}$ and $\vec{z}$
in $F$ have changed. As in the derivations of the UAD part, the
integral $F$ for this new trajectory $\tilde{z}^{\mu}_{B}$ is

\begin{eqnarray}
F&=&\frac{\hbar}{(2\pi)^{3}}\int_{0}^{2\pi}d\phi\int_{-1}^{1}d(\cos\theta)\int_{0}^{\infty}\frac{\omega^{2}d\omega}{2\omega}\int_{-\infty}^{\infty}
\frac{dt}{2\pi}\int_{-\infty}^{\infty}\frac{dt'}{2\pi} e^{i\kappa
t-i\kappa't'-i\omega(z^{0}(t)-z^{0}(t')+i\omega\cos\theta|\vec{z}-\vec{z'}|)}\nonumber\\
&=&\frac{\hbar}{2\pi}\int_{0}^{\infty}
d\omega\int_{-\infty}^{\infty}
\frac{dt}{2\pi}\int_{-\infty}^{\infty}\frac{dt'}{2\pi}e^{i\kappa
t-i\kappa't'-i\omega\left[z^{0}(t)-z^{0}(t')\right]}\frac{\sin(\omega|\vec{z}(t)-\vec{z}(t')|)}{|\vec{z}(t)-\vec{z}(t')|}
\nonumber\\
&=&\frac{\hbar}{(2\pi)^{4}}\int_{-\infty}^{\infty}
dt\int_{-\infty}^{\infty}
dt'\frac{e^{i\kappa(t-\frac{i\epsilon}{2})-i\kappa'(t'+\frac{i\epsilon}{2})}}{|\vec{z}(t-\frac{i\epsilon}{2})-\vec{z}(t'+\frac{i\epsilon}{2})|^{2}-\left[z^{0}(t-\frac{i\epsilon}{2})-z^{0}(t+\frac{i\epsilon}{2})\right]^{2}}\nonumber\\
&=&\frac{\hbar}{(2\pi)^{4}}\int_{-\infty}^{\infty}
dt\int_{-\infty}^{\infty}
dt'\frac{e^{\frac{\epsilon}{2}(\kappa+\kappa')+i\kappa t-i\kappa'
t'}}{\gamma^{2}(\tau-\tau'-i\epsilon)^{2}(v^{2}-1)}\nonumber\\
&=&\frac{\hbar}{(2\pi)^{4}}\int_{-\infty}^{\infty}
dT\int_{-\infty}^{\infty}
d\Delta\frac{e^{\frac{\epsilon}{2}(\kappa+\kappa')+i(\kappa-\kappa')T+\frac{i\Delta}{2}(\kappa+\kappa')}}{-(\Delta-i\epsilon)^{2}}\nonumber\\
&=&\frac{\hbar}{(2\pi)^{3}}\delta(\kappa-\kappa')\int_{-\infty}^{\infty}
d\Delta\frac{e^{\kappa\epsilon}e^{i\kappa\Delta}}{-(\Delta-i\epsilon)^{2}}\nonumber\\
&=&\frac{\hbar}{(2\pi)^{2}}\delta(\kappa-\kappa')\kappa\,,\,\,\kappa\geq0\,\,.\label{Fv}
\end{eqnarray}

Note that $F=0$ when $\kappa<0$. The velocity $v$ is canceled out in
the denominator and does not appear in the integration term $F$,
which implies that, no matter how fast or slow the velocity is, the
result for $F$ is the same.

As in Eq. $(\ref{delint})$, plugging $F$ back into the two-point
function $\langle Q(\tau-\tau_{0})Q(\tau''-\tau''_{0})\rangle_{v}$
and performing the integration of $\tau$, we have the two-point
function for the inertial detector

\begin{eqnarray}
&&\langle Q(\tau-\tau_{0})Q(\tau''-\tau''_{0})\rangle_{v}\nonumber\\
&=&\frac{\lambda_{0}^{2}\hbar}{(2\pi)^{2}m_{0}^{2}}\sum_{j,j'}\int_{0}^{\infty}\kappa
d\kappa\frac{c_{j}c_{j'}^{\ast}e^{-i\kappa(\tau_{0}-\tau''_{0})}}{(w_{j}+i\kappa)(w^{\ast}_{j'}-i\kappa)}
(e^{w_{j}(\tau-\tau_{0})}-e^{-i\kappa(\tau-\tau_{0})})\cdot(e^{w^{\ast}_{j'}(\tau''-\tau''_{0})}-e^{-i\kappa(\tau''-\tau''_{0})})\,.\label{2pvc}
\end{eqnarray}

Performing the $\kappa$ integration and then using the same
calculation steps for the uniformly accelerated detector that we
used previously in the paper, we obtain the following result for the
two-point correlation function of the internal degrees of freedom
$Q$ for the inertial detector Bob

\begin{eqnarray}
& &\langle Q(\eta)^{2}\rangle_{v}\equiv
\lim_{\eta''\rightarrow\eta}\frac{1}{2}\langle \{Q(\eta),Q(\eta'')\}\rangle_{v}\nonumber\\
&=&\frac{2\hbar\gamma}{\pi
m_{0}}\theta(\eta)Re\{\tilde{\Lambda}_{0}+\frac{e^{-2\gamma(\tau-\tau_{0})}}{8\Omega^{2}}
\left[(1-\frac{i\Omega}{\gamma}-e^{2i\Omega(\tau-\tau_{0})})(i\pi-2\log(\gamma-i\Omega)+\Gamma(0,(-\gamma+i\Omega)(\tau-\tau_{0})))\right.\nonumber\\
&
&+\left.(1+\frac{i\Omega}{\gamma}-e^{-2i\Omega(\tau-\tau_{0})})(-3i\pi-2\log(\gamma+i\Omega)+\Gamma(0,-(\gamma+i\Omega)(\tau-\tau_{0})))\right]+\frac{i}{8\Omega\gamma}\nonumber\\
&
&\cdot\left[-2i\pi+2\log(\gamma-i\Omega)-2\log(\gamma+i\Omega)-\Gamma(0,(\gamma-i\Omega)(\tau-\tau_{0}))+\Gamma(0,(\gamma+i\Omega)(\tau-\tau_{0}))\right]\}\,,\label{2pvcform}
\end{eqnarray}

where $\tilde{\Lambda}_{0}$ contains divergent parts [i.e.,
$\Gamma(0,0)$ and $\log(0)$] as $\tau''\rightarrow\tau$ and
$\tau''_{0}\rightarrow\tau_{0}$ and are absorbed into the
renormalized constant or coefficient in the experiment.

Similarly, the result of  $\langle \dot{Q}(\eta)^{2}\rangle$ as
\begin{eqnarray}
& &\langle \dot{Q}(\eta)^{2}\rangle_{v}\equiv
\lim_{\eta''\rightarrow\eta}\frac{1}{2}\langle \{Q(\eta),Q(\eta'')\}\rangle_{v}\nonumber\\
&=&\frac{2\hbar\gamma}{\pi
m_{0}}\theta(\eta)Re\{\tilde{\Lambda}_{0_{v}}+\frac{e^{-2\gamma(\tau-\tau_{0})}}{8\Omega^{2}}\left[
(\frac{(\gamma-i\Omega)(\gamma^{2}+\Omega^{2})}{\gamma}-(\gamma-i\Omega)^{2}e^{2i\Omega(\tau-\tau_{0})})(-i\pi-2\log(\gamma-i\Omega))\right.\nonumber\\
&
&+\left.(\frac{(\gamma+i\Omega)(\gamma^{2}+\Omega^{2})}{\gamma}-(\gamma+i\Omega)^{2}e^{-2i\Omega(\tau-\tau_{0})})(-i3\pi-2\log(\gamma+i\Omega))\right]\nonumber\\
&
&+\frac{e^{(-\gamma+i\Omega)(\tau-\tau_{0})}}{8\Omega^{2}}\left[\gamma-i\Omega-\frac{(\gamma-i\Omega)^{2}}{\gamma}\right]
\left[\frac{2}{\tau-\tau_{0}}-2(\gamma-i\Omega)e^{(\gamma-i\Omega)(\tau-\tau_{0})}\Gamma(0,(\gamma-i\Omega)(\tau-\tau_{0}))\right]\nonumber\\
&
&+\frac{e^{(-\gamma-i\Omega)(\tau-\tau_{0})}}{8\Omega^{2}}\left[\gamma+i\Omega-\frac{(\gamma+i\Omega)^{2}}{\gamma}\right]
\left[\frac{2}{\tau-\tau_{0}}-2(\gamma+i\Omega)e^{(\gamma+i\Omega)(\tau-\tau_{0})}\Gamma(0,(\gamma+i\Omega)(\tau-\tau_{0}))\right]\nonumber\\
&
&+\frac{e^{-\gamma(\tau-\tau_{0})}}{8\Omega^{2}}\left[(-2i\sin(\Omega(\tau-\tau_{0}))+\frac{i\Omega}{\gamma}e^{i\Omega}(\tau-\tau_{0}))
\left[\frac{2(\gamma+i\Omega)}{\tau-\tau_{0}}-2(\gamma+i\Omega)^{2}e^{-(\gamma+i\Omega)(\tau-\tau_{0})}
(i\pi\right.\right.\nonumber\\
&
&\left.-\Gamma(0,-(\gamma+i\Omega)(\tau-\tau_{0})))\right]+(2i\sin(\Omega(\tau-\tau_{0}))-\frac{i\Omega}{\gamma}e^{-i\Omega}(\tau-\tau_{0}))
\left[\frac{2(\gamma-i\Omega)}{\tau-\tau_{0}}\right.\nonumber\\
&
&\left.\left.+2(\gamma-i\Omega)^{2}e^{-(\gamma-i\Omega)(\tau-\tau_{0})}\cdot(i\pi+\Gamma(0,-(\gamma-i\Omega)(\tau-\tau_{0})))\right]\right]+\frac{i}{8\Omega\gamma}
\cdot\nonumber\\
&
&\left[-(\gamma-i\Omega)^{2}(2\log(\gamma-i\Omega)-i\pi)+(\gamma+i\Omega)^{2}(2\log(\gamma+i\Omega)+i\pi)\right]\}\,,\label{vcPBv2}
\end{eqnarray}

where $\tilde{\Lambda}_{0_{v}}$ contains divergent parts[i.e.,
$\Gamma(0,0)$ and $\log(0)$] as $\tau''\rightarrow\tau$ and
$\tau''_{0}\rightarrow\tau_{0}$ that are absorbed into the
renormalized constant or coefficient in the experiment.

Note that the condition $\kappa\geq0$ is very important. If one does
not notice that $\kappa\geq0$ and takes the integration region of
$\kappa$ from $-\infty$ to $\infty$ in Eq. $(\ref{2pvc})$[ i.e.,
$\int_{-\infty}^{+\infty}\kappa d\kappa f(\kappa)$] , one will have
$\langle Q(\eta)^{2}\rangle_{v}=0$ in Eq. $(\ref{2pvcform})$ for the
inertial detector. This implies that the variance from the
background quantum field is zero, which is highly unlikely because a
quantum field always contributes a nonzero variance. Actually, this
strange result is the motivation for our rechecking the two-point
function $\langle Q(\eta)^{2}\rangle_{v}$ for the UAD and inertial
detectors. An interesting point is that Eq. $(\ref{2pvc})$ can be
reshaped in the form of Eq. $(\ref{2p2term1})$[i.e.,
$\int_{0}^{+\infty}\kappa d\kappa
f(\kappa)=\int_{-\infty}^{+\infty}\kappa d\kappa
f(\kappa)-\int_{-\infty}^{0}\kappa d\kappa f(\kappa)$]; thus, the
two-point function $\langle Q(\eta)^{2}\rangle_{v}$ is now nonzero,
and the variance for the inertial detector is nonzero if we insist
on taking the integration region of $\kappa$ as
$\int_{-\infty}^{+\infty}\kappa d\kappa f(\kappa)$ (i.e., this
integration region is what we applied in the previous calculations).
Therefore we think that the second term $\langle
Q(\eta)^{2}\rangle_{v2}$ is the missing term and is important when
we talk about the variance from the background quantum field for the
inertial detector. The missing term also changes the trend for the
two-point function $\langle \dot{Q}(\eta)^{2}\rangle_{v}$ curve in
the UAD case, as shown previously. Later, we will plot the curves of
the two-point functions $\langle Q(\eta)^{2}\rangle_{v}$ and
$\langle \dot{Q}(\eta)^{2}\rangle_{v}$ for the inertial detector.

\begin{figure}[htbp]
\centerline{\psfig{file=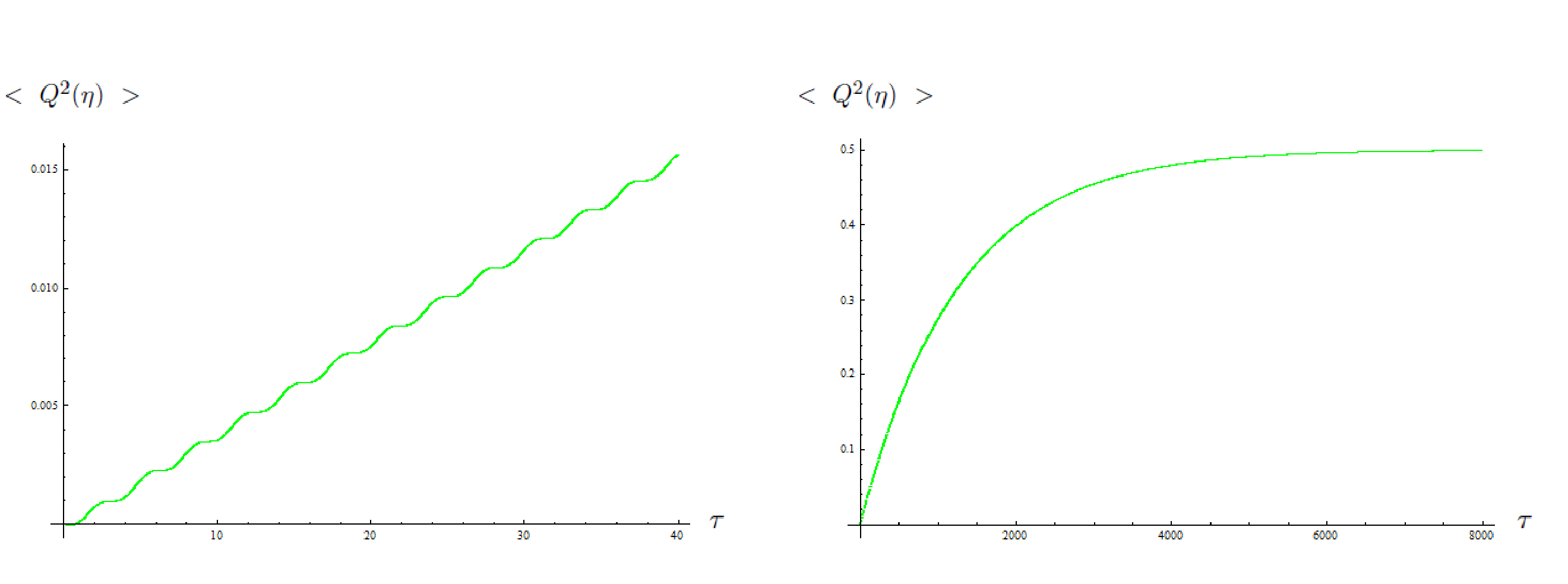, width=18cm}} \caption{ The two-point
correlation function $\langle Q(\eta)^{2}\rangle_{v}$ for an
inertial detector. Here $a=0.001$,
$\gamma=0.000398(\lambda_{0}=0.1)$, $m_{0}=\hbar=1$, and $\Omega=1$.
Two different timescales are shown. The value increases slowly in
the beginning with ripples on the curve, then reaches a saturated
value later.} \label{VD}
\end{figure}

\begin{figure}[htbp]
\centerline{\psfig{file=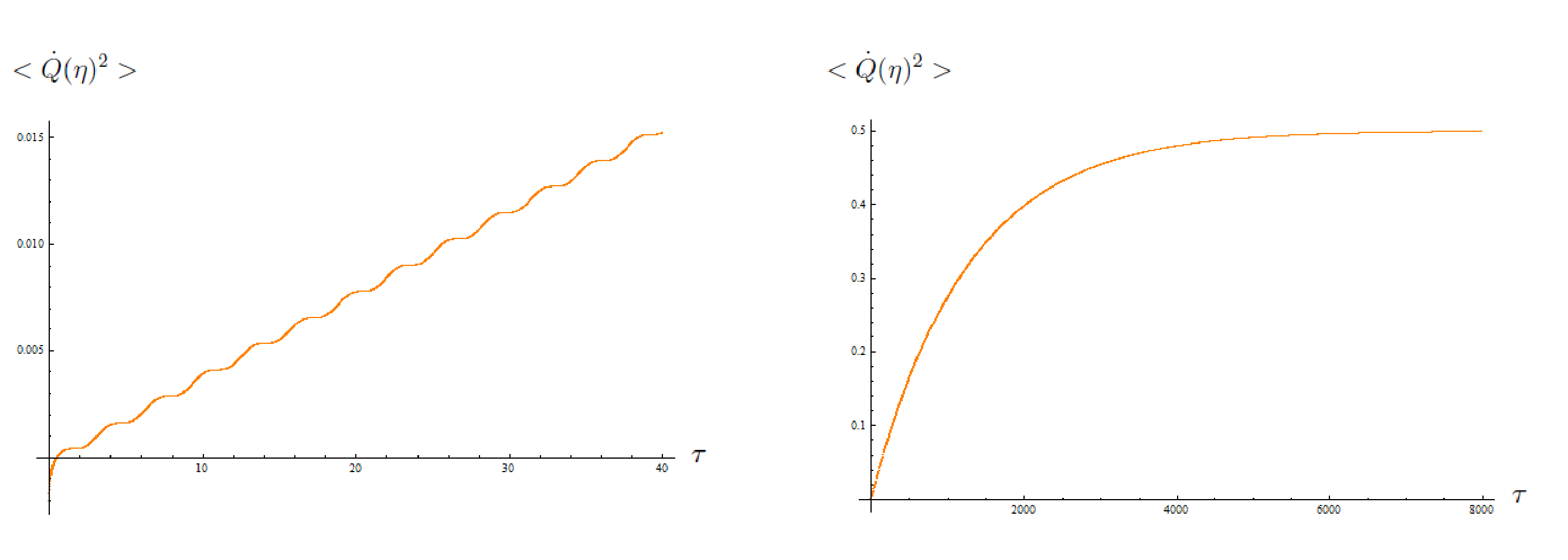, width=18cm}} \caption{The
two-point correlation function $\langle \dot{Q}(\eta)^{2} \rangle$
for an inertial detector. Here $a=0.001$,
$\gamma=0.000398(\lambda_{0}=0.1)$, $m_{0}=\hbar=1$, and $\Omega=1$.
These plots are similar to those  for the $\langle
Q(\eta)^{2}\rangle$ case. Two different timescales are shown. The
value of the two-point function increases slowly in the beginning
with ripples on the curve, and then reaches a saturated value
later.} \label{VD_PB}
\end{figure}

The numerical plots for $\langle Q(\eta)^{2}\rangle_{v}$ and
$\langle \dot{Q}(\eta)^{2}\rangle_{v}$ are shown  in Fig. ~\ref{VD}
and ~\ref{VD_PB}. The values for the two-point functions $\langle
Q(\eta)^{2}\rangle_{v}$ and $\langle \dot{Q}(\eta)^{2}\rangle_{v}$
follow the same trend. The value first increase slowly with ripples
on the curve, and then reache a saturated value. This differs from
the UAD case; for example, in Fig. ~\ref{A_m}, the amplitude of the
ripples gradually becomes small. For the UAD case, the early
amplitude
 is larger than the later amplitudes. Note that the value for the two-point correlation functions
 of UAD changes only slightly relative to the inertial detector
 . The curve for UAD is quite flat. The magnitude of $\langle
Q(\eta)^{2}\rangle_{v}$
 for the inertial detector has an obvious change from the beginning to the end.
 Besides, the ripples on the inertial detector curve of the two-point function $\langle
 Q(\eta)^{2}\rangle_{v}$ has the same oscillating amplitude on the ripples until it reaches the saturated value.
 The effect of acceleration is clear in the early-time region if we compare the UAD detector curve to the inertial detector curve.

Fig. ~\ref{VD_PB} shows $\langle \dot{Q}(\eta)^{2}\rangle$ for an
inertial detector. When we compare it to the Fig. ~\ref{A_PB} [i.e.,
$\langle \dot{Q}(\eta)^{2}\rangle$ for UAD], the curve for UAD
decreases to a saturated value, which differs from the inertial
case. This feature can be seen with the term $\langle
Q(\eta)^{2}\rangle$; for example, in Fig. ~\ref{A_m} the amplitude
of the oscillations gradually become small, which implies that the
changes of $\langle \dot{Q}(\eta)^{2}\rangle$ also become small as
$\tau$ increases. A clear feature of a UAD is that its magnitude of
$\langle \dot{Q}(\eta)^{2}\rangle$ decreases in the early-time
region.

In Figs. ~\ref{A_m}-~\ref{VD_PB} we can see that the difference
between the inertial detector and the UAD is clear in the plots for
the two-point correlation functions $\langle Q(\eta)^{2}\rangle$ and
$\langle \dot{Q}(\eta)^{2} \rangle$. For the two-point function
$\langle Q(\eta)^{2}\rangle$ of the UAD, the curve has larger
oscillations at first and then experiences smaller oscillations, and
the magnitude does not change much from beginning to the end. On the
contrary, the amplitude of the oscillations of the two-point
function $\langle Q(\eta)^{2}\rangle$ for the inertial detector does
not shrink in the beginning, and the magnitude increases from the
beginning until it reaches the saturated region. For the two-point
function $\langle \dot{Q}(\eta)^{2} \rangle$, the difference between
the UAD and the inertial detector is more obvious that the curve for
$\langle \dot{Q}(\eta)^{2} \rangle$ of the UAD decreases to a
saturated value, while the curve for the inertial detector
increases. The effect of proper acceleration is evident in the
two-point correlation functions $\langle Q(\eta)^{2}\rangle$ and
$\langle \dot{Q}(\eta)^{2} \rangle$. We think that the difference is
from the Unruh effect in that the uniformly accelerated detector
would experience a thermal bath at temperature $T_{U}=\hbar a/(2\pi
c\emph{k}_{B})$, where $a$ is the proper acceleration. This thermal
bath changes the two-point correlation functions.

%Fig.10
%%%%%%%%%%%%%%%%%%%%%%%%%%%%%%%%%%%%%%%%%%%%%%%%%%%%%%%%%%%%%%%%%%%%%%%%%%%%%%%%%%%%%%%%%%%%%%%%%%%%%%
%%%%%%%%%%%%%%%%%%%%%%%%%%%%%%%%%%%%%%%%%%%%%%%%%%%%%%%%%%%%%%%%%%%%%%%%%%%%%%%%%%%%%%%%%%%%%%%%%%%%%%

%Fig.11
%%%%%%%%%%%%%%%%%%%%%%%%%%%%%%%%%%%%%%%%%%%%%%%%%%%%%%%%%%%%%%%%%%%%%%%%%%%%%%%%%%%%%%%%%%%%%%%%%%%%%%
%%%%%%%%%%%%%%%%%%%%%%%%%%%%%%%%%%%%%%%%%%%%%%%%%%%%%%%%%%%%%%%%%%%%%%%%%%%%%%%%%%%%%%%%%%%%%%%%%%%%%%

%%%%%%%%%%%%%%%%%%%%%%%%%%%%%%%%%%%%%%%%%%%%%%%%%%%%%%%%%%%%%%%%%%%%%%%%%%%%%%%%%%%%
\section{summary}
%%%%%%%%%%%%%%%%%%%%%%%%%%%%%%%%%%%%%%%%%%%%%%%%%%%%%%%%%%%%%%%%%%%%%%%%%%%%%%%%%%%%

In this paper, we investigate the two moving detectors Alice and Bob
in a quantum field. In this system, Alice is static while Bob either
accelerates uniformly or moves at a constant velocity. We apply two
different types of trajectories for such a setup and calculate the
solutions for the internal degrees of freedom $Q$ for the moving
detector Bob under the influence of the background quantum field. In
this work, we find the following points:

$(1)$ The inertial worldline that we need for two moving detectors
to exchange the signals within a reasonable finite time interval
cannot be replaced by the UAD trajectory in Rindler space
$z^{\mu}_{B}=(a^{-1}\sinh a\tau, a^{-1}\cosh a\tau, 0, 0)$ by
setting the proper acceleration $a=0$. When the proper acceleration
in the Rindler space goes to zero, the UAD worldline is shifted to
very far away such that Alice and Bob no longer exchange signals
within a reasonable time interval. Therefore, we need to apply a
true trajectory $\tilde{z}^{\mu}_{B}=(\gamma\tau, \gamma
v\tau+x_{a}+d, 0, 0)$ for a detector moving at constant velocity. By
using the trajectory $\tilde{z}^{\mu}_{B}$, Bob is separated from
Alice at the distance "$d$ " in the beginning so that they can
exchange the signal in a finite time interval. We can apply this
trajectory to compute the two-point functions $\langle
Q(\eta)^{2}\rangle$ and $\langle \dot{Q}(\eta)^{2} \rangle$ for the
inertial detector, then compare the two-point correlation functions
$\langle Q(\eta)^{2}\rangle$ and $\langle \dot{Q}(\eta)^{2} \rangle$
for the UAD and inertial detector.

$(2)$We find that a term was missing from both two-point correlation
functions $\langle Q(\eta)^{2}\rangle_{v}$ and $\langle
\dot{Q}(\eta)^{2} \rangle_{v}$ in the previous calculations. Without
this term, the variance from the background quantum field part of
the inertial detector is "$0$ " [ i.e., $\langle
Q(\eta)^{2}\rangle_{v}=0$] , which is highly unlikely. However, if
the missing term is included, the variance from the background
quantum field for the inertial detector is nonzero, which is what we
expect when a moving detector interacts with a quantum field. The
missing term also changes the behavior of the two-point correlation
functions $\langle Q(\eta)^{2}\rangle_{v}$ and $\langle
\dot{Q}(\eta)^{2} \rangle_{v}$ for UAD; this point was not noticed
previously.

$(3)$The values of the parameters in this model are in an allowed
region. We apply the perturbation method to obtain the solutions for
$Q$; therefore, the parameters in this model should obey the basic
assumption for perturbations that the next leading order must be
smaller than the leading order. In the previous work, we did not
notice this and took the decay parameter $\gamma$ to be
$\gamma=0.1$, which means that the expansion parameter $\lambda_{0}$
is larger than $1$ (when $\gamma=0.1$, the coupling constant
$\lambda_{0}\simeq 1.58$). This value of $\gamma$ is inconsistent
with the basic assumption of the perturbation and will give us an
artefact. And this will lead to misleading results pertaining to the
effects of proper acceleration.

$(4)$ Including the above considerations, the UAD and inertial
detector result in different behaviors in $\langle
Q(\eta)^{2}\rangle_{v}$ and $\langle \dot{Q}(\eta)^{2} \rangle_{v}$.
In the early-time region the two-point function $\langle
Q(\eta)^{2}\rangle_{v}$ for UAD has a quite flat curve, while the
inertial detector has an increasing curve. The amplitude of those
oscillations on the ripples of the UAD curve gradually shrinks,
while the amplitude of the oscillations on the ripples of the
inertial detector does not change. For the two-point function
$\langle \dot{Q}(\eta)^{2} \rangle_{v}$, the difference is more
clear than it is for $\langle Q(\eta)^{2}\rangle_{v}$. The curve for
$\langle \dot{Q}(\eta)^{2} \rangle_{v}$ of the UAD is high at first
and then decreases until it reaches the saturated value. However,
the curve of $\langle \dot{Q}(\eta)^{2} \rangle_{v}$ for the
inertial detector increases until it reaches the saturated value.
This part is quite different from the previous result. We think that
this implies that the proper acceleration $a$ has some effect on the
vacuum state of $Q$ and thus affects the vacuum fluctuations of the
UAD(the Unruh effect).

The foundations of the calculations were built by Lin and
Hu~\cite{LH2005}, who used the quantum filed theory method and then
 applied it to the two moving detectors system by Lin \emph{et} \emph{al.}~\cite{LCH2008}. These are not easy calculations. Here, based on
their work, we redo the calculations and modify them. Since the
calculation is tricky, we write the detailed calculations here for
those who are interested.

In the future, we would like to apply these results to the quantum
teleportation issue, for example, the two moving detector system in
which Alice and Bob have relativistic motion with each other. We
would like to see whether the Unruh effect may play a role in the
quantum teleportation process for two relatively moving detectors.

%%%%%%%%%%%%%%%%%%%%%%%%%%%%%%%%%%%%%%%%%%%%%%%%%%%%%%%%%%%%%%%%%%%%%%%%%%%%%%%%%%%%
\section{ACKNOWLEDGEMENT}
%%%%%%%%%%%%%%%%%%%%%%%%%%%%%%%%%%%%%%%%%%%%%%%%%%%%%%%%%%%%%%%%%%%%%%%%%%%%%%%%%%%%

We would like to thank Dr. Shih-Yuin Lin, Dr. Chung-Hsien Chou, Dr.
Jen-Tsung Hsiang and Dr. Ron-Chou Hsieh for helpful discussions, and
 Professor Bei-Lok Hu and Professor
Kin-Wang Ng for the encouragement and help. Special thanks are given
to Dr. Shih-Yuin Lin for providing some detailed notes about the
calculations of the two-point functions. This work is supported in
part by the Ministry of Science and Technology, Taiwan under Grants
No. MOST $106$-$2811$-M-$108$-$006$ and No. MOST
$109$-$2112$-M-$001$-$003$.

%%%%%%%%%%%%%%%%%%%%%%%%%%%%%%%%%%%%%%%%%%%%%%%%%%%%%%%%%%%%%%%%%%%%%%%%%%%%%%%%%%%%%
\section*{APPENDIX A: QUANTIZATION}
%%%%%%%%%%%%%%%%%%%%%%%%%%%%%%%%%%%%%%%%%%%%%%%%%%%%%%%%%%%%%%%%%%%%%%%%%%%%%%%%%%%%%

We quantize the field $\Phi$ and the harmonic oscillators $Q_{A}$,
$Q_{B}$ in the Heisenberg picture. The conjugate momentum
$(P(\tau)$,$\Pi(x))$ of these canonical coordinate and momentum
$(Q(\tau),\Phi(x))$ are

\begin{equation} P_{d}(\tau)=\frac{\delta S}{\delta
\dot{Q_{d}}(\tau)}=\dot{Q}_{d}(\tau)\;, d=A,B,
\end{equation}
\begin{equation}
\Pi(\tau)=\frac{\delta S}{\delta
\partial_{t}\Phi(x)}=\partial_{t}\Phi(x)\;.
\end{equation}
The equal time commutation relations of these dynamical variables
are
\begin{eqnarray}
  [ \hat{Q}_{d}(\tau), \hat{P}_{d}(\tau) ] &=& i\hbar, d=A,B ;\label{QPCM} \\
  \left. [ \hat{\Phi}(t,{\bf x}),\hat{\Pi} (t,{\bf x'}) ]\right.
  &=& i\hbar\delta^3 ({\bf x}-{\bf x'}).\label{phipiCM}
\end{eqnarray}
According to the Heisenberg equations of motion, one can write the
equation of motions for $\hat{Q}$ and $\hat{\Phi}$ as
\begin{eqnarray}
  \partial_\tau^2\hat{Q}_{d}(\tau)+\Omega_0^2 \hat{Q}_{d}(\tau) &=&
  \lambda_0\hat{\Phi}_{d}(\tau,{\bf z}(\tau))\,,\, d=A, B, \label{eomq}
 \\
  \left( \partial_t^2-\nabla^2\right)\hat{\Phi}_{d}(x) &=& {\lambda_0}
  \int_0^\infty d\tau \hat{Q}_{d}(\tau)\delta^4(x-z(\tau)).\label{eomPhi}
\end{eqnarray}
The operators $\hat{Q}_{d}(\tau_{d})$ and $\hat{\Phi}_{d}(x_{d})$
are expanded by the mode functions and the creation(annihilation)
operators as

\begin{eqnarray}
  \hat{Q}_i(\tau_i) &=& \sqrt{\hbar\over 2\Omega_r}\sum_j\left[
    q_i^{(j)}(\tau_i)\hat{a}_j^{}+q_i^{(j)*}(\tau_i)\hat{a}_j^\dagger
    \right]
   +\int {d^3 k\over (2\pi)^3}\sqrt{\hbar\over 2\omega}
    \left[q_i^{(+)}(\tau_i,{\bf k})\hat{v}_{\bf k} +
    q_i^{(-)}(\tau_i,{\bf k})\hat{v}_{\bf k}^\dagger\right],
    \nonumber\\ \\
  \hat{\Phi}(x) &=& \sqrt{\hbar\over 2\Omega_r}\sum_j\left[
    f^{(j)}(x)\hat{a}_j^{}+f^{(j)*}(x)\hat{a}_j^\dagger \right]
    +\int {d^3 k\over (2\pi)^3}
    \sqrt{\hbar\over 2\omega}\left[f^{(+)}(x,{\bf k})\hat{v}_{\bf k}
    +f^{(-)}(x,{\bf k})\hat{v}_{\bf k}^\dagger\right],\nonumber\\
\end{eqnarray}
where $i,j = A,B$, $\tau_A = t$, $\tau_B=\tau$, $q_i^{(j)}$,
$q_i^{(\pm)}$, $f^{(j)}$, and $f^{(\pm)}$ are the \emph{c}-number
mode functions. The conjugate momenta are $\hat{P}_A(t)
=\partial_t\hat{Q}_A(t)$, $\hat{P}_B(\tau) =\partial_\tau \hat{Q}_B
(\tau)$, and $\hat{\Pi}(x) =
\partial_t\hat{\Phi}(x)$. The equations of motion for
the mode functions are as follows:
\begin{eqnarray}
    \left( \partial_{\tau_i}^2 + \Omega_0^2\right)q_i^{(j)}(\tau_i) &=&
      \lambda_0 f^{(j)}(z_i^\mu (\tau_i)), \label{eomqAB1}\\
    \left( \partial_t^2 - \nabla^2 \right)f^{(j)}(x) &=& \lambda_0
      \left[\int_0^{\infty} dt\, q_A^{(j)}\delta^4(x -z^{}_A(t))
     +\int_0^{\infty} d\tau\, q_B^{(j)}\delta^4 (x-z^{}_B(\tau)) \right],
     \label{feAB1}\\
  \left(\partial_{\tau_i}^2 + \Omega_0^2\right)q_i^{(+)}(\tau_i,{\bf k}) &=&
      \lambda_0 f^{(+)}(z_i^\mu(\tau_i), {\bf k}), \label{eomq+1} \\
    \left( \partial_t^2 - \nabla^2 \right)f^{(+)}(x,{\bf k}) &=& \lambda_0
      \left[\int_0^{\infty} dt\, q_A^{(+)}(t,{\bf k})\delta^4(x
      -z^{}_A(t))\right.\nonumber\\
     & & \left.+\int_0^{\infty}d\tau\,q_B^{(+)}(\tau,{\bf k})\delta^4(x-z^{}_B(\tau))
     \right] \label{fe+1} .
\end{eqnarray}

%%%%%%%%%%%%%%%%%%%%%%%%%%%%%%%%%%%%%%%%%%%%%%%%%%%%%%%%%%%%%%%%%%%%%%%%%%%%%%%%%%%%%
\section*{APPENDIX B: THE EQUATIONS OF MOTION, MODE FUNCTIONS AND STATES}
%%%%%%%%%%%%%%%%%%%%%%%%%%%%%%%%%%%%%%%%%%%%%%%%%%%%%%%%%%%%%%%%%%%%%%%%%%%%%%%%%%%%%
The equations of motion of one moving detector for the Lagrangian in
Eq.~(\ref{Stot1B}) are as follows:

\begin{eqnarray}
  \partial_\tau^2\hat{Q}(\tau)+\Omega_0^2 \hat{Q}(\tau) &=&
  \lambda_0\hat{\Phi}_{}(\tau,{\bf z}(\tau)), \label{eomq}
 \\
  \left( \partial_t^2-\nabla^2\right)\hat{\Phi}(x) &=& {\lambda_0}
  \int_0^\infty d\tau \hat{Q}_{}(\tau)\delta^4(x-z(\tau)).\label{eomPhi}
\end{eqnarray}
We assume that the system is prepared before $\tau=0$ and that the
coupling is turned on at $\tau=0$ when we allow all the dynamical
variables to begin to interact and evolve under the influence of one
another. The time evolution of $\hat{\Phi}(x)$ is a linear
transformation in the phase space spanned by the orthonormal basis
$( \hat{\Phi}({\bf x}), \hat{\Pi}({\bf x}) , \hat{Q},\hat{P})$, and
$\hat{\Phi}(x)$ can be expressed in the form
\begin{equation}
  \hat{\Phi}(t,{\bf x}) = \int d^3 x'\left[ f^\Phi(t,{\bf x},{\bf x'})
  \hat{\Phi}(0,{\bf x'})+f^\Pi(t,{\bf x},{\bf x'})\hat{\Pi}(0,{\bf x'})
  \right]+f^Q(x)\hat{Q}(0)+ f^P(x)\hat{P}(0). \label{defPhi}
\end{equation}
Here $f^\Phi(x,{\bf x'}), f^\Pi(x,{\bf x'}), f^Q(x)$ and $f^P(x)$
are \emph{c}-number functions. Similarly, the operator
$\hat{Q}(\tau)$ can be expressed as follows:
\begin{equation}
  \hat{Q}(\tau) = \int d^3 x' \left[q^\Phi(\tau,{\bf x'})
    \hat{\Phi}(0,{\bf x'})+q^\Pi(\tau, {\bf x'})\hat{\Pi}(0,{\bf x'})
    \right]+ q^Q(\tau) \hat{Q}(0)+ q^P(\tau) \hat{P}(0), \label{defq}
\end{equation}
with \emph{ c}-number functions $q^Q(\tau), q^P(\tau),
q^\Phi(\tau,{\bf x'})$ and $q^\Pi(\tau,{\bf x'})$.

For the case in which initial operators are the free field
operators, namely, $\hat{\Phi} (0,{\bf x})=\hat{\Phi}_0({\bf x})$,
$\hat{\Pi}(0,{\bf x})=\hat{\Pi}_0({\bf x})$, $\hat{Q}(0)=\hat{Q}_0$
and $\hat{P}(0)=\hat{P}_0$, one can go further by introducing the
following complex operators $\hat{v}_{\bf k}$ and $\hat{a}$:
\begin{eqnarray}
  \hat{\Phi}_0({\bf x}) &=& \int {d^3k\over (2\pi)^3}
    \sqrt{\hbar\over 2\omega}\left[ e^{i{\bf k\cdot x}}\hat{v}_{\bf k}
    + e^{-i{\bf k\cdot x}}\hat{v}^\dagger_{\rm k}\right] ,\\
  \hat{\Pi}_0({\bf x}) &=& \int {d^3k\over (2\pi)^3}
    \sqrt{\hbar\over 2\omega}(-i\omega )\left[ e^{i{\bf k\cdot x}}
    \hat{v}_{\bf k}-e^{-i{\bf k\cdot x}}\hat{v}^\dagger_{\bf k}\right]
\end{eqnarray}
with $\omega \equiv |{\bf k}|$, and
\begin{equation}
  \hat{Q}_0 = \sqrt{\hbar\over 2\Omega_r m_0}(\hat{a}+\hat{a}^\dagger),
  \;\;\;\;\;
  \hat{P}_0 = -i\sqrt{\hbar\Omega_rm_0\over 2}( \hat{a}-\hat{a}^\dagger ).
\end{equation}
Note that, instead of $\Omega_0$, we use the renormalized natural
frequency $\Omega_r$ [to be defined in Eq.~(\ref{renormO})] in the
definition of $\hat{a}$. Then the commutation relations
$(\ref{QPCM})$ and $(\ref{phipiCM})$ give
\begin{equation}
  [ \hat{a},\hat{a}^\dagger ]=1, \;\;\;\;\;
  [ \hat{v}_{\bf k}, \hat{v}_{\bf k'}^\dagger ] =
    (2\pi)^3\delta^3({\bf k}-{\bf k'}),
\end{equation}
and the expressions $(\ref{defPhi})$ and $(\ref{defq})$ can be
rewritten as
\begin{eqnarray}
  \hat{\Phi}(t,{\bf x}) &=&\hat{\Phi}_v(x) +\hat{\Phi}_a(x),\label{phiab}\\
  \hat{Q}(\tau) &=& \hat{Q}_v(\tau) + \hat{Q}_a(\tau)  \label{qab}
\end{eqnarray}
where
\begin{eqnarray}
  \hat{\Phi}_v(x) &=& \int {d^3k\over (2\pi)^3}\sqrt{\hbar\over 2\omega}
    \left[f^{(+)}(t,{\bf x};{\bf k})\hat{v}_{\bf k} +
    f^{(-)}(t,{\bf x};{\bf k})\hat{v}_{\bf k}^\dagger
    \right],\label{Phiv}\\
  \hat{\Phi}_a(x) &=& \sqrt{\hbar \over 2\Omega_r m_0}\left[f^a(t,{\bf x})
    \hat{a}+ f^{a*}(t,{\bf x})\hat{a}^\dagger \right], \label{PhiB}\\
  \hat{Q}_v(\tau) &=& \int {d^3 k\over (2\pi)^3}\sqrt{\hbar\over 2\omega}
  \left[q^{(+)}(\tau,{\bf k})\hat{v}_{\bf k} +
    q^{(-)}(\tau,{\bf k})\hat{v}_{\bf k}^\dagger\right],\label{Qv} \\
  \hat{Q}_a(\tau) &=& \sqrt{\hbar\over 2\Omega_r m_0}\left[
    q^a(\tau)\hat{a}+q^{a*}(\tau)\hat{a}^\dagger \right] .\label{QB}
\end{eqnarray}
The entire problem, therefore, can be transformed by solving
c-number functions $f(x)$ and $q(\tau)$ from Eq.~(\ref{eomq}) and
~(\ref{eomPhi}) with suitable initial conditions. Since $\hat{Q}$
and $\hat{\Phi}$ are Hermitian, one has $f^{(-)}= (f^{(+)})^*$ and
$q^{(-)}= (q^{(+)})^*$. Hence, it is sufficient to solve the
\emph{c}-number functions $f^{(+)}(t,{\bf x};{\bf k})$,
$q^{(+)}(\tau,{\bf k})$, $f^a(t,{\bf x})$ and $q^a(\tau)$. To place
this in a more general setting, let us perform a Lorentz
transformation shifting $\tau =0$ to $\tau=\tau_0$, and let us
define
\begin{equation}
  \eta \equiv \tau -\tau_0.
\end{equation}
Now the coupling between the detector and the field would be turned
on at $\tau=\tau_0$. We are looking for solutions with the initial
conditions such as the following:
\begin{eqnarray}
 && f^{(+)}(t(\tau_0),{\bf x};{\bf k}) = e^{i{\bf k\cdot x}},\;\;\;
    \partial_t f^{(+)}(t(\tau_0),{\bf x};{\bf k})=
      -i\omega e^{i{\bf k\cdot x}}, \;\;\;
q^{(+)}(\tau_0;{\bf k})= \dot{q}^{(+)}(\tau_0;{\bf k}) = 0, \label{IC+}\\
 && f^a (t(\tau_0),{\bf x}) =\partial_t f^a (t(\tau_0),{\bf x}) =0,
 \;\;\; q^a(\tau_0) = 1, \;\;\; \dot{q}^a(\tau_0)= -i\Omega_r. \label{ICa}
\end{eqnarray}

The solutions for $f^{(+)}_{0}$, $f^{(+)}$, $q^{(+)}$, $f^{a}$, and
$q^{a}$ are as follows (detailed calculations for one moving
detector Bob are written in Appendix C). The general solution for
$f^{(+)}$ reads
\begin{equation}
  f^{(+)}(x;{\bf k}) = f_0^{(+)}(x;{\bf k})
  + f_1^{(+)}(x;{\bf k}), \label{Phi+}
\end{equation}
where
\begin{equation}
  f_0^{(+)}(x;{\bf k})\equiv  e^{-i\omega t+i{\bf k\cdot x}}
  \label{freephi}
\end{equation}
is the free field solution and
\begin{equation}
  f^{(+)}_1(z(\tau);{\bf k}) =%\lim_{\Lambda\to\infty}
  {\lambda_0\over 4\pi}\left[
  \Lambda \zeta q^{(+)}(\tau;{\bf k})
  -\partial_\tau q^{(+)}(\tau;{\bf k}) + O(\Lambda^{-1})\right],
\end{equation}
where $\zeta = 2^{7/4} \Gamma ( 5/4 )/\sqrt{\pi}$ and $\Lambda$ is
about the regularization scheme.

The mode function of the internal degrees of freedom $\hat{Q}$ about
the vacuum fluctuations part is
\begin{equation}
  q^{(+)}(\tau;{\bf k}) = {\lambda_0\over m_0 }\sum_{j=+,-}
  \int^\tau_{\tau_0} d\tau' c_j e^{w_j(\tau-\tau')}
  f^{(+)}_0(z(\tau');{\bf k}),\label{q+1}
\end{equation}
where $c_\pm = \pm {1\over 2i\Omega},\;\;w_\pm = -\gamma \pm
i\Omega,$ with $\Omega \equiv \sqrt{\Omega_r^2 -\gamma^2}$:

\begin{equation}
  f^a(x)= {\lambda_0\theta(\eta_-)  \over 2\pi a X}q^a(\tau_-),
  \label{Phibfinal}
\end{equation}
where $X$ is as defined in Appendix C.

The mode functions of the internal degrees of freedom $\hat{Q}$
about the intrinsic part are
\begin{equation}
  q^a(\tau) = {1\over 2}\theta(\eta)e^{-\gamma\eta}
  \left[\left(1-{\Omega_r+ i\gamma\over\Omega}\right)e^{i\Omega\eta} +
  \left(1+{\Omega_r+ i\gamma\over\Omega}\right)e^{-i\Omega\eta}\right].
\label{qb}
\end{equation}

Above is the general form of the solutions. The explicit solutions
will depend on the specific worldline of the detector(the trajectory
for the moving detector in spacetime).

As is shown above, when $\hat{Q}$ evolves,  some nonzero terms
proportional to $\hat{\Phi}$ and $\hat{\Pi}$ will be generated.
Suppose that the detector is initially prepared in a state that can
be factorized into the quantum state $\left|\right.q\left. \right>$
for $Q$ and the Minkowski vacuum $|\left. 0_M\right> $ for the
scalar field $\Phi$, that is,
\begin{equation}
  | \tau_0 \left.\right> =  \left|\right.q\left. \right> |\left.
  0_M\right>.
\label{initQS}
\end{equation}
The two-point function of $Q$ will then split into two parts,
\begin{eqnarray}
  \left<\right. Q(\tau)Q(\tau')\left.\right>  &=&  \left<\right. 0_M|
  \left< \right. q\left.\right| \left[\hat{Q}_v(\tau) + \hat{Q}_a(\tau)\right]
  \left[\hat{Q}_v(\tau) + \hat{Q}_a(\tau)\right]\left|\right.q\left. \right>
  |0_M\left.\right> \nonumber\\ &=& \left<\right.q \,|\, q\left.\right>
  \left< \right. Q(\tau)Q(\tau')\left.\right>_{\rm v}+ \left< \right.
  Q(\tau)Q(\tau')\left.\right>_{\rm a}\left< 0_M| 0_M\right>, \label{splitQQ}
\end{eqnarray}
where, from Eq.~(\ref{qab}),
\begin{eqnarray}
  \left< \right. Q(\tau)Q(\tau')\left.\right>_{\rm v} &=&  \left< 0_M\right.|
    \hat{Q}_v(\tau)\hat{Q}_v(\tau')|\left. 0_M\right>, \label{defQQv}\\
  \left< \right. Q(\tau)Q(\tau')\left.\right>_{\rm a} &=&
  \left< \right. q\left.\right|\hat{Q}_a(\tau)\hat{Q}_a(\tau)
  \left|\right.q\left. \right>. \label{defQQb}
\end{eqnarray}
Similar splitting happens for every two-point function of
$\hat{\Phi}(x)$.

Observe that $\left< \right. Q(\tau)Q(\tau')\left.\right>_{\rm v}$
depends on the initial state of the field, or the Minkowski vacuum,
while $\left< \right. Q(\tau)Q(\tau') \left.\right>_{\rm a}$ depends
on the initial state of the detector only. One can thus interpret
$\left< \right. Q(\tau)Q(\tau')\left. \right>_{\rm v} $ as
accounting for the response to the vacuum fluctuations, while
$\left< \right. Q(\tau)Q(\tau')\left. \right>_{\rm a}$ corresponds
to the intrinsic quantum fluctuations in the detector. Here we will
focus on the $\left< \right. Q(\tau)Q(\tau')\left. \right>_{\rm v} $
part (the response to the vacuum fluctuations) and demonstrate the
explicit forms of the two-point correlation functions.

%%%%%%%%%%%%%%%%%%%%%%%%%%%%%%%%%%%%%%%%%%%%%%%%%%%%%%%%%%%%%%%%%%%%%%%%%%%%%%%%%%%%%%%%%%%
\section*{APPENDIX C: SOLVING FOR $f^{(+)}_{0}$, $f^{(+)}$, $q^{(+)}$, $f^{a}$, AND $q^{a}$}
%%%%%%%%%%%%%%%%%%%%%%%%%%%%%%%%%%%%%%%%%%%%%%%%%%%%%%%%%%%%%%%%%%%%%%%%%%%%%%%%%%%%%%%%%%%

The method for obtaining $f$ and $q$ is analogous to what we did in
classical field theory~\cite{LH2005}. We first find an expression
relating the harmonic oscillator to the field amplitude right at the
detector. Substituting this relation into the equation of motions
for the oscillator, we then obtain the equation of motion for $q$
using the information from the field. We then solve this equation of
motion for $q$ and, from its solution, determine the field $f$
consistently.

Eq.$(\ref{eomPhi})$ implies that
\begin{equation}
  (\partial_t^2-\nabla^2)f^{(+)}(x;{\bf k}) =
    {\lambda_0}\int_{\tau_0}^\infty d\tau
    \delta^4(x-z(\tau))q^{(+)}(\tau;{\bf k}).
\label{eomPhi+}
\end{equation}
The general solution for $f^{(+)}$ reads
\begin{equation}
  f^{(+)}(x;{\bf k}) = f_0^{(+)}(x;{\bf k})
  + f_1^{(+)}(x;{\bf k}), \label{Phi+}
\end{equation}
where
\begin{equation}
  f_0^{(+)}(x;{\bf k})\equiv  e^{-i\omega t+i{\bf k\cdot x}}
  \label{freephi}
\end{equation}
is the free field solution and
\begin{equation}
  f_1^{(+)}(x;{\bf k})\equiv {\lambda_0} \int_{\tau_0}^\infty d\tau
    G_{\rm ret}(x;z(\tau))q^{(+)}(\tau;{\bf k})
\label{fretard}
\end{equation}
is the retarded solution, which looks like the retarded field in
classical field theory. Here $\omega=|{\bf k}|$ and the retarded
Green's function $G_{\rm ret}$ in Minkowski space is given by
\begin{equation}
  G_{\rm ret}(x,x') = {1\over 4\pi}\delta(\sigma)\theta(t-t') \label{Gret}
\end{equation}
with $\sigma\equiv -(x_\mu-x'_\mu)(x^\mu-x'^\mu)/2$. Applying the
explicit form of the retarded Green's function, one can go further
to write
\begin{equation}
  f_1^{(+)}(x;{\bf k}) =
    {\lambda_0\theta(\eta_-)  \over 2\pi a X}q^{(+)}(\tau_-;{\bf k}),
    \label{Phi+1}
\end{equation}
where
\begin{eqnarray}
  X &\equiv& \sqrt{(-UV+\rho^2+a^{-2})^2+4a^{-2}UV}, \label{defX}\\
  \tau_- &\equiv& -{1\over a}\ln {a\over 2|V|}\left(X-UV+\rho^2+a^{-2}
  \right),\label{deftaum}\\
  \eta_- &\equiv& \tau_- -\tau_0,
\end{eqnarray}
with $\rho \equiv \sqrt{x_2{}^2 +x_3{}^2}$, $U\equiv t-x^1$ and $V
\equiv t+x^1$.

The formal retarded solution $(\ref{Phi+1})$ is singular on the
trajectory of the detector. To deal with the singularity, note that
the UAD detector here is a quantum mechanical object, and also that
the detector number would always be $1$. This means that at the
energy threshold of detector creations, there is a natural cutoff of
the frequency, which sets an upper bound on the resolution to be
explored in our theory. Thus, it is justified to assume here that
the detector has a finite extent $O(\Lambda^{-1})$, which will
introduce the backreaction on the detector.

Let us regularize the retarded Green's function by invoking the
essence of effective field theory:
\begin{equation}
  G^\Lambda_{\rm ret}(x,x') =
%{\Lambda \over 4\pi\sqrt{\sigma}} \theta(t-t')
%\theta(\sigma) J_1(\Lambda\sqrt{\sigma}),
  {1\over 4\pi}\sqrt{8\over \pi}\Lambda^2 e^{-2\Lambda^4\sigma^2}
  \theta(t-t') . \label{GretL}
\end{equation}
(For more details on this regularization scheme, see
Refs.\cite{JH2002,GH1}.)
%where $J_1(s)$ is the Bessel function.
Taking this, right on the trajectory, the retarded solution for
large $\Lambda$ is
\begin{equation}
  f^{(+)}_1(z(\tau);{\bf k}) =%\lim_{\Lambda\to\infty}
  {\lambda_0\over 4\pi}\left[
  \Lambda \zeta q^{(+)}(\tau;{\bf k})
  -\partial_\tau q^{(+)}(\tau;{\bf k}) + O(\Lambda^{-1})\right],
\end{equation}
where $\zeta = 2^{7/4} \Gamma ( 5/4 )/\sqrt{\pi}$. Substituting the
above expansion into Eq.~(\ref{eomq}) and neglecting the
$O(\Lambda^{-1})$ terms, one obtains the following equation of
motion for $q^{(+)}$ with backreaction:
\begin{equation}
  (\partial_\tau^2 +2\gamma\partial_\tau+ \Omega_r^2)q^{(+)}(\tau;{\bf k}) =
  {\lambda_0\over m_0} f^{(+)}_0(z(\tau);{\bf k}).\label{eomq2}
\end{equation}
Fortunately, there is no higher derivative of $q$ present in the
above equation of motion. Now $q^{(+)}$ behaves like a damped
harmonic oscillator driven by the vacuum fluctuations of the scalar
field, with the damping constant
\begin{equation}
  \gamma \equiv {\lambda_0^2\over 8\pi m_0},
\end{equation}
and the renormalized natural frequency
\begin{equation}
  \Omega_r^2 \equiv \Omega_0^2-{\lambda_0^2\Lambda\zeta \over 4\pi m_0}.
  \label{renormO}
\end{equation}

In Eq.~(\ref{eomq2}), the solution for $q^{(+)}$ compatible with the
initial conditions $q^{(+)}(\tau_0;{\bf
k})=\dot{q}^{(+)}(\tau_0;{\bf k})=0$ is
\begin{equation}
  q^{(+)}(\tau;{\bf k}) = {\lambda_0\over m_0 }\sum_{j=+,-}
  \int^\tau_{\tau_0} d\tau' c_j e^{w_j(\tau-\tau')}
  f^{(+)}_0(z(\tau');{\bf k}),\label{q+1}
\end{equation}
where $f^{(+)}_0$ was given in Eq.~(\ref{freephi}), $c_\pm$ and
$w_\pm$ are defined as
\begin{equation}
  c_\pm = \pm {1\over 2i\Omega}, \;\;\;
  w_\pm = -\gamma \pm i\Omega,
\end{equation}
with
\begin{equation}
  \Omega \equiv \sqrt{\Omega_r^2 -\gamma^2}. \label{bigW}
\end{equation}
Throughout this paper we consider only the underdamped case with
$\gamma^2 < \Omega_r^2$, so $\Omega$ is always real.

Similarly, from Eq. $(\ref{eomq})$, $(\ref{eomPhi})$,
$(\ref{phiab})$ and $(\ref{qab})$, the equations of motion for $f^a$
and $q^a$ read
\begin{eqnarray}
  &&   \left( \partial_t^2-\nabla^2\right)f^a(x)= \lambda_0\int d\tau
    \delta^4(x-z(\tau))q^a(\tau),\\
  && (\partial_\tau^2 +\Omega_0^2)q^a (\tau) = {\lambda_0\over m_0}
    f^a (z(\tau)). \label{eomqb}
\end{eqnarray}
The general solution for $f^a$, as in Eq. $(\ref{Phi+})$, is
\begin{equation}
  f^a(x) = f_0^a(x) +{\lambda_0}\int_{\tau_0}^\infty d\tau
  G_{\rm ret}(x;z(\tau)) q^a(\tau_-) \label{Phib}\,.
\end{equation}
However, according to the initial condition $(\ref{ICa})$, one has
$f^a_0 = 0$; hence,
\begin{equation}
  f^a(x)= {\lambda_0\theta(\eta_-)  \over 2\pi a X}q^a(\tau_-).
  \label{Phibfinal}
\end{equation}
Again, the value of $f^a$ is singular right at the position of the
detector. Performing the same regularization as was  given for
$q^{(+)}$, Eq.$(\ref{eomqb})$ becomes
\begin{equation}
 \left(\partial_\tau^2 + 2\gamma \partial_\tau + \Omega_r^2\right)
 q^a (\tau) = 0, \label{noforceeom}
\end{equation}
which describes a damped harmonic oscillator free of driving force.
The solution consistent with the initial condition $q^a(\tau_0)=1$
and $\dot{q}^b(\tau_0)= -i\Omega_r$ reads
\begin{equation}
  q^a(\tau) = {1\over 2}\theta(\eta)e^{-\gamma\eta}
  \left[\left(1-{\Omega_r+ i\gamma\over\Omega}\right)e^{i\Omega\eta} +
  \left(1+{\Omega_r+ i\gamma\over\Omega}\right)e^{-i\Omega\eta}\right].
\label{qb}
\end{equation}

%%%%%%%%%%%%%%%%%%%%%%%%%%%%%%%%%%%%%%%%%%%%%%%%%%%%%%%%%%%%%%%%%%%%%%%%%%%%%%%%%%%%%%%%%%%
\section*{APPENDIX D: THE INTEGRATION OF $\kappa$ }
%%%%%%%%%%%%%%%%%%%%%%%%%%%%%%%%%%%%%%%%%%%%%%%%%%%%%%%%%%%%%%%%%%%%%%%%%%%%%%%%%%%%%%%%%%%

\begin{eqnarray}
& &\langle Q(\tau-\tau_{0})Q(\tau''-\tau''_{0})\rangle_{v}\nonumber\\
&=&\frac{\lambda_{0}^{2}\hbar}{2m_{0}^{2}(2\pi)^{2}}\left[\sum_{j,j'=\pm}C_{j}C^{\ast}_{j'}\int_{-\infty}^{\infty}\frac{\kappa
d\kappa
e^{-i\kappa(\tau_{0}-\tau''_{0})}(e^{w_{j}(\tau-\tau_{0})}-e^{i\kappa(\tau_{0}-\tau)})(e^{w^{\ast}_{j'}(\tau''-\tau''_{0})}-e^{i\kappa(\tau''-\tau''_{0})})}
{(1-e^{-2\pi\kappa/a})(w_{j}+i\kappa)(w^{\ast}_{j'}-i\kappa)}\right.\nonumber\\
&
&\left.-\sum_{j,j'=\pm}C_{j}C^{\ast}_{j'}\int_{-\infty}^{0}\frac{\kappa
d\kappa
e^{-i\kappa(\tau_{0}-\tau''_{0})}(e^{w_{j}(\tau-\tau_{0})}-e^{i\kappa(\tau_{0}-\tau)})(e^{w^{\ast}_{j'}(\tau''-\tau''_{0})}-e^{i\kappa(\tau''-\tau''_{0})})}
{(w_{j}+i\kappa)(w^{\ast}_{j'}-i\kappa)}\right]\nonumber\\
&=&\langle QQ\rangle_{v1}-\langle QQ\rangle_{v2}\,,
\end{eqnarray}

\begin{eqnarray}
& &\langle QQ\rangle_{v1}\nonumber\\
&=&\frac{\lambda_{0}^{2}\hbar}{2m_{0}^{2}(2\pi)^{2}}\sum_{j,j'=\pm}C_{j}C^{\ast}_{j'}\int_{-\infty}^{\infty}\frac{\kappa
d\kappa
e^{-i\kappa(\tau_{0}-\tau''_{0})}(e^{w_{j}(\tau-\tau_{0})}-e^{i\kappa(\tau_{0}-\tau)})(e^{w^{\ast}_{j'}(\tau''-\tau''_{0})}-e^{i\kappa(\tau''-\tau''_{0})})}
{(1-e^{-2\pi\kappa/a})(w_{j}+i\kappa)(w^{\ast}_{j'}-i\kappa)}\nonumber\\
&=&\frac{\lambda_{0}^{2}\hbar}{2m_{0}^{2}(2\pi)^{2}}\sum_{j,j'=\pm}\frac{C_{j}C^{\ast}_{j'}}{w_{j}+w^{\ast}_{j'}}
\int_{-\infty}^{\infty}\frac{d\kappa}{1-e^{-2\pi\kappa/a}}\left(\frac{w_{j}}{\kappa-iw_{j}}+\frac{w^{\ast}_{j'}}{\kappa+iw^{\ast}_{j'}}\right)
(e_{1}+e_{2}+e_{3}+e_{4})\nonumber\\
&=&P_{1}+P_{2}+P_{3}+P_{4}\,,
\end{eqnarray}

\begin{equation}
e_{1}=e^{-i\kappa(\tau_{0}-\tau''_{0})+w_{j}(\tau-\tau_{0})+w^{\ast}_{j'}(\tau''-\tau''_{0})}\,,
\end{equation}

\begin{equation}
e_{2}=-e^{w_{j}(\tau-\tau_{0})+i\kappa(\tau''-\tau_{0})}\,,
\end{equation}

\begin{equation}
e_{3}=-e^{w^{\ast}_{j'}(\tau''-\tau''_{0})-i\kappa(\tau-\tau''_{0})}\,,
\end{equation}

\begin{equation}
e_{4}=e^{i\kappa(\tau''-\tau)}\,.
\end{equation}

where
\begin{eqnarray}
P_{1}&=&\frac{\lambda_{0}^{2}\hbar}{2m_{0}^{2}(2\pi)^{2}}\sum_{j,j'=\pm}\frac{C_{j}C^{\ast}_{j'}}{w_{j}+w^{\ast}_{j'}}
\int_{-\infty}^{\infty}\frac{d\kappa}{1-e^{-2\pi\kappa/a}}\left(\frac{w_{j}}{\kappa-iw_{j}}+\frac{w^{\ast}_{j'}}{\kappa+iw^{\ast}_{j'}}\right)
e^{-i\kappa(\tau_{0}-\tau''_{0})+w_{j}(\tau-\tau_{0})+w^{\ast}_{j'}(\tau''-\tau''_{0})}\nonumber\\
&=&\frac{\lambda_{0}^{2}\hbar}{2m_{0}^{2}(2\pi)^{2}}\sum_{j,j'=\pm}\frac{C_{j}C^{\ast}_{j'}}{w_{j}+w^{\ast}_{j'}}
e^{w_{j}(\tau-\tau_{0})+w^{\ast}_{j'}(\tau''-\tau''_{0})}\cdot(-2\pi
i)\cdot\nonumber\\
& &\left[\frac{a}{2\pi}\sum^{-1}_{n=-\infty}\left(\frac{w_{j}}{i n
a-iw_{j}}+\frac{w^{\ast}_{j'}}{i n
a+iw^{\ast}_{j'}}\right)e^{-na(\tau_{0}-\tau''_{0})}+\frac{w_{j}e^{w_{j}(\tau_{0}-\tau''_{0})}}{1-e^{-2i\pi
w^{\ast}_{j}}/a}\right]\nonumber\\
&=&\frac{\lambda_{0}^{2}\hbar}{2m_{0}^{2}(2\pi)^{2}}\sum_{j,j'=\pm}\frac{C_{j}C^{\ast}_{j'}}{w_{j}+w^{\ast}_{j'}}
e^{w_{j}(\tau-\tau_{0})+w^{\ast}_{j'}(\tau''-\tau''_{0})}\cdot\nonumber\\
&
&\left[\frac{w_{j}e^{-a(\tau_{0}-\tau''_{0})}}{1+w_{j}/a}F_{w_{j}}(e^{-a(\tau_{0}-\tau''_{0})})
+\frac{w^{\ast}_{j'}e^{-a(\tau_{0}-\tau''_{0})}}{1-w^{\ast}_{j'}/a}F_{-w^{\ast}_{j'}}(e^{-a(\tau_{0}-\tau''_{0})})
-\frac{2\pi iw_{j}e^{w_{j}(\tau_{0}-\tau''_{0})}}{1-e^{-2i\pi
w_{j}}/a}\right]\,.\label{accp1}
\end{eqnarray}

\begin{eqnarray}
P_{2}&=&\frac{-\lambda_{0}^{2}\hbar}{2m_{0}^{2}(2\pi)^{2}}\sum_{j,j'=\pm}\frac{C_{j}C^{\ast}_{j'}}{w_{j}+w^{\ast}_{j'}}
\int_{-\infty}^{\infty}\frac{d\kappa}{1-e^{-2\pi\kappa/a}}\left(\frac{w_{j}}{\kappa-iw_{j}}+\frac{w^{\ast}_{j'}}{\kappa+iw^{\ast}_{j'}}\right)
e^{w_{j}(\tau-\tau_{0})+i\kappa(\tau''-\tau_{0})}\nonumber\\
&=&\frac{-\lambda_{0}^{2}\hbar}{2m_{0}^{2}(2\pi)^{2}}\sum_{j,j'=\pm}\frac{C_{j}C^{\ast}_{j'}}{w_{j}+w^{\ast}_{j'}}
e^{w_{j}(\tau-\tau_{0})}\cdot(2\pi
i)\cdot\nonumber\\
& &\left[\frac{a}{2\pi}\sum^{\infty}_{0}\left(\frac{w_{j}}{i n
a-iw_{j}}+\frac{w^{\ast}_{j'}}{i n
a+iw^{\ast}_{j'}}\right)e^{-na(\tau'-\tau_{0})}+\frac{w^{\ast}_{j'}e^{w^{\ast}_{j'}(\tau''-\tau_{0})}}{1-e^{2i\pi
w^{\ast}_{j'}}/a}\right]\nonumber\\
&=&\frac{-\lambda_{0}^{2}\hbar}{2m_{0}^{2}(2\pi)^{2}}\sum_{j,j'=\pm}\frac{C_{j}C^{\ast}_{j'}}{w_{j}+w^{\ast}_{j'}}
e^{w_{j}(\tau-\tau_{0})}\cdot\left[\frac{w_{j}e^{-a(\tau''-\tau_{0})}}{1-w_{j}/a}F_{-w_{j}}(e^{-a(\tau''-\tau_{0})})\right.\nonumber\\
& &\left.
+\frac{w^{\ast}_{j'}e^{-a(\tau''-\tau_{0})}}{1+w^{\ast}_{j'}/a}F_{w^{\ast}_{j'}}(e^{-a(\tau''-\tau_{0})})
+\frac{2\pi
iw^{\ast}_{j'}e^{w^{\ast}_{j'}(\tau''-\tau_{0})}}{1-e^{2i\pi
w^{\ast}_{j'}}/a}\right]\,.\label{accp2}
\end{eqnarray}

Since $e_{3}=e^{\ast}_{2}\mid_{\tau\leftrightarrow\tau',
\tau_{0}\leftrightarrow \tau'_{0}}$, we have
$P_{3}=P^{\ast}_{2}\mid_{\tau\leftrightarrow\tau',
\tau_{0}\leftrightarrow \tau'_{0}}$. And

\begin{eqnarray}
P_{4}&=&\frac{\lambda_{0}^{2}\hbar}{2m_{0}^{2}(2\pi)^{2}}\sum_{j,j'=\pm}\frac{C_{j}C^{\ast}_{j'}}{w_{j}+w^{\ast}_{j'}}
\cdot\left[\frac{w_{j}e^{-a(\tau_{0}-\tau''_{0})}}{1+w_{j}/a}F_{w_{j}}(e^{-a(\tau_{0}-\tau''_{0})})\right.\nonumber\\
& &\left.
+\frac{w^{\ast}_{j'}e^{-a(\tau_{0}-\tau''_{0})}}{1-w^{\ast}_{j'}/a}F_{-w^{\ast}_{j'}}(e^{-a(\tau_{0}-\tau''_{0})})
-\frac{2\pi iw_{j'}e^{w_{j}(\tau_{0}-\tau''_{0})}}{1-e^{-2i\pi
w_{j}}/a}\right]\,.\label{accp4}
\end{eqnarray}

we use the following formula to show our results:

\begin{equation}
\sum^{\infty}_{n=1}=\frac{e^{-nx}}{n+y}=\frac{e^{-x}}{1+y}\,\,
_2F_1(1+y,1,2+y,e^{-x})\equiv\frac{e^{-x}}{1+y}F_{ay}(e^{-x})\,.
\end{equation}

When we combine $P_{1}, P_{2}, P_{3}$ and $P_{4}$ and define
$\eta\equiv\tau-\tau_{0}$, $\eta''\equiv\tau''-\tau''_{0}$, the two
point function $\langle Q(\eta)Q(\eta'')\rangle_{v1}$ is

\begin{eqnarray}
\langle Q(\eta),Q(\eta'')\rangle_{v1}\equiv \frac{1}{2}\langle
Q(\eta)Q(\eta'')+Q(\eta'')Q(\eta)\rangle_{v1}=Re\{P_{1}+P_{2}+P_{3}+P_{4}\}
\end{eqnarray}

\begin{eqnarray}
& &\langle Q(\eta)^{2}\rangle_{v1}\equiv
\lim_{\eta''\rightarrow\eta}\frac{1}{2}\langle
\{Q(\eta),Q(\eta'')\}\rangle_{v1}=
\lim_{\eta''\rightarrow\eta}Re\{P_{1}+P_{2}+P_{3}+P_{4}\}\nonumber\\
&=&\frac{\hbar\gamma}{\pi
m_{0}\Omega^{2}}\theta(\eta)Re\{(\Lambda_{0}-\ln\frac{a}{\Omega})e^{-2\gamma\eta}\sin^{2}\Omega\eta\nonumber\\
&
&+\frac{a}{2}e^{-(\gamma+a)\eta}\left[\frac{F_{\gamma+i\Omega}(e^{-a\eta})}{\gamma+i\Omega+a}\left(\frac{-i\Omega}{\gamma}\right)e^{-i\Omega\eta}+
\frac{F_{-\gamma-i\Omega}(e^{-a\eta})}{\gamma+i\Omega-a}\left(\left(1+\frac{i\Omega}{\gamma}\right)e^{i\Omega\eta}-e^{-i\Omega\eta}\right)\right]\nonumber\\
&
&-\frac{1}{4}\left[\left(\frac{i\Omega}{\gamma}+e^{-2\gamma\eta}\left(\frac{i\Omega}{\gamma}+1-e^{-2i\Omega\eta}\right)\right)(\psi_{\gamma+i\Omega}+\psi_{-\gamma-i\Omega})\right.\nonumber\\
&
&\,\,\,\,\,\,\left.-\left(\frac{-i\Omega}{\gamma}+e^{-2\gamma\eta}\left(\frac{i\Omega}{\gamma}+1-e^{-2i\Omega\eta}\right)\right)i\pi\coth\frac{\pi}{a}(\Omega-i\gamma)\right]\}\,.
\end{eqnarray}

Here $\psi_{s}\equiv\psi(1+\frac{s}{a})$ and
$\Lambda_{0}\equiv-\gamma_{E}-\ln\Omega|\tau_{0}-\tau'_{0}|$ as
$\eta'\rightarrow\eta$

\begin{eqnarray}
& &-\langle QQ\rangle_{v2}\nonumber\\
&=&\frac{-\lambda_{0}^{2}\hbar}{2m_{0}^{2}(2\pi)^{2}}\sum_{j,j'=\pm}c_{j}c^{\ast}_{j'}\int_{-\infty}^{0}\frac{\kappa
d\kappa
e^{-i\kappa(\tau_{0}-\tau''_{0})}(e^{w_{j}(\tau-\tau_{0})}-e^{i\kappa(\tau_{0}-\tau)})(e^{w^{\ast}_{j'}(\tau''-\tau''_{0})}-e^{i\kappa(\tau''-\tau''_{0})})}
{(w_{j}+i\kappa)(w^{\ast}_{j'}-i\kappa)}\nonumber\\
&=&\frac{-\lambda_{0}^{2}\hbar}{2m_{0}^{2}(2\pi)^{2}}\sum_{j,j'=\pm}\frac{c_{j}c^{\ast}_{j'}}{w_{j}+w^{\ast}_{j'}}
\int_{-\infty}^{0}d\kappa\left(\frac{w_{j}}{\kappa-iw_{j}}+\frac{w^{\ast}_{j'}}{\kappa+iw^{\ast}_{j'}}\right)
(e_{1}+e_{2}+e_{3}+e_{4})\nonumber\\
&=&-\tilde{P}_{1}-\tilde{P}_{2}-\tilde{P}_{3}-\tilde{P}_{4}\,,
\end{eqnarray}

\begin{eqnarray}
\tilde{P}_{1}&=&\frac{\lambda_{0}^{2}\hbar}{2m_{0}^{2}(2\pi)^{2}}\sum_{j,j'=\pm}\frac{c_{j}c^{\ast}_{j'}}{w_{j}+w^{\ast}_{j'}}
\int_{-\infty}^{0}d\kappa\left(\frac{w_{j}}{\kappa-iw_{j}}+\frac{w^{\ast}_{j'}}{\kappa+iw^{\ast}_{j'}}\right)
e^{-i\kappa(\tau_{0}-\tau''_{0})+w_{j}(\tau-\tau_{0})+w^{\ast}_{j'}(\tau''-\tau''_{0})}\nonumber\\
&=&
\frac{\lambda_{0}^{2}\hbar}{2m_{0}^{2}(2\pi)^{2}}\left[\left(\frac{c_{+}c^{\ast}_{+}}{w_{+}+w^{\ast}_{+}}w_{+}e^{w_{+}(\tau-\tau''_{0})+w^{\ast}_{+}(\tau''-\tau''_{0})}+
\frac{c_{+}c^{\ast}_{-}}{w_{+}+w^{\ast}_{-}}w_{+}e^{w_{+}(\tau-\tau''_{0})+w^{\ast}_{-}(\tau''-\tau''_{0})}\right)\right.\nonumber\\
&
&\,\,\cdot\left(-i2\pi-\Gamma(0,w_{+}(\tau_{0}-\tau''_{0}))+\log(w_{+})+\log(\tau_{0}-\tau''_{0})-\log(w_{+}(\tau_{0}-\tau''_{0}))\right)\nonumber\\
&
&+\left(\frac{c_{-}c^{\ast}_{+}}{w_{-}+w^{\ast}_{+}}w_{-}e^{w_{-}(\tau-\tau''_{0})+w^{\ast}_{+}(\tau''-\tau''_{0})}+
\frac{c_{-}c^{\ast}_{-}}{w_{-}+w^{\ast}_{-}}w_{-}e^{w_{-}(\tau-\tau''_{0})+w^{\ast}_{-}(\tau''-\tau''_{0})}\right)\nonumber\\
&
&\,\,\cdot\left(-\Gamma(0,w_{-}(\tau_{0}-\tau''_{0}))+\log(w_{-})+\log(\tau_{0}-\tau''_{0})-\log(w_{-}(\tau_{0}-\tau''_{0}))\right)\nonumber\\
&
&-\left(\frac{c_{+}c^{\ast}_{+}}{w_{+}+w^{\ast}_{+}}w^{\ast}_{+}e^{w_{+}(\tau-\tau_{0})+w^{\ast}_{+}(\tau''-\tau_{0})}+
\frac{c_{-}c^{\ast}_{+}}{w_{-}+w^{\ast}_{+}}w^{\ast}_{+}e^{w_{-}(\tau-\tau_{0})+w^{\ast}_{+}(\tau''-\tau_{0})}\right)\nonumber\\
&
&\,\,\cdot\left(\Gamma(0,-w^{\ast}_{+}(\tau_{0}-\tau''_{0}))+\log(-1/w^{\ast}_{+})-\log(\tau_{0}-\tau''_{0})+\log(-w^{\ast}_{+}(\tau_{0}-\tau''_{0}))\right)\nonumber\\
&
&-\left(\frac{c_{+}c^{\ast}_{-}}{w_{+}+w^{\ast}_{-}}w^{\ast}_{-}e^{w_{+}(\tau-\tau_{0})+w^{\ast}_{-}(\tau''-\tau_{0})}+
\frac{c_{-}c^{\ast}_{-}}{w_{-}+w^{\ast}_{-}}w^{\ast}_{-}e^{w_{-}(\tau-\tau_{0})+w^{\ast}_{-}(\tau''-\tau_{0})}\right)\nonumber\\
&
&\cdot\left.\left(\Gamma(0,-w^{\ast}_{-}(\tau_{0}-\tau''_{0}))+\log(-1/w^{\ast}_{-})-\log(\tau_{0}-\tau''_{0})+\log(-w^{\ast}_{-}(\tau_{0}-\tau''_{0}))\right)\right]
\,,\label{accp1tilde}
\end{eqnarray}

\begin{eqnarray}
\tilde{P}_{2}&=&\frac{\lambda_{0}^{2}\hbar}{2m_{0}^{2}(2\pi)^{2}}\sum_{j,j'=\pm}\frac{c_{j}c^{\ast}_{j'}}{w_{j}+w^{\ast}_{j'}}
\int_{-\infty}^{0}d\kappa\left(\frac{w_{j}}{\kappa-iw_{j}}+\frac{w^{\ast}_{j'}}{\kappa+iw^{\ast}_{j'}}\right)
e_{2}\nonumber\\
&=& \frac{-\lambda_{0}^{2}\hbar}{2m_{0}^{2}(2\pi)^{2}}
\left[-\left(\frac{c_{+}c^{\ast}_{+}}{w_{+}+w^{\ast}_{+}}+\frac{c_{+}c^{\ast}_{-}}{w_{+}+w^{\ast}_{-}}\right)w_{+}
e^{w_{+}(\tau-\tau'')}\,\right.\nonumber\\
&
&\,\,\,\,\,\,\,\,\cdot\left(\Gamma(0,-w_{+}(\tau''-\tau_{0}))+\log(-1/w_{+})+\log(-w_{+}(\tau''-\tau_{0}))-\log(\tau''-\tau_{0})\right)\nonumber\\
&
&\,\,\,\,\,\,\,\,\,\,\,\,-\left(\frac{c_{-}c^{\ast}_{+}}{w_{-}+w^{\ast}_{+}}+\frac{c_{-}c^{\ast}_{-}}{w_{-}+w^{\ast}_{-}}\right)w_{-}
e^{w_{-}(\tau-\tau'')}\nonumber\\
&
&\,\,\,\,\,\,\,\,\,\cdot\left(\Gamma(0,-w_{-}(\tau''-\tau_{0}))+\log(-1/w_{-})+\log(-w_{-}(\tau''-\tau_{0}))-\log(\tau''-\tau_{0})\right)\nonumber\\
&
&\,\,\,\,\,\,\,\,\,+\left(\frac{c_{+}c^{\ast}_{+}}{w_{+}+w^{\ast}_{+}}w^{\ast}_{+}e^{w_{+}(\tau-\tau_{0})}
+\frac{c_{-}c^{\ast}_{+}}{w_{-}+w^{\ast}_{+}}w^{\ast}_{+}e^{w_{-}(\tau-\tau_{0})}\right)e^{w^{\ast}_{+}(\tau''-\tau_{0})}\nonumber\\
& &\,\,\,\,\,\,\,\,\,\cdot\left(2\pi
i-\Gamma(0,w^{\ast}_{+}(\tau''-\tau_{0}))+\log(w^{\ast}_{+})-\log(w^{\ast}_{+}(\tau''-\tau_{0}))+\log(\tau''-\tau_{0})\right)\nonumber\\
&
&\,\,\,\,\,\,\,\,\,-\left(\frac{c_{+}c^{\ast}_{-}}{w_{+}+w^{\ast}_{-}}w^{\ast}_{-}e^{w_{+}(\tau-\tau_{0})}
+\frac{c_{-}c^{\ast}_{-}}{w_{-}+w^{\ast}_{-}}w^{\ast}_{-}e^{w_{-}(\tau-\tau_{0})}\right)e^{w^{\ast}_{-}(\tau''-\tau_{0})}\nonumber\\
&
&\,\,\,\,\,\,\,\,\left.\cdot\left(\Gamma(0,w^{\ast}_{-}(\tau''-\tau_{0}))-\log(w^{\ast}_{-})+\log(w^{\ast}_{-}(\tau''-\tau_{0}))-\log(\tau''-\tau_{0})\right)
\right]\,,\label{accp2tilde}
\end{eqnarray}

\begin{eqnarray}
\tilde{P}_{3}&=&\frac{\lambda_{0}^{2}\hbar}{2m_{0}^{2}(2\pi)^{2}}\sum_{j,j'=\pm}\frac{c_{j}c^{\ast}_{j'}}{w_{j}+w^{\ast}_{j'}}
\int_{-\infty}^{0}d\kappa\left(\frac{w_{j}}{\kappa-iw_{j}}+\frac{w^{\ast}_{j'}}{\kappa+iw^{\ast}_{j'}}\right)
e_{3}\nonumber\\
&=& \frac{-\lambda_{0}^{2}\hbar}{2m_{0}^{2}(2\pi)^{2}}
\left[-e^{w_{+}(\tau-\tau''_{0})}\left(\frac{c_{+}c^{\ast}_{+}}{w_{+}+w^{\ast}_{+}}w_{+}e^{w^{\ast}_{+}(\tau''-\tau''_{0})}
+\frac{c_{+}c^{\ast}_{-}}{w_{+}+w^{\ast}_{-}}w_{+}e^{w^{\ast}_{-}(\tau''-\tau''_{0})}\right)\right.\nonumber\\
& &\,\,\,\,\,\,\,\,\,\cdot\left(-2\pi
i+\Gamma(0,w_{+}(\tau-\tau''_{0}))-\log(w_{+})-\log(\tau-\tau''_{0})+\log(w_{+}(\tau-\tau''_{0}))\right)\nonumber\\
&
&\,\,\,\,\,\,\,\,\,-e^{w_{-}(\tau-\tau''_{0})}\left(\frac{c_{-}c^{\ast}_{+}}{w_{-}+w^{\ast}_{+}}w_{-}e^{w^{\ast}_{+}(\tau''-\tau''_{0})}
+\frac{c_{-}c^{\ast}_{-}}{w_{-}+w^{\ast}_{-}}w_{-}e^{w^{\ast}_{-}(\tau''-\tau''_{0})}\right)\nonumber\\
&
&\,\,\,\,\,\,\,\,\,\cdot\left(\Gamma(0,w_{-}(\tau-\tau''_{0}))-\log(w_{-})-\log(\tau-\tau''_{0})+\log(w_{-}(\tau-\tau''_{0}))\right)\nonumber\\
&
&\,\,\,\,\,\,\,\,-\left(\frac{c_{+}c^{\ast}_{+}}{w_{+}+w^{\ast}_{+}}w^{\ast}_{+}e^{w^{\ast}_{+}(\tau''-\tau)}
+\frac{c_{-}c^{\ast}_{+}}{w_{-}+w^{\ast}_{+}}w^{\ast}_{+}e^{w^{\ast}_{+}(\tau''-\tau)}\right)\nonumber\\
&
&\cdot\left(\Gamma(0,-w^{\ast}_{+}(\tau-\tau''_{0}))+\log(-1/w^{\ast}_{+})-\log(\tau-\tau''_{0})+\log(-w^{\ast}_{+}(\tau-\tau''_{0}))\right)
\nonumber\\&
&\,\,\,\,\,\,\,\,\,-\left(\frac{c_{+}c^{\ast}_{-}}{w_{+}+w^{\ast}_{-}}w^{\ast}_{-}e^{w^{\ast}_{-}(\tau''-\tau)}
+\frac{c_{-}c^{\ast}_{-}}{w_{-}+w^{\ast}_{-}}w^{\ast}_{-}e^{w^{\ast}_{-}(\tau''-\tau)}\right)\nonumber\\
&
&\,\,\,\,\,\,\,\,\,\left.\cdot\left(\Gamma(0,-w^{\ast}_{-}(\tau-\tau''_{0}))+\log(-1/w^{\ast}_{-})-\log(\tau-\tau''_{0})+\log(-w^{\ast}_{-}(\tau-\tau''_{0}))\right)
\right]\,,\label{accp3tilde}
\end{eqnarray}

\begin{eqnarray}
\tilde{P}_{4}&=&\frac{\lambda_{0}^{2}\hbar}{2m_{0}^{2}(2\pi)^{2}}\sum_{j,j'=\pm}\frac{c_{j}c^{\ast}_{j'}}{w_{j}+w^{\ast}_{j'}}
\int_{-\infty}^{0}d\kappa\left(\frac{w_{j}}{\kappa-iw_{j}}+\frac{w^{\ast}_{j'}}{\kappa+iw^{\ast}_{j'}}\right)
e_{4}\nonumber\\
&=& \frac{\lambda_{0}^{2}\hbar}{2m_{0}^{2}(2\pi)^{2}}
\left[-\left(\frac{c_{+}c^{\ast}_{+}}{w_{+}+w^{\ast}_{+}}
+\frac{c_{+}c^{\ast}_{-}}{w_{+}+w^{\ast}_{-}}\right)w_{+}e^{-w_{+}(\tau''-\tau)}\right.\nonumber\\
& &\,\,\,\,\,\,\,\,\,\cdot\left(\Gamma(0,-w_{+}(\tau''-\tau))+\log(-1/w_{+})-\log(\tau''-\tau)+\log(-w_{+}(\tau''-\tau))\right)\nonumber\\
&
&\,\,\,\,\,\,\,\,\,-\left(\frac{c_{-}c^{\ast}_{+}}{w_{-}+w^{\ast}_{+}}
+\frac{c_{-}c^{\ast}_{-}}{w_{-}+w^{\ast}_{-}}\right)w_{-}e^{-w_{-}(\tau''-\tau)}\nonumber\\
&
&\,\,\,\,\,\,\,\,\,\cdot\left(\Gamma(0,-w_{-}(\tau''-\tau))+\log(-1/w_{-})-\log(\tau''-\tau)+\log(-w_{-}(\tau''-\tau))\right)\nonumber\\
&
&\,\,\,\,\,\,\,\,\,+\left(\frac{c_{+}c^{\ast}_{+}}{w_{+}+w^{\ast}_{+}}
+\frac{c_{-}c^{\ast}_{+}}{w_{-}+w^{\ast}_{+}}\right)w^{\ast}_{+}e^{w^{\ast}_{+}(\tau''-\tau)}\nonumber\\
& &\,\,\,\,\,\,\,\,\,\cdot\left(2\pi
i-\Gamma(0,w^{\ast}_{+}(\tau''-\tau))+\log(w^{\ast}_{+})+\log(\tau''-\tau)-\log(w^{\ast}_{+}(\tau''-\tau))\right)\nonumber\\
&
&\,\,\,\,\,\,\,\,\,-\left(\frac{c_{+}c^{\ast}_{-}}{w_{+}+w^{\ast}_{-}}
+\frac{c_{-}c^{\ast}_{-}}{w_{-}+w^{\ast}_{-}}\right)w^{\ast}_{-}e^{w^{\ast}_{-}(\tau''-\tau)}\nonumber\\
&
&\,\,\,\,\,\,\,\,\,\left.\cdot\left(\Gamma(0,w^{\ast}_{-}(\tau''-\tau))-\log(w^{\ast}_{-})-\log(\tau''-\tau)+\log(w^{\ast}_{-}(\tau''-\tau))\right)
\right]\,.\label{accp4tilde}
\end{eqnarray}

As $\eta\rightarrow\eta''$(that is, $\tau''\rightarrow\tau$ and
$\tau''_{0}\rightarrow\tau_{0}$),

\begin{eqnarray}
& &-\langle
Q^{2}(\eta)\rangle_{v2}=-\lim_{\eta''\rightarrow\eta}\langle
\{Q(\eta),Q(\eta'')\}\rangle_{v2}
=-\lim_{\eta''\rightarrow\eta}Re\{\tilde{P_{1}}+\tilde{P_{2}}+\tilde{P_{3}}+\tilde{P_{4}}\}\nonumber\\
&=&\frac{\lambda_{0}^{2}\hbar}{2m_{0}^{2}(2\pi)^{2}}\theta(\eta)Re\{\Lambda_{0_{v2}}-
\left(\frac{\gamma-i\Omega}{8\Omega^{2}\gamma}e^{-2\gamma(\tau-\tau_{0})}-\frac{1}{8\Omega^{2}}e^{-2\gamma(\tau-\tau_{0})+2i\Omega(\tau-\tau_{0})}\right)
\nonumber\\&
&\cdot\left(-2i\pi+\log(-\gamma+i\Omega)-\log(\frac{1}{\gamma-i\Omega})\right)
-\left(\frac{\gamma+i\Omega}{8\Omega^{2}\gamma}e^{-2\gamma(\tau-\tau_{0})}-\frac{1}{8\Omega^{2}}e^{-2\gamma(\tau-\tau_{0})-2i\Omega(\tau-\tau_{0})}\right)
\nonumber\\&
&\cdot\left(\log(-\gamma-i\Omega)+\log(\frac{1}{\gamma+i\Omega})\right)+\left(\frac{-\gamma+i\Omega}{8\Omega^{2}\gamma}+\frac{1}{8\Omega^{2}}\right)
\cdot\left[\Gamma(0,(\gamma-i\Omega)(\tau-\tau_{0}))\right.\nonumber\\
&
&\left.-\log(\gamma-i\Omega)-\log(\tau-\tau_{0})+\log((\gamma-i\Omega)(\tau-\tau_{0}))\right]
+\left(\frac{-\gamma-i\Omega}{8\Omega^{2}\gamma}+\frac{1}{8\Omega^{2}}\right)\nonumber\\
&
&\,\,\cdot\left[\Gamma(0,(\gamma+i\Omega)(\tau-\tau_{0}))-\log(\gamma+i\Omega)-\log(\tau-\tau_{0})+\log((\gamma+i\Omega)(\tau-\tau_{0}))\right]
\nonumber\\
&
&+\left(\frac{\gamma+i\Omega}{8\Omega^{2}\gamma}e^{-2\gamma(\tau-\tau_{0})}-\frac{1}{8\Omega^{2}}e^{-2(\gamma+i\Omega)(\tau-\tau_{0})}\right)
\cdot\left[2\pi
i-\Gamma(0,(-\gamma-i\Omega)(\tau-\tau_{0}))+\log(-\gamma-i\Omega)\right.\nonumber\\
&
&\left.+\log(\tau-\tau_{0})-\log((-\gamma-i\Omega)(\tau-\tau_{0}))\right]
+\left(\frac{-\gamma+i\Omega}{8\Omega^{2}\gamma}e^{-2\gamma(\tau-\tau_{0})}+\frac{1}{8\Omega^{2}}e^{2(-\gamma+i\Omega)(\tau-\tau_{0})}\right)
\nonumber\\
&
&\cdot\left[\Gamma(0,(-\gamma+i\Omega)(\tau-\tau_{0}))-\log(-\gamma+i\Omega)-\log(\tau-\tau_{0})+\log((-\gamma+i\Omega)(\tau-\tau_{0}))\right]
\nonumber\\&
&+\left(\frac{-\gamma+i\Omega}{8\Omega^{2}\gamma}e^{-2\gamma(\tau-\tau_{0})}+\frac{1}{8\Omega^{2}}e^{2(-\gamma+i\Omega)(\tau-\tau_{0})}\right)
\cdot\left[2\pi
i+\Gamma(0,(-\gamma+i\Omega)(\tau-\tau_{0}))-\log(-\gamma+i\Omega)\right.\nonumber\\
&
&\left.-\log(\tau-\tau_{0})+\log((-\gamma+i\Omega)(\tau-\tau_{0}))\right]+
\left(\frac{-\gamma-i\Omega}{8\Omega^{2}\gamma}e^{-2\gamma(\tau-\tau_{0})}+\frac{1}{8\Omega^{2}}e^{-2(\gamma+i\Omega)(\tau-\tau_{0})}\right)
\nonumber\\&
&\cdot\left[\Gamma(0,(-\gamma-i\Omega)(\tau-\tau_{0}))-\log(-\gamma-i\Omega)-\log(\tau-\tau_{0})+\log((-\gamma-i\Omega)(\tau-\tau_{0}))\right]
\nonumber\\
&
&+\frac{i}{8\Omega\gamma}\cdot\left[-\Gamma(0,(\gamma+i\Omega)(\tau-\tau_{0}))+\log(\gamma+i\Omega)+\log(\tau-\tau_{0})
-\log((\gamma+i\Omega)(\tau-\tau_{0}))\right.\nonumber\\
&
&\left.+\Gamma(0,(\gamma-i\Omega)(\tau-\tau_{0}))-\log(\gamma-i\Omega)-\log(\tau-\tau_{0})
+\log((\gamma-i\Omega)(\tau-\tau_{0}))\right]\nonumber\\
&
&+\left(\frac{1}{8\Omega^{2}}-\frac{(\gamma-i\Omega)}{8\Omega^{2}\gamma}\right)\cdot\left(-\log(\gamma-i\Omega)-\log(-\gamma-i\Omega)\right)
+\left(\frac{1}{8\Omega^{2}}-\frac{(\gamma+i\Omega)}{8\Omega^{2}\gamma}\right)\nonumber\\
& &\cdot\left(-2\log(\gamma+i\Omega)-2\pi i\right)\}\,,
\end{eqnarray}

where $\Lambda_{0_{v2}}$ contains the divergent parts $\Gamma(0,0)$
and $\log(0)$ as $\tau''\rightarrow\tau$ and
$\tau''_{0}\rightarrow\tau_{0}$ and is absorbed into the
renormalized constant or coefficient in the experiment.

%%%%%%%%%%%%%%%%%%%%%%%%%%%%%%%%%%%%%%%%%%%%%%%%%%%%%%%%%%%%%%%%%%%%%%%%%%%%%%%%%%%%%%

\end{document}